\documentclass[aps, prc, tightenlines, letterpaper, amsmath, amssymb, preprintnumbers, floatfix, longbibliography, nofootinbib]{revtex4-2}
\usepackage{latexsym,amsfonts,bm}
\usepackage{xspace}
\usepackage{cancel}
\usepackage{graphicx}    
\usepackage{microtype}   
\usepackage{amsmath}     
\usepackage{amsthm}    
\usepackage{amssymb}   
\usepackage{siunitx}     
\usepackage{booktabs}    
\usepackage{hyperref}
\hypersetup{colorlinks=true,citecolor=blue}
\usepackage[capitalize]{cleveref}
\usepackage{tensor}

\usepackage{adjustbox}    
\usepackage{subcaption}
\usepackage{xcolor}
\usepackage{mathtools} 
\usepackage{bm}
\usepackage{dsfont}
\usepackage[shortlabels]{enumitem}  
\usepackage{physics}   

\providecommand{\ie}{\emph{i.e.}}
\providecommand{\eg}{\emph{e.g.}}

\providecommand{\CA}{\mathcal{A}}
\providecommand{\CC}{\mathcal{C}}
\providecommand{\CE}{\mathcal{E}}

\providecommand{\CO}{\mathcal{O}}

\providecommand{\CD}{\mathcal{D}}

\providecommand{\nn}{\nonumber}

\providecommand{\sutwo}{$\mathrm{SU}(2)$\xspace}
\providecommand{\suthree}{$\mathrm{SU}(3)$\xspace}

\providecommand{\g}{{\mathfrak{g}}}
\providecommand{\h}{{\mathfrak{h}}}

\providecommand{\sutwo}{$\mathrm{SU}(2)$\xspace}

\providecommand{\df}{{\rm d}}

\newcommand{\NL}{{n_\kappa}} 

\newcommand{\LOri}[1]{{L_{#1} }}
\newcommand{\MOri}[1]{{M_{#1} }}
\newcommand{\NSeq}{n }
\newcommand{\LSeq}[1]{{\ell_{#1} }}
\newcommand{\MSeq}[1]{{m_{#1} }}
\newcommand{\LTot}{L}
\newcommand{\MTot}{M}
\newcommand{\NTot}{N}

\newcommand{\LOp}{{\hat{\mathbf{L}}}}
\newcommand{\LOpSq}{{\hat{L}^2}}
\newcommand{\LOpComp}[1]{{\hat{L}^{#1}}}
\newcommand{\LTotOp}{{\hat{\mathbf{L}}_\text{Tot}}}
\newcommand{\LpTotOp}{{\hat{\mathbf{L}}'_\text{Tot}}}
\newcommand{\LTotOpSq}{{\hat{L}_\text{Tot}^2}}
\newcommand{\LpTotOpSq}{{\hat{L}'^2_\text{Tot}}}
\newcommand{\LTotOpComp}[1]{{\hat{L}_\text{Tot}^{#1}}}
\newcommand{\LpTotOpComp}[1]{{\hat{L}'^{#1}_\text{Tot}}}
\newcommand{\LSeqOp}{{\hat{\mathbf{\Lambda}}}}
\newcommand{\LSeqOpSq}{{\hat{\Lambda}^2}}
\newcommand{\LSeqOpComp}[1]{{\hat{\Lambda}^{#1}}}
\newcommand{\LSigmaOp}{{\hat{\boldsymbol{\Lambda}}_{\sigma}}}
\newcommand{\LSigmaOpComp}[1]{{\hat{\Lambda}_\sigma^{#1}}}
\newcommand{\LSigmaOpSq}{{\hat{\Lambda}^2_{\sigma}}}

\newcommand{\NSeqSq}{\hat{\cal N}^2}

\newcommand{\thetaDef}{\Theta}
\newcommand{\thetaOri}[1]{\vartheta_{#1}}
\newcommand{\phiOri}[1]{\varphi_{#1}}
\newcommand{\thetaSeq}[1]{\theta_{#1}}
\newcommand{\phiSeq}[1]{\phi_{#1}}

\newcommand{\RotE}{{\mathbf{R}}}
\newcommand{\RotEIndex}[2]{R^{#1 #2}}

\newcommand{\nSeqBody}[1]{{\boldsymbol{\eta}_{#1}}}

\newcommand{\ALPSymbol}{{\cal P}}

\newcommand{\WDSymbol}{{\cal D}}
\newcommand{\WD}{\WDSymbol^{\LTot}_{\MTot \NTot}(\alpha, \beta, \gamma)}
\newcommand{\WDpDag}{\WDSymbol^{\LTot'}_{\MTot' \NTot'}(\alpha, \beta, \gamma)}
\newcommand{\WDDagP}{\WDSymbol^{\LTot}_{\MTot \NTot}(\alpha', \beta', \gamma')}

\newcommand{\YLMSymbol}{{Y}}
\newcommand{\YLM}[1]{\YLMSymbol_{\LSeq{#1} \MSeq{#1}}(\thetaSeq{\mu}, \phiSeq{\mu})}

\newcommand{\OpCoeff}[4]{{\CC^{#1,#2}_{#3, #4}}}
\newcommand{\cas}[1]{{\CC_{r}(#1)}}

\newcommand{\cotw}[1]{{\,\cot\left(\frac{\omega_{#1}}{2}\right)}\,}
\newcommand{\crossvL}[2]{\left(\nSeqBody{#1}\times \LSeqOp_{#1}\right)}
\begin{document}

\title{\boldmath A Fully Gauge-Fixed  \texorpdfstring{\sutwo}{SU(2)} Hamiltonian for Quantum Simulations}

\author{Dorota M. Grabowska}
\email{grabow@uw.edu}
\affiliation{InQubator for Quantum Simulation (IQuS), Department of Physics, University of Washington, Seattle, WA 98195}

\author{Christopher F. Kane}
\email{cfkane24@umd.edu}
\affiliation{Department of Physics, University of Arizona, Tucson, AZ 85719, US}
\affiliation{Maryland Center for Fundamental Physics and Department of Physics, University of Maryland, College Park, MD 20742, USA}

\author{Christian W. Bauer}
\email{cwbauer@lbl.gov}
\affiliation{Physics Division, Lawrence Berkeley National Laboratory, Berkeley, CA 94720, USA}
\affiliation{Department of Physics, University of California, Berkeley, Berkeley, CA 94720}

\preprint{IQuS@UW-21-087}
\date{\today}

\begin{abstract}
We demonstrate how to construct a fully gauge-fixed lattice Hamiltonian for a pure SU(2) gauge theory. 
Our work extends upon previous work, where a formulation of an SU(2) lattice gauge theory was developed that is efficient to simulate at all values of the gauge coupling.
That formulation utilized maximal-tree gauge, where all local gauge symmetries are fixed and a residual global gauge symmetry remains.
By using the geometric picture of an SU(2) lattice gauge theory as a system of rotating rods, we demonstrate how to fix the remaining global gauge symmetry.
In particular, the quantum numbers associated with total charge can be isolated by rotating between the lab and body frames using the three Euler angles.
The Hilbert space in this new `sequestered' basis partitions cleanly into sectors with differing total angular momentum, which makes gauge-fixing to a particular total charge sector trivial, particularly for the charge-zero sector.  
In addition to this sequestered basis inheriting the property of being efficient at all values of the coupling, we show that, despite the global nature of the final gauge-fixing procedure, this Hamiltonian can be simulated using quantum resources scaling only polynomially with the lattice volume.
\end{abstract}

\maketitle

\tableofcontents

\section{Introduction}

Quantum computers hold the promise of enabling first-principles simulations of the non-perturbative phenomena of quantum field theories, including real-time dynamics and systems with non-zero chemical potential, that are intractable on classical computers.
One class of theories that are of particular physical relevance are gauge theories, which describe condensed matter systems, nuclear physics, and interactions between fundamental particles in the Standard Model of particle physics.
Since the original proposal by Jordan, Lee and Preskil~\cite{Jordan:2012xnu}, there has been dramatic progress in simulating gauge theories (for recent reviews, see~\cite{Bauer:2022hpo, Davoudi:2022bnl, DiMeglio:2023nsa, Bauer:2023qgm,Watson:2023oov}).  
After early pioneering work~\cite{Byrnes:2005qx}, a range of different gauge theories have been considered, such as Abelian $Z_2$~\cite{Mildenberger:2022jqr,Pardo:2022hrp}, and 
${\rm U}(1)$~\cite{Banerjee:2012pg,Hauke:2013jga,Zohar:2013zla,Kuhn:2014rha,Kasper:2015cca,Zohar:2015hwa,Martinez:2016yna,Yang:2016hjn,Kokail:2018eiw,Klco:2018kyo,Lu:2018pjk,Kaplan:2018vnj,Mil:2019pbt,Davoudi:2019bhy,Surace:2019dtp,Haase:2020kaj,Luo:2019vmi,Shaw:2020udc,Yang:2020yer,Ott:2020ycj,Paulson:2020zjd,Nguyen:2021hyk,Zhou:2021kdl,Riechert:2021ink,Bauer:2021gek,Kane:2022ejm,Grabowska:2022uos,zhang2023observation,Farrell:2023fgd,Nagano:2023uaq}, theories, as well as 
non-Abelian ${\rm SU}(2)$~\cite{Zohar:2012xf,Stannigel:2013zka,Mezzacapo:2015bra,Raychowdhury:2018osk,Raychowdhury:2019iki,Klco:2019evd,Dasgupta:2020itb,Davoudi:2020yln,Atas:2021ext,ARahman:2021ktn,Osborne:2022jxq,halimeh2022gauge,ARahman:2022tkr,zache2023quantum,Alexandru:2023qzd,DAndrea:2023qnr,Turro:2024pxu} 
and ${\rm SU}(3)$~\cite{Anishetty:2009nh,Alexandru:2019nsa,Ciavarella:2021nmj,Farrell:2022wyt,Farrell:2022vyh,Atas:2022dqm,Ciavarella:2021lel,Ciavarella:2023mfc,hayata2023qdeformedformulationhamiltonian,Farrell:2024fit,Ciavarella:2024fzw}.
There has also been progress on tensor network simulations~\cite{Pichler:2015yqa,Banuls:2017ena,Banuls:2018jag,Banuls:2022vxp}, 
algorithmic studies~\cite{Kan:2021xfc,Alexandru:2022son,tong2022provably,Davoudi:2022xmb,Kane:2023jdo,Hariprakash:2023tla,Rhodes:2024zbr,Du:2024ixj,Li:2024lrl,Kane:2024odt}, formulations of 
quantum parton showers~\cite{Bauer:2019qxa,Bauer:2023ujy,Chigusa:2022act} 
and more~\cite{Tagliacozzo:2012df,Bazavov:2015kka,Jordan:2017lea,Gonzalez-Cuadra:2017lvz,Gorg:2018xyc,Lamm:2019bik,Zohar:2019ygc,Buser:2020cvn,Barata:2020jtq,Stryker:2021asy,kreshchuk2021light,kreshchuk2022quantum,kreshchuk2023simulatingscatteringcompositeparticles}. 
However, further theoretical developments are imperative to fully utilize future quantum devices.

Constructing a Hamiltonian suitable for quantum simulations requires two largely independent steps. 
One step is to regulate the theory by constructing a Hamiltonian on a spatial lattice.
The standard Lattice Gauge Theory (LGT) Hamiltonian is the Kogut-Susskind Hamiltonian with staggered fermions~\cite{PhysRevD.11.395}, but improved Hamiltonians with reduced discretization errors~\cite{Carena:2022kpg, Gustafson:2023aai}, alternative fermionic discretizations~\cite{Zache:2018jbt, Creutz:2001wp, Hayata:2023zuk}, triamond lattices~\cite{Kavaki:2024ijd},
and light-front formulations~\cite{kreshchuk2021light, kreshchuk2022quantum, kreshchuk2023simulatingscatteringcompositeparticles, PhysRevD.109.076004, PhysRevD.106.074013} have also been proposed and studied. 
The second step is to choose a formulation of the theory, which requires selecting a basis to represent the operators, and a truncation scheme to \emph{digitize} the infinite dimensional bosonic Hilbert space.
To avoid the possibility of charge-violating transitions, any digitization must be gauge-invariant.
Additionally, in the interest of performing calculations close to the continuum, the ideal digitization should be efficient at small lattice spacings. 
For lower-dimensional gauge theories as well as theories with asymptotic freedom including non-Abelian theories like quantum chromodynamics (QCD), this corresponds to the limit of weak gauge coupling $g$. 
It turns out, however, that constructing a digitization of non-Abelian gauge theories that is simultaneously gauge-invariant, efficient at weak coupling, and systematically improvable is a difficult task.

Constructing a Hamiltonian requires at least partial gauge fixing, and LGT Hamiltonians are generally expressed in the temporal gauge.
Written as a sum of an electric Hamiltonian $H_E$ and a magnetic Hamiltonian $H_B$, digitization methods typically work in the electric basis (where $H_E$ is diagonal), or in the magnetic basis (where $H_B$ is diagonal).
In the standard convention, $H_E \sim g^2$ and $H_B \sim 1/g^2$, which implies that simulations near the continuum are more efficient in the magnetic basis.
Despite becoming more expensive as one approaches the continuum, electric-basis formulations have the advantage that, even in the partially gauge-fixed temporal gauge, they can be digitized in a gauge-invariant, systematically improvable way by simply placing a cutoff on the maximum value of the electric field. Electric basis approached have been used in recent formulations of non-Abelian gauge theories~\cite{Mathur:2004kr, mathur2006loop, mathur2007loop, Raychowdhury:2018osk, Raychowdhury:2019iki, Kadam:2022ipf, Kadam:2024zkj, Ciavarella:2021nmj, Klco:2019evd, Davoudi:2020yln, Byrnes:2005qx, Anishetty:2009nh, Banuls:2017ena}.

Magnetic basis digitizations in the temporal gauge, however, require additional care. 
In particular, one must ensure that digitizing the continuous group elements does not lead to Gauss law violations.
One such method is to work with discrete subgroups of the continuous group. 
While this poses no limitation for Abelian groups, for non-Abelian groups, which have been studied in Refs.~\cite{Alam:2021uuq, Alexandru:2021jpm, Gustafson:2022xdt, Gustafson:2023kvd, Gustafson:2024kym, Assi:2024pdn, Lamm:2024jnl, Muarari:2024dqx}, there exists only a finite number of discrete subgroups for a given non-Abelian group.
As the bare coupling of the theory tends to zero, the lowest lying states of the theory get localized close to the identity of the gauge group; discrete subgroups are unable to adequately sample densely enough near the identity, which places a theoretical limit on what values of the lattice spacing can be simulated, see, \eg, Ref.~\cite{Gustafson:2023kvd}. 
Despite the difficulties of using electric bases and discrete subgroup magnetic bases near the continuum, these methods are likely useful tools for early quantum simulations, as resource limitations will dictate the need to use relatively large lattice spacings.
However, as quantum computers improve and precision calculations requiring state of the art values of the lattice spacing become feasible, it could become difficult to extract precise results using these methods.

The limitations in the regime of small lattice spacings can be overcome, at least in principle, by gauge-fixing.
In a gauge-fixed theory, the Hilbert space is restricted to the physical charge sector and digitizing can no longer lead to charge-violating transitions.
This implies one is no longer limited to using discrete subgroups in the magnetic basis.
For U(1) gauge theories, such a gauge-fixing procedure has led to dual basis formulations~\cite{Kaplan:2018vnj, Bender:2020ztu, Haase:2020kaj}, which can be simulated efficiently at all values of the gauge coupling~\cite{Bauer:2021gek}.
While applying the same techniques to non-Abelian theories is not possible due to the non-Abelian Gauss law, magnetic-basis formulations of SU(2) gauge theory have been developed~\cite{DAndrea:2023qnr, Romiti:2023hbd}.
The formulation in Ref.~\cite{DAndrea:2023qnr} works in the maximal-tree gauge, which enables a gauge-invariant digitization that is efficient at all values of the coupling. This is achieved by working in a mixed-basis representation; efficient variational state preparation techniques for this formulation have been studied in Ref.~\cite{Fontana:2024rux}.
The formulation in Ref.~\cite{Romiti:2023hbd} uses a similar approach to Ref.~\cite{DAndrea:2023qnr}, except that, because no gauge fixing was performed, gauge-invariance is recovered only in the limit of removing the digitization; the ramifications of this residual gauge-violation are currently being explored~\cite{Garofalo:2023zkd}.

While the formulation in Ref.~\cite{DAndrea:2023qnr} is both efficient at all values of the coupling and systematically improvable, there is a residual global gauge-symmetry, with different gauge sectors corresponding to states with different total charge.
This total charge can be thought of as total angular momentum, due to $\text{S}^3$ being diffeomorphic to \sutwo.
Preparing states with definite angular momentum, while possible in principle, involves the highly non-trivial task of dealing with triangle inequalities that arise when adding the angular momentum at each site. 
Alternatively, one could project onto the physical gauge-sector using the methods in Refs~\cite{Dong:2022mmq, Kane:2023jdo}, albeit with a potentially large cost due to the relatively small number of physical states in the full Hilbert space~\cite{Carena:2024dzu}.
These difficulties with preparing initial states in the physical gauge-sector, however, could be avoided by fully gauge-fixing and removing the remaining global gauge symmetry.

In this paper, we extend upon the work in Ref.~\cite{DAndrea:2023qnr} and present a fully gauge-fixed formulation of an SU(2) LGT Hamiltonian, using the so-called `sequestered-basis'.
By using the geometric picture of an SU(2) LGT as a system of rotating rods, the quantum numbers associated with total angular momentum can be isolated by rotating between the lab and body frames, which partitions the Hilbert space cleanly into sectors with total angular momentum.
Crucially, we show that, despite the global nature of the final gauge-fixing procedure, the degree of coupling (DoC) of this Hamiltonian, defined as the maximum number of sites a given operator acts on, is independent of the volume; this fact implies that this Hamiltonian can be simulated using quantum resources that grow only polynomially with the volume.

The rest of this work is organized as follows.
In \cref{sec:review}, we review the work in Ref.~\cite{DAndrea:2023qnr}, including a review of the maximal-tree gauge fixing procedure and the mixed-basis formulation.
Next, in \cref{sec:seq_basis} we discuss the construction of the sequestered magnetic basis, which is a reparameterization of the basis originally constructed in terms of axis-angle coordinates. In this section, and related appendices, we construct the form of all possible operators that can appear in the electric and magnetic Hamiltonians, regardless of the maximal-tree gauge fixing convention that is chosen. In \cref{sec:mixed}, and related appendices, we demonstrate how to change into the mixed basis, from the magnetic basis, using the sequestered basis results; this is analogous to what was done Ref.~\cite{DAndrea:2023qnr}. Lastly, in \cref{sec:ResourceScaling}, we discuss the resource scaling for simulating this Hamiltonian on digital quantum devices. We conclude in \cref{sec:Conclusions}.

\section{Review of Gauge Theories in Maximal-Tree Gauge \label{sec:review}}

In this section we provide a brief discussion of the formulation of gauge theories in maximal-tree gauge fixing.
While we keep the discussion agnostic to the gauge group at the beginning, several details are specific to \sutwo gauge theories. 
This section serves as a review of Ref.~\cite{DAndrea:2023qnr} in which the formulation of \sutwo in the mixed basis was worked out. 
The goal of this section is to only introduce the basic points required for this paper, and for a more pedagogical introduction we refer the reader to Ref.~\cite{DAndrea:2023qnr} and references therein.

\subsection{Hamiltonian in maximal-tree gauge}
The Hilbert space of a pure Yang-Mills LGT is given as the tensor product of the Hilbert spaces for each link on the lattice.
The Hilbert space at a single link is given by the space of square-integrable wave functions, $L^2(G,\dd{\g})$, over the Lie group $G$ with respect to the Haar measure $\dd{\g}$.
This single-link Hilbert space is spanned by the eigenstates $\ket{\g_\ell}$ of the link operators
\begin{equation}
\hat U_{mm'}(\ell) \equiv \left[\mathcal{P}\exp( i\int_{n}^{n+e_i} \!\!\dd{x^\mu} A_{\mu}(x) )\right]_{mm'}
\,,
\end{equation}
where the link $\ell \equiv \ell(n, e_i)$ originates at lattice site $n$ and goes in direction $e_i$, ending at site $n + e_i$. 
The eigenvalues are the matrix of the group element $\g_\ell$ in the fundamental represenation $u_{mm'}(\g_\ell)$, such that one can write
\begin{equation}
\label{eq:def_link_op}
\hat{U}_{mm'}(\ell) \equiv \int\!\dd{\g_\ell} u_{mm'}(\g_\ell) \op{\g_\ell} 
\,.
\end{equation}
One can also define link operators in different representations similarly as 
\begin{equation}
\label{eq:def_link_op_gen}
\hat{U}^{(r)}_{mm'}(\ell) \equiv \int\!\dd{\g_\ell} u^{(r)}_{mm'}(\g_\ell) \op{\g_\ell} 
\,,
\end{equation}
where $u^{(r)}_{mm'}(\g_\ell)$ is the matrix of the group element $\g$ in the representation $r$. 
The Hilbert space for the full lattice containing $n_\ell$ links is then spanned by the states
\begin{equation}
\label{eq:generalState}
    \ket{\g_1, \ldots, \g_{n_\ell}} = \ket{\g_1}\otimes \cdots \otimes \ket{\g_{n_\ell}}
    \,.
\end{equation}

A formulation of the Hamiltonian requires electric operators as well. 
They are called $\hat E_L$ and $\hat E_R$ and furnish two independent Lie algebras of the group
\begin{equation}
\label{eq:EEcommutators}
\comm{\hat{E}_L^a}{\hat{E}_L^b} = -i f^{abc} \hat{E}_L^c \,, \qquad 
\comm{\hat{E}_R^a}{\hat{E}_R^b} = i f^{abc} \hat{E}_R^c \,, \qquad
\comm{\hat{E}_L^a}{\hat{E}_R^b} = 0 \,,
\end{equation}
and satisfy the commutation relation with the link operators $\hat U$
\begin{equation}\label{eq:EUcommutators}
  \begin{split}
    \comm{\hat{E}_L^a}{\hat{U}^{(j)}_{mn}} = T^{{(j)}a}_{mm'}\hat U^{(j)}_{m'n} \, \qquad
    \comm{\hat{E}_R^a}{\hat{U}^{(j)}_{mn}} = \hat{U}^{(j)}_{mn'}T^{{(j)}a}_{n'n} \,.
  \end{split}
\end{equation}
These electric operators define left and right translation operators, parameterized by a group element $\h$
\begin{equation}\label{eq:E_def}
\hat{\Theta}_{L\h} = e^{i \phi^a(\h) \hat E_L^a} \qc\hat{\Theta}_{R\h} = e^{i \phi^a(\h) \hat E_R^a} \,,
\end{equation}
where $\phi^a(\h)$ is a set of $\dim(G)$ parameters, called normal coordinates. 
These operators act on a basis state $\ket{\g}$ at each link $\ell$ via left and right group composition as
\begin{equation}
\label{eq:link_trafo}
\hat{\Theta}_{L\h} \ket{\g} = \ket{\h^{-1}\g} \qc\hat{\Theta}_{R\h} \ket{\g} = \ket{\g\h^{-1}} \,.
\end{equation}
Note that transforming by the combination $\hat{\Theta}_{L\h}\hat{\Theta}^\dagger_{R\h}$ of left and right translations performs a rotation such that 
\begin{align}
    e^{-i\phi^a(\h) \hat{L}^a} = \hat{\Theta}_{L\h}\hat{\Theta}^\dagger_{R\h}
    \,,
\end{align}
and therefore the difference of right and left electric operators is equal to the angular momentum operator 
\begin{align}
\label{eq:Ldef}
\hat E^a_R - \hat E^a_L = \LOpComp{a}
\,.
\end{align}
This will become important later.

The discussion so far has not mentioned gauge invariance. In a lattice theory, gauge invariance implies that states that are related through gauge transformations at any lattice site are equivalent to one another. 
A gauge transformation $\Omega$ at lattice site $n$ is mediated by the operator
\begin{equation}
    \hat{\Theta}_\Omega(n) = \exp(i \phi^a(\Omega) \hat{G}^a(n)) \,,
\end{equation}
where $\phi(\Omega)$ are normal coordinates for $\Omega$ and $\hat{G}^a(n)$ is the Hermitian operator
\begin{equation}\label{eq:Ga_def}
    \hat{G}^a(n) = \sum_{i=1}^d \left[\hat{E}_R^a(n-e_i, e_i) - \hat{E}_L^a(n, e_i) \right] \,.
\end{equation}
From \cref{eq:link_trafo} one finds that a gauge transformation at a given lattice site $n$ will change the states of all links connected to that site. 
The physical subspace satisfies
\begin{equation}\label{eq:gauss_law}
    \hat{G}^{a}(n) \ket{\Psi_\text{phys}} = 0 \,\qquad  \forall n\,,
\end{equation}
and is a particular linear combination of the states $\ket{\g_0, \ldots, \g_{n_\ell-1}}$. 

This gauge redundancy requires a careful treatment, such that physically equivalent states are not treated as distinct from one another. One way with dealing with this is through gauge fixing. 
Gauge fixing chooses a particular gauge transformation at each lattice site, and through this process, keeps only a single state from all gauge-equivalent states as a representative of the gauge-invariant state.
This eliminates the requirement of dealing with gauge invariance entirely, but comes at the price of introducing a non-locality into the system, as we will discuss. 

Gauge fixing is typically performed by choosing gauge transformations that set the state on as many links as possibility to the identity. 
The total number of independent gauge transformations possible is equal to the number of vertices $n_v$ in the lattice. 
One can show that using $n_v-1$ such gauge transformations, all links on a so-called maximal tree of links can be set to the identity~\cite{PhysRevD.15.1128}. 
A tree is a collection of links that do not form closed loops, and a maximal tree is a tree for which adding any additional link would create a closed loop
Since links on the maximal tree are gauge-fixed to the identity we will refer to them as unphysical links in the rest of this paper, while those not on the maximal tree will be called physical.
The collection of all $\NL$ physical links will be labeled by $\kappa$. 
Since we have used up $n_v-1$ gauge transformations to fix the states on the links on the maximal tree, the gauge transformation of a single vertex in the lattice has not been used up by this procedure. We define this vertex as the origin of the lattice and call this vertex $n_0$.

The dynamical degrees of freedom remaining in the theory are the physical links. 
To write the Hamiltonian, it is useful to perform a change of variables to those that have definite transformation properties with regards to the final gauge transformation at the origin. 
This is accomplished by parallel transporting the endpoints of all physical links to the origin of the lattice.
Since the maximal tree traverses every lattice site, the path ${\cal P}(n)$ for such a parallel transport can be chosen along the maximal tree, composed of only unphysical links.
While this does not alter the value of any parallel transformed object if unphysical gauge links are fixed to the identity, it does enforce that all physical links transform under the remaining gauge transformation, making it a global transformation. 
We define the parallel transport operator along this path as
\begin{align}
    \hat W(n) = \prod_{\ell \in {\cal P}(n)} \hat U(\ell)^{\sigma_\ell}
    \,,
\end{align}
where $\sigma_\ell$ is $+1$ if the link $\ell$ is traversed in the positive orientation, and $-1$ if it is traversed in the negative orientation. 
Each physical link $\kappa \equiv \kappa(n, e_i)$ can therefore be transported to the origin of the lattice, resulting in a loop operator
\begin{align}
    \hat X(\kappa) = \hat W(n) \, \hat U(\kappa) \,\hat W(n+e_i)^\dagger
    \,.
\end{align}

The Hilbert of a single loop is spanned by the eigenstates $\ket{\g_\kappa}$ of the loop operator 
\begin{equation}
\hat{X}_{mm'}(\kappa) \equiv \int\dd{\g_\kappa} u_{{mm'}}(\g_\kappa) \op{\g_\kappa} 
\,,
\end{equation}
and the full gauge-fixed Hilbert space is spanned by
\begin{equation}
    \ket{\g_1, \ldots, \g_{\NL}} = \ket{\g_1}\otimes \cdots \otimes \ket{\g_{\NL}}
    \,.
\end{equation}
Each loop operator transforms under the gauge transformation $\hat{\Theta}_\Omega(n_0)$ at the origin of the lattice as
\begin{align}
    \hat{\Theta}_\Omega(n) \ket{\g_\kappa} = \ket{\Omega \, \g_\kappa \, \Omega^{-1}}\,.
\end{align}
One can also view this as a transformation of the operators, given by
\begin{align}
\label{eq:global_gauge_y}
    \hat X(\kappa) \to \Omega(n_0) \hat X(\kappa) \Omega(n_0)^\dagger
    \,.
\end{align}

To write the Hamiltonian, one requires operators conjugate to $\hat X$.
They are defined as
\begin{align}
    [\hat{\mathcal{E}}_L^a(\kappa), \hat X(\kappa')] = T^a \hat X(\kappa) \delta_{\kappa,\kappa'} \,,\qquad 
    [ \hat{\mathcal{E}}_R^a(\kappa), \hat X(\kappa')] = \hat X(\kappa) T^a \delta_{\kappa,\kappa'}
    \, .
\end{align}
Using the same argument that led to \cref{eq:Ldef}, these operators are related to the the angular momentum operator 
\begin{align}
\label{eq:Ldef2}
\hat\CE^a_R(\kappa) - \hat\CE^a_L(\kappa) = \LOpComp{a}(\kappa)
\,,
\end{align}
which implies that one can write
\begin{align}
\label{eq:ELRdef}
\hat\CE^a_R(\kappa) = \frac{1}{2}\left(\hat\Sigma^a(\kappa) + \LOpComp{a}(\kappa)\right)
\,,\qquad
\hat\CE^a_L(\kappa) = \frac{1}{2}\left(\hat\Sigma^a(\kappa) - \LOpComp{a}(\kappa)\right)
\,.
\end{align}
This provides the defining equations for $\hat\Sigma^a(\kappa)$.

To relate these operators to the original electric operators, one first defines parallel transported electric operators at each link $\ell \equiv \ell(n, e_i)$ as
\begin{align}
    \hat{\mathcal{J}}(\ell) &\equiv \hat W^\dagger(n) \hat E_L(\ell) \hat W(n)
    \,.
\end{align}
and relates these operators to the operators $\CE^a_{L/R}$ as
\begin{align}
\label{eq:Ja_relation}
    \hat{\mathcal{J}}^a(\ell) = \sum_{\kappa \in t_+(\kappa)} \hat{\mathcal{E}}^a_{L}(\kappa) - \sum_{\kappa \in t_{-}(\kappa)} \hat{\mathcal{E}}^{a}_{R}(\kappa)
    \,.
\end{align}
Here $t_+(\kappa)$ ($t_-(\kappa)$) is the set of all physical links $\kappa$ such that $\ell$ is contained in path $P(\kappa)$ as a positive (negative) link. 

The gauge-fixed Hamiltonian $H = H_E + H_B$ can now be written as
\begin{align}
\label{eq:HB_maxTree}
    H_B &= \frac{1}{2g^2 a}\sum_p \Tr \left( I - \prod_{\kappa\in p} \hat X(\kappa)^{\sigma(\kappa)} \right) + \mathrm{h.c.}\nn\\
    H_E &= \frac{g^2}{2 a} \sum_{\ell} \left( \sum_{\kappa \in t_+(\ell)} \hat{\mathcal{E}}^a_{L}(\kappa) - \sum_{\kappa \in t_{-}(\ell)} \hat{\mathcal{E}}^{a}_{R}(\kappa) \right)^2,
\end{align}
where $\sum_p$ and $\sum_\ell $ denotes sums over all plaquettes and all links of the lattice, respectively.
In the remainder of this paper we will usually denote the dependence of the various operators on $\kappa$ through a subscript, providing a more compact notation. 

\subsection{Axis-angle and mixed-basis representation for \sutwo}
\label{subsec:mixed_basis}
A group element $\g$ of the \sutwo gauge group can be characterized by spherical coordinates $(\thetaOri{},\phiOri{})$ defining a rotation axis $\hat{n} = (\cos\phiOri{} \sin\thetaOri{}, \sin\phiOri{}\sin\thetaOri{}, \cos\thetaOri{})$ and a rotation angle about this axis $\omega$.
The ranges of these coordinates are $\thetaOri{} \in [0,\pi]$, $\phiOri{} \in [0,2\pi]$ and $\omega \in [0,2\pi]$. 
The Haar measure in axis-angle coordinates is
\begin{align}
    \dd{\g} = 4 \sin^2 \frac{\omega}{2} \sin\thetaOri{}\, \dd{\omega} \dd{\thetaOri{}} \dd{\phiOri{}} .
\end{align}
For each physical link, the basis states can therefore parameterized by these three variables
\begin{align}
    \ket{\g} \equiv \ket{\omega, \thetaOri{}, \phiOri{}}
    \,,
\end{align}
and the Hilbert space of the lattice is therefore spanned by
\begin{equation}
    \ket{\omega_1,\thetaOri{1}, \phiOri{1}, \ldots \omega_{\NL}, \thetaOri{\NL}, \phiOri{\NL}} =  \ket{\omega_1, \thetaOri{1}, \phiOri{1}}\otimes \cdots \otimes \ket{\omega_{\NL}, \thetaOri{\NL}, \phiOri{\NL}}
    \,.
\end{equation}

In this axis-angle representation, the loop operator can be written as
\begin{align}
\label{eq:XOrig}
\hat X_\kappa &=\left(
\begin{array}{cc}\cos \frac{\omega_\kappa}{2}-i \cos \thetaOri{\kappa}  \sin \frac{\omega_\kappa}{2} & -i \sin \thetaOri{\kappa} \sin \frac{\omega_\kappa}{2} \left(\cos \phiOri{\kappa} -i \sin \phiOri{\kappa}\right) \\
-i \sin\thetaOri{\kappa} \sin \frac{\omega_\kappa}{2} \left(\cos \phiOri{\kappa} +i \sin\phiOri{\kappa}\right) & \cos \frac{\omega_\kappa}{2}+i \cos\thetaOri{\kappa} \sin \frac{\omega_\kappa}{2} \\
\end{array}
\right)\,,
\end{align}
the angular momentum operator takes the well known form
\begin{align}
\label{eq:Lop}
\LOpComp{x}_\kappa &= i \left( \sin\phiOri{\kappa} \pdv{}{\thetaOri{\kappa}} + \cot\thetaOri{\kappa} \cos\phiOri{\kappa} \pdv{}{\phiOri{\kappa}}\right)\nonumber\\
\LOpComp{y}_\kappa &= i \left( -\cos\phiOri{\kappa} \pdv{}{\thetaOri{\kappa}} + \cot\thetaOri{\kappa} \sin\phiOri{\kappa} \pdv{}{\phiOri{\kappa}}\right)\nonumber\\
\LOpComp{z}_\kappa &= -i \pdv{}{\phiOri{\kappa}}\nonumber\\
\LOpSq_\kappa &=-\left(\pdv[2]{}{\thetaOri{\kappa}}+\cot \thetaOri{\kappa}\pdv{}{\thetaOri{\kappa}}+\csc \thetaOri{\kappa} \pdv[2]{}{\phiOri{\kappa}}\right)
\,.
\end{align}
and the $\hat \Sigma_\kappa$ operators are given by~\cite{DAndrea:2023qnr}
\begin{align}
\label{eq:Sigmaop}
\hat \Sigma_\kappa^x &= 2i \sin\thetaOri{\kappa} \cos\phiOri{\kappa} \frac{\partial}{{\partial} \omega_\kappa} + i \cot\frac{\omega}{2}  \left(\cos\thetaOri{\kappa} \cos\phiOri{\kappa} \frac{\partial}{{\partial} \thetaOri{\kappa}} - \csc\thetaOri{\kappa} \sin\phiOri{\kappa} \frac{\partial}{{\partial} \phiOri{\kappa}}\right)
    \nonumber\\
\hat \Sigma_\kappa^y &= 2i \sin\thetaOri{\kappa} \sin\phiOri{\kappa} \frac{\partial}{{\partial} \omega_\kappa} + i \cot\frac{\omega}{2} \left(\cos\thetaOri{\kappa} \sin\phiOri{\kappa} \frac{\partial}{{\partial} \thetaOri{\kappa}} + \csc\thetaOri{\kappa} \cos\phiOri{\kappa} \frac{\partial}{{\partial} \phiOri{\kappa}}\right)
    \nonumber\\
\hat \Sigma_\kappa^z &= 2i \cos\thetaOri{\kappa} \frac{\partial}{{\partial} \omega_\kappa}- i \cot\frac{\omega}{2}\sin\thetaOri{\kappa} \frac{\partial}{{\partial} \thetaOri{\kappa}}
    \, .
\end{align}
Note that the $\hat \Sigma_\kappa$ operators can be written in the compact form
\begin{align}
\label{eq:sigmadef}
    \boldsymbol{\Sigma_\kappa} &= 2 i \mathbf{n}_{\kappa} \pdv{}{\omega_\kappa}+ \cot \left(\frac{\omega_\kappa}{2}\right)\left(\mathbf{n}_{\kappa}\cross \LOp_\kappa\right)
    \,.
\end{align}

One can switch from this magnetic basis $\ket{\omega,\theta,\phi}$ to a mixed basis $\ket{\omega, \LOri{}, \MOri{}}$ by representing the angular dependence of the rotation axis through spherical harmonics via
\begin{equation}
\label{eq:mixedoriginalOverlap}
    \bra{\omega' \theta \phi}\ket{\omega \LOri{} \MOri{}}
      = \frac{\delta(\omega-\omega')}{2 \sin\frac{\omega}{2}} \YLMSymbol_{\LOri{}\MOri{}}(\theta,\phi) \,,
\end{equation}
where we omit the dependence on $\kappa$ for now.
The factor $1/(2\sin(\omega/2))$ cancels the corresponding factor in the Haar measure.
The name mixed basis arises because $\omega$ denotes a magnetic quantum number, while $\LOri{}$ and $\MOri{}$ have properties of electric quantum numbers. 
In this mixed basis, the Hilbert space is therefore spanned by a tensor product of the states $\ket{\omega \LOri{} \MOri{}}$ for each physical link $\kappa$ 
\begin{equation}
    \ket{\omega_1 \LOri{1} \MOri{1} \ldots \omega_{\NL}\LOri{\NL} \MOri{\NL}} = \ket{\omega_1 \LOri{1} \MOri{1}}\otimes \cdots \otimes \ket{\omega_{\NL}\LOri{\NL} \MOri{\NL}}
    \,.
\end{equation}

Note that the set of quantum numbers $\LOri{\kappa}$ and $\MOri{\kappa}$ are the eigenvalues of the complete set of commuting observables (CSCO) formed by $\LOpSq_\kappa$ and $\LOpComp{z}_\kappa$.
The CSCO for the angular information in the mixed basis is therefore the set operators 
\begin{align}
\{\text{CSCO}\}_O = \left\{\LOpSq_1,\LOpComp{z}_1,\LOpSq_2,\LOpComp{z}_2,\dots \LOpSq_\NL,\LOpComp{z}_\NL \right\},
\end{align}
where we include the subscript $O$ to denote that this was the original mixed-bases used in Ref.~\cite{DAndrea:2023qnr}.

One can obtain matrix elements of vector operators using the Wigner-Eckart theorem. It states that the matrix element of a vector operator $\hat V^a$ at a physical lattice site is given by (again suppressing the $\kappa$ subscripts)
\begin{align}
\label{eq:vec_ops}
    \bra{\omega'\LOri{}'\MOri{}'}\hat V^q \ket{\omega \LOri{} \MOri{}} = \langle \omega'\LOri{}' \vert \vert \bm{V} \vert\vert \omega \LOri{} \rangle \bra{\LOri{}' \MOri{}'}\ket{\LOri{} 1 \MOri{} q}
    \,.
\end{align}
Here the operators $\hat V^q$ with $q=+1,0,-1$ are related to the operators $V^a$ with $a = 1, 2, 3$ as
\begin{align}
    \hat{V}^0 &= \hat{V}^3 \,, \qquad \hat{V} ^{\pm} = \mp \frac{1}{\sqrt{2}} \left( \hat{V}^1  \pm i \hat{V}^2  \right)
    \,,
\end{align}
and $\bra{\LOri{}' \MOri{}'}\ket{\LOri{} 1 \MOri{} q}$ is a Clebsch–Gordan coefficient.

Using this, the reduced matrix elements for the electric operator are given by
\begin{align}\label{eq:elec_mat_elements}
\begin{split}
    & \langle \omega'\LOri{}' \vert \vert \hat{\bm{\mathcal{E}}}_{R(L)} \vert\vert \omega \LOri{} \rangle = \delta(\omega'-\omega)
    \begin{cases}
    -i \sqrt{\frac{\LOri{}}{2\LOri{}'+1}} \big( \frac{\df }{\df \omega} + \frac{\LOri{}}{2}\cot\frac{\omega}{2} \big) & \LOri{}'=\LOri{}-1 \\[15pt]
    +(-)\frac{\sqrt{\LOri{}(\LOri{}+1)}}{2} & \LOri{}'=\LOri{} \\[15pt]
    -i \sqrt{\frac{\LOri{}'}{2\LOri{}'+1}} \big( -\frac{\df }{\df \omega} + \frac{\LOri{}'}{2}\cot\frac{\omega}{2} \big) & \LOri{}'= \LOri{}+1
    \end{cases}
    \,.
\end{split}
\end{align}

To write the matrix elements of the loop operators $\hat X$, we define
\begin{align}
    \hat{X} = \hat{S} I + \hat{X}^a T^a \,, \qquad \hat{S} \equiv \tfrac{1}{2}\Tr \hat{X} \,, \qquad \hat{X}^a = 2 \Tr T^a \hat{X}
    \,.
\end{align}
The matrix element of the scalar operator $\hat S$ is therefore given by
\begin{align}\label{eq:scalar_mat_elements}
    \bra{\omega' \LOri{}' \MOri{}'}\hat{S}\ket{\omega \LOri{} \MOri{}} = \delta_{\LOri{}'\LOri{}}\delta_{\MOri{}'\MOri{}}\delta(\omega'-\omega) \cos\frac{\omega}{2}
    \,.
\end{align}
while the matrix elements of the vector operator $\hat X^a$ can be obtained by \cref{eq:vec_ops} and the reduced matrix element
\begin{align}\label{eq:mag_mat_elements}
    \langle \omega'\LOri{}' \vert \vert \bm{X} \vert\vert \omega \LOri{} \rangle = 
    \delta(\omega-\omega')\begin{cases}
    i \sqrt{\frac{\LOri{}}{2\LOri{}'+1}} \sin\frac{\omega}{2} & \LOri{}' = \LOri{}-1 \\
    0 & \LOri{}' = \LOri{} \\
    -i \sqrt{\frac{\LOri{}'}{2\LOri{}'+1}} \sin\frac{\omega}{2} & \LOri{}'=\LOri{}+1
    \end{cases}
    \,.
\end{align}
These relations can be used to obtain the Hamiltonian in the mixed basis.
For details see Ref.~\cite{DAndrea:2023qnr}.

\subsection{Residual Global Gauge Symmetry}
As mentioned above, the maximal-tree gauge fixing procedure has removed the redundancy due to gauge transformations at all but a single lattice site, which we called the origin of the lattice. 
All operators have been defined to transform in a simple manner under this remaining gauge transformation.
The generator of this gauge transformation is equal to the total angular momentum operator
\begin{align}
\label{eq:Gadef}
   \hat G^a(n_0) = -\LTotOpComp{a} = - \sum_\kappa\LOpComp{a}(\kappa)
    \,.
\end{align}
This can be seen using \cref{eq:Ga_def,eq:Ldef2,eq:Ja_relation} together with the fact that all paths $P(\kappa)$ start and end at the origin of the lattice.

An intuition for how to remove this last (global) gauge redundancy can developed by noticing that a basis state in the axis-angle representation has a very simple geometric interpretation~\cite{Littlejohn:1997qb}.
Since each physical link is represented by a rotation angle $\omega$ and an axis parameterized by polar angles $n(\vartheta, \varphi)$, each physical link can be interpreted as a rod with length $\omega$ in direction $n(\vartheta, \varphi)$. 
The Hilbert space of the complete lattice therefore consists of $n_\kappa$ such rods.
In this geometric interpretation, the residual global gauge redundancy implies that the remaining gauge transformation is a generator of rotation of the system of rods. Furthermore, any two systems of rods that are related by an overall rotation are equivalent to one another. 
In other words, only the shape of the system of rods, described by the length of the rods and the relative angles between them, is physical, while the overall orientation of the system of rods is unphysical. The reminder of this paper is dedicated to puting this intuition onto firmer mathematical footing, as well as deriving the explicit form of the Hamiltonian in a basis where it is trivial to eliminate the last gauge redundancy present in the system.

\section{Construction of Sequestered Magnetic Basis \label{sec:seq_basis}}
\begin{figure}[t]
\includegraphics[width=0.9\textwidth]{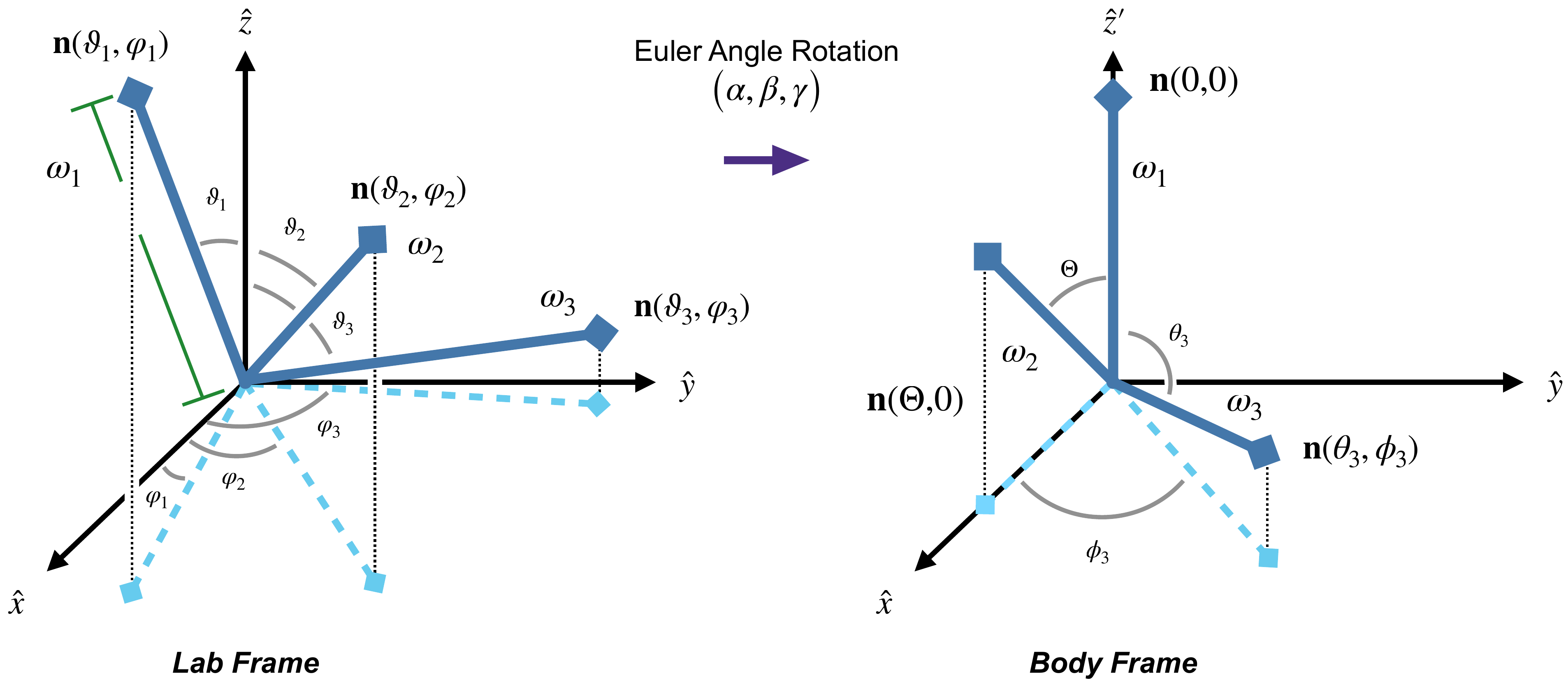}
\caption{The left and right coordinate systems show the lab and body frames, respectively.
Transforming between the lab frame and the body frame is done via the rotation matrix in Eq.~\eqref{eq:RotE}, which is parameterized by the Euler angles $\alpha, \beta, \gamma$.
The body frame is defined to be the frame where Rod 1 is aligned with the $\hat z'$ axis and the $\hat x' - \hat z'$ plane is defined via the relative orientation of Rod 1 and Rod 2; Rod 2 is always orthogonal to $\hat y'$. 
Note that this is not a unique definition of the body frame, as any two rods can be used to define the coordinate system in this manner.
The expressions derived in this paper can be converted for different choices in a straightforward way by simply swapping the variables to match the chosen convention.}
\label{fig:EulerRotV}
\end{figure}

As discussed in \cref{sec:review}, a group element of \sutwo can be characterized by hyperspherical coordinates, with $\omega_\kappa$ being the radial coordinate and $\thetaOri{\kappa}, \phiOri{\kappa}$ the angular ones. 
This allows us to interpret the degrees of freedom of the full system as those of a system of rigid rods, connected together at the origin. 
The dynamics of the system can cause each rod to change in length ($\delta \omega_\kappa$) and in direction ($\delta \thetaOri{\kappa}, \delta \phiOri{\kappa}$).
The rods remain fixed at the origin and cannot curve. 
This system is shown in \cref{fig:EulerRotV}.

There are two ways to describe the dynamics of this system of rigid rods.\footnote{While there may be alternative useful parameterizations, we will restrict ourselves to these two in this paper.}
The first is use a fixed coordinate system; 
this is the lab frame, with the axes labeled as $\hat x, \hat y, \hat z$. 
The dynamics of the system are therefore fully determined in terms of the variables $\{\omega_\kappa, \thetaOri{\kappa}, \phiOri{\kappa}\}$, with $i$ ranging from $1$ to $\NL$. 
Each rod's direction is given by
\begin{align}
\label{eq:nkappa_orig}
\mathbf{n}_{\kappa} = \left(\sin \thetaOri{\kappa}\cos \phiOri{\kappa}, \sin \thetaOri{\kappa}\sin \phiOri{\kappa}, \cos \thetaOri{\kappa}\right) \, ,
\end{align}
using the hyperspherical interpretation of \sutwo, with each rod's respective length being given by $\omega_\kappa$. 
We call this representation the `original basis', as developed in Ref.~\cite{DAndrea:2023qnr}. 
 
The second approach is to use a coordinate system that depends on the orientation of the system of rods;  this is the body frame, with axes labeled as $\hat x',\hat y', \hat z'$. For simplicity, we choose this coordinate system to align $z'$ with Rod 1 and define the $\hat x'-\hat z'$ plane using Rod 1 and Rod 2. 
The shape of the system is then defined by the lengths $\omega_\kappa$ of all the rods, the angle $\thetaDef$ between Rods 1 and 2, and the angles $\thetaSeq{\mu}$ and $\phiSeq{\mu}$, with $\mu = 3, \ldots, \NL$, specifying the directions of Rods $\mu$ in the body frame.
The orientation of the system is then characterized by the Euler angles $\alpha, \beta, \gamma$ used to rotate between the lab and body frame.
The full system can described in terms of the union of these variables. 
The directions $\mathbf{n}_\kappa$ can be obtained through a rotation from body to lab frame
\begin{align}
\label{eq:nkappa_seq}
\mathbf{n}_{\kappa} = \RotE \cdot \nSeqBody{\kappa} \quad \text{where}\qquad \nSeqBody{\kappa} &= \begin{cases}\left(0,0,1\right) \quad &\kappa = 1 \\
\left(\sin \thetaDef,0,\cos \thetaDef\right) \quad &\kappa = 2 \\
\left(\sin \thetaSeq{\mu}\cos \phiSeq{\mu}, \sin \thetaSeq{\mu}\sin \phiSeq{\mu}, \cos \thetaSeq{\mu}\right) \quad &\kappa = \mu
\end{cases}
\,,
\end{align}
where $\RotE$ is the rotation matrix that relates the two frames, written in terms of Euler angles
\begin{align}
\label{eq:RotE}
\RotE= \left(\begin{array}{ccc}
\cos \alpha \cos \beta \cos \gamma-\sin \alpha \sin \gamma & -\sin\alpha \cos \gamma-\cos \alpha \cos \beta \sin \gamma & \cos \alpha \sin \beta  \\
 \sin \alpha \cos \beta \cos \gamma+\cos \alpha \sin \gamma & \cos \alpha \cos \gamma-\sin \alpha \cos \beta \sin \gamma & \sin \alpha \sin\beta  \\
 -\sin \beta \cos \gamma & \sin \beta \sin \gamma & \cos \beta \\
\end{array}
\right) \, .
\end{align} 
We call this representation the `sequestered basis' for reasons that will become clear later. 
The range of the angular variables in this basis is
\begin{align}
\thetaDef = \left[0, \pi\right], \quad\thetaSeq{\mu} = \left[0, \pi\right], \quad \phiSeq{\mu} = \left[0, 2\pi\right], \qquad \beta = \left[0, \pi\right], \quad \alpha = \left[0, 2\pi\right], \quad \gamma = \left[0, 2\pi\right] \, . 
\end{align}
The Hilbert space of the lattice in the sequestered basis is spanned by the states 
\begin{equation}
\ket{\{\omega_\kappa\}; \thetaDef, \{\thetaSeq{\mu}, \phiSeq{\mu}\}; \alpha, \beta, \gamma} \equiv \left(\prod_{\kappa=1}^{\NL}\ket{\omega_\kappa}\right) \otimes \ket{\thetaDef} \otimes \left(\prod_{\mu=3}^{\NL}\ket{\thetaSeq{\mu}, \phiSeq{\mu}} \right) \otimes \ket{\alpha, \beta, \gamma} \,,
\end{equation}
where our notation is to be understood as
\begin{equation}
     \ket{\{\omega_\kappa\}} \equiv \prod_{\kappa=1}^{\NL}\ket{\omega_\kappa} = \ket{\omega_1} \otimes \dots \otimes \ket{\omega_{\NL}}, \qquad \ket{\{\thetaSeq{\mu}, \phiSeq{\mu}\}} \equiv \prod_{\mu=3}^{\NL}\ket{\thetaSeq{\mu}, \phiSeq{\mu}} = \ket{\thetaSeq{3}, \phiSeq{3}} \otimes \dots \otimes \ket{\thetaSeq{\NL}, \phiSeq{\NL}} \,.
\end{equation}
The benefit of the sequestered basis is that it is easy to identify the total global charge of the system, and to perform the gauge fixing that removes the gauge redundancy at the origin.
In particular, the quantum numbers of the total charge are associated with the Euler angle variables, $\alpha, \beta, \gamma $.

The Hamiltonian can be written in either basis, and Ref.~\cite{DAndrea:2023qnr} provides the form of the Hamiltonian in the original basis.
In this work we carry out the change of basis from the original basis to the sequestered. 
The first step in carrying out this change of basis is relating the angles in both bases. 
The lengths of the rods, $\omega_\kappa$, are unchanged when rotating into a different frame. 
The angular coordinates can be related by equating \cref{eq:nkappa_orig,eq:nkappa_seq}. 
Solving this system of equations, the angles  in the sequestered basis are related to the angles in the original basis via
\begin{align}
\label{eq:SeqToOri}
\cos \thetaDef =& \cos \thetaOri{1}\cos \thetaOri{2}+\cos \left(\phiOri{1}- \phiOri{2}\right)\sin \thetaOri{1}\sin \thetaOri{2} \nonumber\\
\cos \thetaSeq{\mu} =& \cos \thetaOri{1}\cos \thetaOri{\mu}+\cos \left(\phiOri{1}- \phiOri{\mu}\right)\sin \thetaOri{1}\sin \thetaOri{\mu} \nonumber \\
\cos \phiSeq{\mu} =&\frac{-\cos \thetaDef \cos \thetaSeq{\mu}+\cos \thetaOri{2}\cos \thetaOri{\mu}+\cos \left(\phiOri{2}- \phiOri{\mu}\right)\sin \thetaOri{2}\sin \thetaOri{\mu}}{\csc \thetaDef \csc \thetaSeq{\mu}}\nonumber\\
=& \frac{1}{\sin \thetaDef \sin \thetaSeq{\mu}}\bigg[\cos \thetaOri{2}\cos \thetaOri{\mu}+\cos \left(\phiOri{2}- \phiOri{\mu}\right)\sin \thetaOri{2}\sin \thetaOri{\mu} \nonumber \\
&-\left(\cos \thetaOri{1}\cos \thetaOri{2}+ \cos \left(\phiOri{1}-\phiOri{2}\right)\sin \thetaOri{1}\sin\thetaOri{2}\right) \times \left(\cos \thetaOri{1}\cos \thetaOri{\mu}+ \cos \left(\phiOri{1}-\phiOri{\mu}\right)\sin\thetaOri{1}\sin\thetaOri{\mu}\right) \bigg] \nonumber \\
\cos \gamma =& -\frac{\sin \thetaOri{1}\cos \thetaOri{2}- \cos \left(\phiOri{1}-\phiOri{2}\right)\cos \thetaOri{1}\sin \thetaOri{2}}{\sin \thetaDef} \qquad \beta =\thetaOri{1} \qquad \alpha =\phiOri{1} \, .
\end{align}
Note that expressions for $\phi_\mu$ and $\gamma$ have implicit dependence on sequestered basis variables ($\cos \gamma$ depends on $\sin\phiSeq{\mu}$ and $\cos \phiSeq{\mu}$ depends on $\csc \thetaDef$ and $\csc \theta_\mu$).
This implicit dependence is kept in order to avoid potential pitfalls with determining the principle value of $\arccos(x)$.

To convert the first-order differential operators from the original to the sequestered basis, we will use of the following chain rules,
\begin{align}
\label{eq:FirstOrderChainRule}
\pdv{}{\thetaOri{\kappa}} &=\pdv{\alpha}{\thetaOri{\kappa}}\pdv{}{\alpha}+\pdv{\beta}{\thetaOri{\kappa}}\pdv{}{\beta}+\pdv{\gamma}{\thetaOri{\kappa}}\pdv{}{\gamma}+\sum_{\mu}\pdv{\thetaSeq{\mu}}{\thetaOri{\kappa}}\pdv{}{\thetaSeq{\mu}}+\sum_{\mu}\pdv{\phiSeq{\mu}}{\thetaOri{\kappa}}\pdv{}{\phiSeq{\mu}}, \nonumber \\
\pdv{}{\phiOri{\kappa}} &= \pdv{\alpha}{\phiOri{\kappa}}\pdv{}{\alpha}+\pdv{\beta}{\phiOri{\kappa}}\pdv{}{\beta}+\pdv{\gamma}{\phiOri{\kappa}}\pdv{}{\gamma}+\sum_{\mu}\pdv{\thetaSeq{\mu}}{\phiOri{\kappa}}\pdv{}{\thetaSeq{\mu}}+\sum_{\mu}\pdv{\thetaSeq{\mu}}{\phiOri{\kappa}}\pdv{}{\phiSeq{\mu}}
\,.
\end{align}
One complication that arises in carrying out this change of basis is that several of the relations in \cref{eq:SeqToOri} are given in terms of cosines of the sequestered basis variables. 
However, this is easily addressed by making use of the identities
\begin{align}
\pdv{\thetaDef}{\thetaOri{\kappa}}= -\frac{1}{\sin \thetaDef}\pdv{\cos\thetaDef}{\thetaOri{\kappa}} \qquad \, \pdv{\thetaDef}{\phiOri{\kappa}}= -\frac{1}{\sin \thetaDef}\pdv{\cos\thetaDef}{\phiOri{\kappa}} \, ,
\end{align}
with analogous expressions for derivatives of $\thetaSeq{\mu}$ and $\phiSeq{\mu}$ with respect to original-basis angular variables.
Evaluating \cref{eq:FirstOrderChainRule} using the expressions in \cref{eq:SeqToOri} results in expressions that contain derivatives with respect to sequestered basis variables, but pre-factors that still depend on original basis variables. 
We choose to eliminate this residual dependence not by inverting the expressions in \cref{eq:SeqToOri}, but instead by directly equating \cref{eq:nkappa_orig,eq:nkappa_seq} to obtain 
\begin{align}
\label{eq:OriToSeqViaN}
\cos\phiOri{\kappa} &= \frac{1}{\sin \thetaOri{\kappa}}\left(\RotE \cdot \nSeqBody{\kappa}\right)\cdot \hat{\mathbf{x}} \qquad \sin\phiOri{\kappa}= \frac{1}{\sin \thetaOri{\kappa}}\left(\RotE \cdot \nSeqBody{\kappa}\right)\cdot \hat{\mathbf{y}} \qquad \cos\thetaOri{\kappa}= \left(\RotE\cdot \nSeqBody{\kappa}\right)\cdot \hat{\mathbf{z}}
\,.
\end{align}

Carrying out these algebraic manipulations, the six unique first-order derivatives that appear are
\begin{align}
\label{eq:FirstDerivOriToSeq}
\pdv{}{\thetaOri{1}} =&\pdv{}{\beta}+\sin\gamma \cot \thetaDef \pdv{}{\gamma} - \cos \gamma \pdv{}{\thetaDef}+\sum_{\mu}\left(\sin \gamma \sin \phiSeq{\mu}-\cos \gamma \cos \phiSeq{\mu}\right)\pdv{}{\thetaSeq{\mu}} \nonumber \\
&+ \sum_\mu \left(-\sin \gamma \cot \thetaDef+ \cot \thetaSeq{\mu}\left(\sin \gamma\cos \phiSeq{\mu}+\cos \gamma\sin \phiSeq{\mu}\right)\right)\pdv{}{\phiSeq{\mu}} \nonumber \\
\pdv{}{\phiOri{1}} =& \pdv{}{\alpha}-\left(\cos \beta +\sin \beta \cos \gamma  \cot \thetaDef\right)\pdv{}{\gamma} - \sin \beta \sin \gamma \pdv{}{\thetaDef}-\sum_{\mu}\sin \beta\left(\sin \gamma \cos \phiSeq{\mu}+\cos \gamma \sin \phiSeq{\mu}\right)\pdv{}{\thetaSeq{\mu}} \nonumber \\
&+\sum_\mu \sin \beta \left(\cos \gamma \cot \thetaDef+ \cot \thetaSeq{\mu}\left(\sin \gamma\sin \phiSeq{\mu}-\cos \gamma\cos \phiSeq{\mu}\right)\right)\pdv{}{\phiSeq{\mu}} \nonumber \\
\pdv{}{\thetaOri{2}}=& \frac{\sin \beta\cos \gamma \cos \thetaDef+\cos \beta \sin \thetaDef}{\sin \thetaOri{2}}\pdv{}{\thetaDef}-\frac{\sin \beta \sin \gamma \csc\thetaDef}{\sin \thetaOri{2}}\left(\pdv{}{\gamma}-\sum_\mu \pdv{}{\phiSeq{\mu}}\right)\nonumber \\
\pdv{}{\phiOri{2}}=&\sin \beta \sin \gamma \pdv{}{\thetaDef} +\left(\cos \beta+\sin \beta \cos \gamma \cot \thetaDef\right)\left(\pdv{}{\gamma}-\sum_\mu \pdv{}{\thetaSeq{\mu}}\right) \nonumber 
\\
\pdv{}{\thetaOri{\mu}}=&\frac{\cos \beta\sin \thetaSeq{\mu}+\sin \beta \cos \thetaSeq{\mu}\left(\cos \gamma \cos \phiSeq{\mu}- \sin \gamma \sin \phiSeq{\mu}\right)}{\sin \thetaOri{\mu}}\pdv{}{\thetaSeq{\mu}}\nonumber
\\
&-\frac{\sin \beta \csc \thetaSeq{\mu}\left(\sin \gamma \cos \phiSeq{\mu}+ \cos \gamma \sin \phiSeq{\mu}\right)}{\sin \thetaOri{\mu}}\pdv{}{\phiSeq{\mu}} \nonumber \\
\pdv{}{\phiOri{\mu}}=& \sin\beta\left(\sin\gamma \cos \phiSeq{\mu}+\cos\gamma \sin \phiSeq{\mu}\right)\pdv{}{\thetaSeq{\mu}}+\left(\cos \beta +\sin \beta \cot \thetaSeq{\mu}\left(\cos\gamma \cos \phiSeq{\mu}-\sin\gamma \sin \phiSeq{\mu}\right)\right)\pdv{}{\phiSeq{\mu}}
\,,
\end{align}

These relations can be used to carry out a change of variable on the operators in the Hamiltonian. 
While this is somewhat tedious, it only has to be done once, as there is only a finite number of different types of operators that can appear in the Hamiltonian, regardless of the number of lattice sites. 
The remainder of this section will be dedicated to determining the form of magnetic and electric components of the Hamiltonian derived in Ref.~\cite{DAndrea:2023qnr}, written in the sequestered basis, as well as the generators of global gauge transformations. 
The latter will prove useful for not only demonstrating the utility of the sequestered basis, but also for determining a useful set of basis vectors for the mixed-basis, which will be done in \cref{sec:mixed}.

\subsection{Angular Momentum and Generators of Global Gauge Transformations}
\label{sec:angular_sequestered}
Before deriving the Hamiltonian in the sequestered basis, we will first look at angular momentum operators and the generators of the global gauge transformations. 
As discussed in \cref{sec:review}, they are given by the total angular momentum operator
\begin{align}
    \hat G^a(n_0) = -\LTotOpComp{a} = - \sum_\kappa\LOpComp{a}(\kappa) \,.
\end{align}
The angular momentum operators in the original basis are given in \cref{eq:Lop}. 
In order to convert these operators into sequestered-basis operators, we need to transform the first-order differential operators that appear this expression

Utilizing the expressions above, in particular \cref{eq:FirstDerivOriToSeq,eq:OriToSeqViaN} one can show that
\begin{align}
\label{eq:TotalChargeGen}
\LTotOpComp{x} &= i \cos \alpha \cot \beta \pdv{}{\alpha}+ i \sin \alpha \pdv{}{\beta}-i \cos \alpha \csc \beta \pdv{}{\gamma} \nonumber \\
\LTotOpComp{y} &=  i \sin \alpha \cot \beta \pdv{}{\alpha}- i \cos \alpha \pdv{}{\beta}-i \sin \alpha \csc \beta \pdv{}{\gamma} \nonumber \\
\LTotOpComp{z} &= -i\pdv{}{\alpha}
\,.
\end{align}
This demonstrates why we have chosen to call the basis of the second description of the rod system the sequestered basis: 
all of the dependence on the global charge is now fully encoded into three Euler angles: $\alpha, \beta, \gamma$, as is expected. After all, the total angular momentum is related to the overall rotation of the system.

It will also prove convenient to define the angular momentum operators $\LSeqOp_\kappa$  in the body frame
\begin{align}
\label{eq:LSeqOpdef}
    \LOp_\kappa &= \RotE \cdot \LSeqOp_\kappa\, .
\end{align}
For the directions $\mu = 3 \ldots \NL$, which are not used to define the orientation of the body frame, one recovers the canonical angular momentum operators when written in the sequestered basis
\begin{align}
\LSeqOpComp{x}_{\mu} &= i \left( \sin\phiSeq{\mu} \pdv{}{\thetaSeq{\mu}} + \cot\thetaSeq{\mu} \cos\phiSeq{\mu} \pdv{}{\phiSeq{\mu}}\right)\nonumber\\
\LSeqOpComp{y}_{\mu} &= i \left( -\cos\phiSeq{\mu} \pdv{}{\thetaSeq{\mu}} + \cot\thetaSeq{\mu} \sin\phiSeq{\mu} \pdv{}{\phiSeq{\mu}}\right)\nonumber\\
\LSeqOpComp{z}_{\mu} &= -i \pdv{}{\phiSeq{\mu}}\nonumber\\
\LSeqOpSq_\mu &= -\left(\pdv[2]{}{\thetaSeq{\mu}}+\cot \thetaSeq{\mu}\pdv{}{\thetaSeq{\mu}}+\csc \thetaSeq{\mu} \pdv[2]{}{\phiSeq{\mu}}\right)
    \,,\
\end{align}

To obtain the expressions for $\LSeqOp_1$ and $\LSeqOp_2$ is slightly more complicated, since they are used to define the orientation of the body frame. In particular, they require knowledge of the total angular momentum operators in the body frame $\LpTotOp$, given by 
\begin{align}
\LTotOp&= \RotE \cdot \LpTotOp\,,
\end{align}
which can be written as
\begin{align}
    \LpTotOpComp{x} &= -i \cos\gamma \cot\beta\frac{\partial}{\partial\gamma} + i \cos \gamma \csc\beta \frac{\partial}{\partial\alpha} - i \sin\gamma \frac{\partial}{\partial\beta} \nonumber\\
    \LpTotOpComp{y} &= i \sin\gamma \cot\beta\frac{\partial}{\partial\gamma} - i \sin \gamma \csc\beta \frac{\partial}{\partial\alpha} - i \cos\gamma \frac{\partial}{\partial\beta} \nonumber\\
    \LpTotOpComp{z} &= -i \frac{\partial}{\partial\gamma}
    \,.
\end{align}
Note that $\LTotOpSq = \LpTotOpSq$ and that the angular momentum operators $\LpTotOpComp{a}$ satisfy anomalous commutation relations 
\begin{align}
    [\LpTotOpComp{a}, \LpTotOpComp{b}] = -i \epsilon^{abc} \LpTotOpComp{c}
    \,,
\end{align}
\text{i.e.} they carry an additional minus sign.

Given this total angular momentum operator in the lab frame, one can obtain the remaining two angular momentum operators in the body frame, which can be written as 
\begin{align}
\label{eq:LSeq12def}
\LSeqOp_1 &= \LSigmaOp+\csc\thetaDef\left[i \left(\nSeqBody{1}\times\nSeqBody{2}\right)\pdv{}{\thetaDef}+\csc\thetaDef \, \left(\left(\nSeqBody{1}\times \nSeqBody{2}\right)\times \nSeqBody{2}\right)\LSigmaOpComp{z}\right]\nonumber\\
\LSeqOp_2 &= -\csc\thetaDef\left[ i  \left(\nSeqBody{1}\times \nSeqBody{2}\right) \pdv{}{\thetaDef}+ \csc\thetaDef  \left(\left(\nSeqBody{1}\times \nSeqBody{2}\right)\times \nSeqBody{2}\right)\LSigmaOpComp{z}\right]
\,,
\end{align}
where we have defined
\begin{align}
\LSigmaOp \equiv \LSeqOp_1 + \LSeqOp_2 = \left(\LpTotOp-\sum_\mu\LSeqOp_\mu\right) \, .
\end{align}

\subsection{Electric Hamiltonian}
\label{subsec:HE}
The electric operators in the original basis were given in \cref{eq:ELRdef,eq:Lop,eq:Sigmaop}.
Using the results for the first derivatives in addition to \cref{eq:OriToSeqViaN}, the $\boldsymbol{\Sigma}$ and $\boldsymbol{L}$ operators can be written in terms of the sequestered basis variables. 
In order to avoid possible pitfalls of taking the principle values of $\arccos \thetaOri{\kappa}$, we use the identity $\sin^2 \thetaOri{\kappa} \equiv 1- \cos^2\thetaOri{\kappa}$, plus the last expression in \cref{eq:OriToSeqViaN} to remove all dependence on original basis variables. 

The angular momentum operators in the sequestered basis were already discussed in \cref{sec:angular_sequestered} and can be written as
\begin{align}
\label{eq:Ldefsec3}
    \LOp_{\kappa} &= \RotE \cdot \LSeqOp_\kappa
    \,,
\end{align}
with the expressions for $\LSeqOp_\mu$ provided there.
What is left is therefore expressions for the $\boldsymbol{\hat\Sigma}_\kappa$ operators in the sequestered basis.
These can be written as
\begin{align}
\label{eq:Sigmadef}
\boldsymbol{\Sigma_\kappa} &= \RotE \cdot \left(2 i \nSeqBody{\kappa} \pdv{}{\omega_\kappa}+ \cot \left(\frac{\omega_\kappa}{2}\right)\left(\nSeqBody{\kappa}\cross \LSeqOp_\kappa\right)\right)
\,.
\end{align}
Not unsurprisingly, this operator has a very similar structure to the $\boldsymbol{\Sigma_\kappa}$ in the original frame \cref{eq:sigmadef}. 

With these forms for $\boldsymbol{\Sigma}_\kappa$ and $\mathbf{L}_\kappa$, it is relatively simple (though tedious in bookkeeping) to evaluate all possible bilinears. 
For example, the bilinear that just involves rod $\mu$ can be easily found to be
\begin{align}
\boldsymbol{\CE}_{\mu L} \cdot \boldsymbol{\CE}_{\mu L} &= \left(\frac{\boldsymbol{\Sigma}_\mu - \mathbf{L}_\mu}{2}\right)\cdot \left(\frac{\boldsymbol{\Sigma}_\mu - \mathbf{L}_\mu}{2}\right) \nonumber \\
&=\frac{1}{4} \left(\RotE\cdot \left\{2i\,  \nSeqBody{\mu}\pdv{}{\omega_\mu}+ \cot \left(\frac{\omega_\mu}{2}\right)\left(\nSeqBody{\mu}\times \LSeqOp_\mu\right)-\LSeqOp_\mu\right\}\right)\cdot \left(\RotE\cdot \left\{2i\,  \nSeqBody{\mu}\pdv{}{\omega_\mu}+ \cot \left(\frac{\omega_\mu}{2}\right)\left(\nSeqBody{\mu}\times \LSeqOp_\mu\right)-\LSeqOp_\mu\right\}\right)\nonumber \\
&=\frac{1}{4} \left(2i\,  \nSeqBody{\mu}\pdv{}{\omega_\mu}+ \cot \left(\frac{\omega_\mu}{2}\right)\left(\nSeqBody{\mu}\times \LSeqOp_\mu\right)-\LSeqOp_\mu\right)\cdot \left(2i\,  \nSeqBody{\mu}\pdv{}{\omega_\mu}+ \cot \left(\frac{\omega_\mu}{2}\right)\left(\nSeqBody{\mu}\times \LSeqOp_\mu\right)-\LSeqOp_\mu\right) \nonumber \\
&= \frac{1}{4}\left(-4\pdv[2]{}{\omega_\mu}+ 2 i \cot \left(\frac{\omega_\mu}{2}\right)\left(\nSeqBody{\mu}\cross \LSeqOp_{\mu}\right)\cdot \nSeqBody{\mu}\right)\pdv{}{\omega_\mu}+ \cot^2\left( \frac{\omega_\mu}{2}\right)\left(\nSeqBody{\mu}\times \LSeqOp_\mu\right)\cdot \left(\nSeqBody{\mu}\times \LSeqOp_\mu\right)+ \LSeqOpSq_\mu \nonumber \\
&= - \pdv[2]{}{\omega_\mu} - \cot\left(\frac{\omega_\mu}{2}\right)\pdv{}{\omega_\mu}+ \frac{1}{4}\csc^2\left( \frac{\omega_\mu}{2}\right)\LSeqOpSq_\mu
\end{align}
where $\mu$ is not summed over in these expressions and we have used that
\begin{gather}
\LSeqOp_{\mu} \cdot \nSeqBody{\mu} = 0 \, ,\quad
\nSeqBody{\mu}  \cdot \LSeqOp_{\mu} =0 \, ,\quad
\nSeqBody{\mu}  \cdot \left(\nSeqBody{\mu}\cross  \LSeqOp_{\mu}\right) = 0 \, ,\quad
\left(\nSeqBody{\mu}\cross  \LSeqOp_{\mu}\right)\cdot \nSeqBody{\mu}  = 2 i \, ,\nonumber \\
\LSeqOp_{\mu}  \cdot \left(\nSeqBody{\mu}\cross  \LSeqOp_{\mu}\right) = 0 \, ,\quad
\left(\nSeqBody{\mu}\cross  \LSeqOp_{\mu}\right)\cdot \LSeqOp_{\mu}  = 0 \, ,\quad
\left(\nSeqBody{\mu}\times \LSeqOp_\mu\right)\cdot \left(\nSeqBody{\mu}\times \LSeqOp_\mu\right) = \LSeqOpSq_\mu \, .
\end{gather}
It is possible to evaluate the more complicated bilinears, for arbitrary lattice volumes, using the same method. 
Special care must be taken when evaluating the bilinears involving Rod 1 and Rod 2, due to non-commuting terms:
\begin{align}
\left[\LTotOpComp{a}, \RotEIndex{b}{c}\right] = i \epsilon^{abd}\RotEIndex{d}{c} \,, \quad\left[\LpTotOpComp{a}, \RotEIndex{b}{c}\right] = -i \epsilon^{acd}\RotEIndex{b}{d} \,.
\end{align}
Since the expressions themselves are not incredibly illuminating, we will simply tabulate them in \cref{sec:EEResults}. 

There are two key observations from \cref{sec:EEResults} that need to be commented on. 
The first is that the total number of bilinears that need to be calculated is independent of the size of the lattice considered. 
In particular, for bilinears that involve only one physical link, there are only six bilinear types that need to be evaluated:
\begin{align}
\boldsymbol{\CE}_{1L} \cdot \boldsymbol{\CE}_{1L},\quad \boldsymbol{\CE}_{1L} \cdot \boldsymbol{\CE}_{1R},\quad \boldsymbol{\CE}_{2L} \cdot \boldsymbol{\CE}_{2L},\quad\boldsymbol{\CE}_{2L} \cdot \boldsymbol{\CE}_{2R}, \quad \boldsymbol{\CE}_{\mu L} \cdot \boldsymbol{\CE}_{\mu L},\quad\boldsymbol{\CE}_{\mu L} \cdot \boldsymbol{\CE}_{\mu R}
\end{align}
where $\mu$ is not summed over in these expressions and the bilinears with different parity are related to these via the relations
\begin{align}
\boldsymbol{\CE}_{\kappa R} \cdot \boldsymbol{\CE}_{\kappa R} = \boldsymbol{\CE}_{\kappa L} \cdot \boldsymbol{\CE}_{\kappa L} \qquad \boldsymbol{\CE}_{\kappa R} \cdot \boldsymbol{\CE}_{\kappa L} = \boldsymbol{\CE}_{\kappa L} \cdot \boldsymbol{\CE}_{\kappa R}.
\end{align}
For bilinears that involve two different rods, there are now four classes of bilinears that must evaluate, each with four parity types. These four classes are 
\begin{gather}
\boldsymbol{\CE}_{1\zeta_1} \cdot \boldsymbol{\CE}_{2\zeta_2},\quad \boldsymbol{\CE}_{1\zeta_1} \cdot \boldsymbol{\CE}_{\mu \zeta_\mu},\quad \boldsymbol{\CE}_{2\zeta_2} \cdot \boldsymbol{\CE}_{\mu \zeta_\mu},\quad  \boldsymbol{\CE}_{\mu \zeta_\mu} \cdot \boldsymbol{\CE}_{\nu \zeta_\nu}\qquad \text{where} \quad \zeta_\kappa = L, R
\end{gather}
and their exact form (and dependence on $\zeta_\kappa$) is provided in \cref{sec:EEResults}. 

The second key observation is related to the Degree of Connectivity (DoC)~\cite{Grabowska:2022uos} in any one bilinear, as well as the total number of terms in any bilinear. In particular, the DoC is volume-independent, as can be seen by noting that the terms that are the most non-local depend on the variables of three physical links. Additionally, the number of terms in any one bilinear grows no faster than $\NL^2$. The ramifications of this, in particular with regards to resource scaling and non-locality, shall be discussed in greater detail in \cref{sec:ResourceScaling}.

In order to construct the electric Hamiltonian, the various bilinears must be combined together with particular coefficients,
\begin{align}
H_E = \frac{g^2}{2}\sum_{\kappa \kappa'}\sum_{\zeta_\kappa \zeta_{\kappa'}}{\cal C}^{\kappa \kappa'}_{\zeta_{\kappa} \zeta_{\kappa'}} \,\boldsymbol{\CE}_{\kappa \zeta_{\kappa}}\cdot \boldsymbol{\CE}_{\kappa' \zeta_{\kappa'}}
\end{align}
where, again, $\boldsymbol{\CE}_{\kappa \zeta_{\kappa}}\cdot \boldsymbol{\CE}_{\kappa' \zeta_{\kappa'}}$ are all given in \cref{sec:EEResults}. The coefficients ${\cal C}^{\kappa \kappa'}_{\zeta_{\kappa} \zeta_{\kappa'}}$ are integers that depend on the specific maximal-tree gauge-fixing convention; some of them will be zero. It is interesting to note that depending on which physical links are used to determine the body frame \ie which physical links are called Rod 1 and Rod 2, it might be possible to decrease the number of terms that have the maximum DoC. This depends heavily on the convention chosen for the maximal-tree gauge fixing procedure, so we will not expand on this further.

\subsection{Magnetic Hamiltonian}
The magnetic Hamiltonian is constructed out of loop variables, which are traces over various combinations of link variables. 
In two dimensions, any one loop operator can contain at most two physical links; in three dimensions, this increases to four. 
Furthermore, all possible link operators can be written in terms of dot and cross products of $\mathbf{n}_{i}$, which can be shown by explicit computation. 

Starting from the gauge link operator in the original basis in \cref{eq:XOrig} one obtains the traces over one, two, three and four physical links
\begin{align}
\text{Tr}\, \hat X_\kappa =&\, 2 \cos \frac{\omega_\kappa}{2}\\
\text{Tr}\, \hat X_\kappa \hat X_\rho =&\, 2 \cos \frac{\omega_\kappa}{2}\cos \frac{\omega_\rho}{2} - 2\left(\cos \thetaOri{\kappa}\cos \thetaOri{\rho}+ \cos \left(\phiOri{\kappa}-\phiOri{\rho}\right)\sin \thetaOri{\kappa}\sin \thetaOri{\rho}\right)\sin \frac{\omega_\kappa}{2}\sin \frac{\omega_\rho}{2} \\
=&\,2 \cos \frac{\omega_\kappa}{2}\cos \frac{\omega_\rho}{2} - 2\left(\mathbf{n}_{\kappa}\cdot \mathbf{n}_{\rho}\right)\sin \frac{\omega_\kappa}{2}\sin \frac{\omega_\rho}{2} \nonumber \\
\text{Tr}\, \hat X_\kappa \hat X_\rho \hat X_\lambda =&\, 2 \cos \frac{\omega_\kappa}{2}\cos \frac{\omega_\rho}{2}\cos \frac{\omega_\lambda}{2}-2 \left(\mathbf{n}_{\kappa}\cdot\left(\mathbf{n}_{\rho}\times \mathbf{n}_{\lambda}\right)\right)\sin \frac{\omega_\kappa}{2}\sin \frac{\omega_\rho}{2}\sin \frac{\omega_\lambda}{2}\nonumber \\
&-2\left(\left(\mathbf{n}_{\kappa}\cdot \mathbf{n}_{\rho}\right)\sin \frac{\omega_\kappa}{2}\sin \frac{\omega_\rho}{2}\cos \frac{\omega_\lambda}{2}+\left(\mathbf{n}_{\kappa}\cdot \mathbf{n}_{\lambda}\right)\sin \frac{\omega_\kappa}{2}\cos \frac{\omega_\rho}{2}\sin \frac{\omega_\lambda}{2}+\left(\mathbf{n}_{\rho}\cdot \mathbf{n}_{\lambda}\right)\cos \frac{\omega_\kappa}{2}\sin \frac{\omega_\rho}{2}\sin\frac{\omega_\lambda}{2}\right) \nonumber \\
\text{Tr}\, \hat X_\kappa \hat X_\rho \hat X_\lambda \hat X_\tau =&\,2 \cos \frac{\omega_\kappa}{2}\cos \frac{\omega_\rho}{2}\cos \frac{\omega_\lambda}{2}\cos \frac{\omega_\tau}{2} - 2 \left[\left(\mathbf{n}_{\kappa}\cdot \mathbf{n}_{\rho}\right)\sin \frac{\omega_\kappa}{2}\sin \frac{\omega_\rho}{2}\cos \frac{\omega_\lambda}{2}\cos \frac{\omega_\tau}{2}\right.\nonumber \\
&+\left(\mathbf{n}_{\kappa}\cdot \mathbf{n}_{\lambda}\right)\sin \frac{\omega_\kappa}{2}\cos \frac{\omega_\rho}{2}\sin \frac{\omega_\lambda}{2}\cos \frac{\omega_\tau}{2}+\left(\mathbf{n}_{\kappa}\cdot \mathbf{n}_{\tau}\right)\sin \frac{\omega_\kappa}{2}\cos \frac{\omega_\rho}{2}\cos \frac{\omega_\lambda}{2}\sin \frac{\omega_\tau}{2}\nonumber \\
&+\left(\mathbf{n}_{\rho}\cdot \mathbf{n}_{\lambda}\right)\cos \frac{\omega_\kappa}{2}\sin \frac{\omega_\rho}{2}\sin \frac{\omega_\lambda}{2}\cos \frac{\omega_\tau}{2}+\left(\mathbf{n}_{\rho}\cdot \mathbf{n}_{\tau}\right)\cos \frac{\omega_\kappa}{2}\sin \frac{\omega_\rho}{2}\cos \frac{\omega_\lambda}{2}\sin \frac{\omega_\tau}{2}\nonumber \\
&+\left.\left(\mathbf{n}_{\lambda}\cdot \mathbf{n}_{\tau}\right)\cos \frac{\omega_\kappa}{2}\cos \frac{\omega_\rho}{2}\sin \frac{\omega_\lambda}{2}\sin \frac{\omega_\tau}{2}\right] -2 \left[\left(\mathbf{n}_{\rho}\cdot\left(\mathbf{n}_{\lambda}\times \mathbf{n}_{\tau}\right)\right)\cos \frac{\omega_\kappa}{2}\sin \frac{\omega_\rho}{2}\sin \frac{\omega_\lambda}{2}\sin \frac{\omega_\tau}{2}\right. \nonumber \\
&+\left(\mathbf{n}_{\kappa}\cdot\left(\mathbf{n}_{\lambda}\times \mathbf{n}_{\tau}\right)\right)\sin \frac{\omega_\kappa}{2}\cos \frac{\omega_\rho}{2}\sin \frac{\omega_\lambda}{2}\sin \frac{\omega_\tau}{2}+\left(\mathbf{n}_{\kappa}\cdot\left(\mathbf{n}_{\rho}\times \mathbf{n}_{\tau}\right)\right)\sin \frac{\omega_\kappa}{2}\sin \frac{\omega_\rho}{2}\cos \frac{\omega_\lambda}{2}\sin \frac{\omega_\tau}{2} \nonumber \\
&+\left.\left(\mathbf{n}_{\kappa}\cdot\left(\mathbf{n}_{\rho}\times \mathbf{n}_{\lambda}\right)\right)\sin \frac{\omega_\kappa}{2}\sin \frac{\omega_\rho}{2}\sin \frac{\omega_\lambda}{2}\cos \frac{\omega_\tau}{2}\right]+2 \left[\left(\mathbf{n}_{\kappa}\cdot \mathbf{n}_{\rho}\right)\left(\mathbf{n}_{\lambda}\cdot \mathbf{n}_{\tau}\right)\right.\nonumber \\
&\left.-\left(\mathbf{n}_{\kappa}\cdot \mathbf{n}_{\lambda}\right)\left(\mathbf{n}_{\rho}\cdot \mathbf{n}_{\tau}\right)+\left(\mathbf{n}_{\kappa}\cdot \mathbf{n}_{\tau}\right)\left(\mathbf{n}_{\rho}\cdot \mathbf{n}_{\lambda}\right)\right]\sin \frac{\omega_\kappa}{2}\sin \frac{\omega_\rho}{2}\sin \frac{\omega_\lambda}{2}\sin \frac{\omega_\tau}{2}
\,.
\end{align}
Here $\kappa, \rho, \lambda$ and $\tau$ are the index of any of the $\NL$ rods. 
If the traces included conjugate transposes of link variables, the general structure shown above remains the same and the only change is in relative signs. 
For example,
\begin{align}
\text{Tr} U_\kappa U^\dagger_\rho &=2 \cos \frac{\omega_\kappa}{2}\cos \frac{\omega_\rho}{2} + 2\left(\mathbf{n}_{\kappa}\cdot \mathbf{n}_{\rho}\right)\sin \frac{\omega_\kappa}{2}\sin \frac{\omega_\rho}{2}
\,.
\end{align}

All angular dependence is included in the dot and cross products of the directions $\mathbf{n}_i$, and switching to the sequestered basis therefore just requires expressions for those. 
We find
\begin{align}
\mathbf{n}_{1}\cdot \mathbf{n}_{2}=&\, \cos \thetaDef \nonumber 
\\
\mathbf{n}_{1}\cdot \mathbf{n}_{\mu} =&\, \cos \thetaSeq{\mu}\nonumber 
\\
\mathbf{n}_{2}\cdot \mathbf{n}_{\mu} =&\,\cos \thetaDef \cos \thetaSeq{\mu}+\cos \phiSeq{\mu}\sin \thetaDef \sin \thetaSeq{\mu} \nonumber 
\\
\mathbf{n}_{\mu}\cdot \mathbf{n}_{\nu} =&\,\cos \thetaSeq{\mu} \cos \thetaSeq{\nu}+\cos\left( \phiSeq{\mu}-\phiSeq{\nu}\right)\sin \thetaSeq{\mu} \sin \thetaSeq{\nu} 
\\
\mathbf{n}_{1}\cdot \left(\mathbf{n}_{2}\times \mathbf{n}_{\mu}\right)=&\,\sin \thetaDef \sin \thetaSeq{\mu}\sin \phiSeq{\mu}\nonumber \\
\mathbf{n}_{1}\cdot \left(\mathbf{n}_{\mu}\times \mathbf{n}_{\nu}\right)=&\,-\sin \thetaSeq{\mu}\sin \thetaSeq{\nu}\sin\left( \phiSeq{\mu}-\phiSeq{\nu}\right) \nonumber \\
\mathbf{n}_{2}\cdot \left(\mathbf{n}_{\mu}\times \mathbf{n}_{\nu}\right)=&\,\sin \thetaDef\left( \sin \thetaSeq{\mu}\sin \phiSeq{\mu} \cos\thetaSeq{\nu} -\cos\thetaSeq{\mu}\sin \thetaSeq{\nu}  \sin\phiSeq{\nu}\right)-\cos \thetaDef\sin\thetaSeq{\mu}  \sin\thetaSeq{\nu} \sin (\phiSeq{\mu}-\phiSeq{\nu})\nonumber \\
\mathbf{n}_{\mu}\cdot \left(\mathbf{n}_{\nu}\times \mathbf{n}_{\lambda}\right)=&\,-\cos \thetaSeq{\mu}\sin \thetaSeq{\nu}\sin \thetaSeq{\lambda}\sin \left(\phiSeq{\nu}-\phiSeq{\lambda}\right)+\sin \thetaSeq{\mu}\cos \thetaSeq{\nu}\sin \thetaSeq{\lambda}\sin \left(\phiSeq{\mu}-\phiSeq{\lambda}\right) -\sin \thetaSeq{\mu}\sin \thetaSeq{\nu}\cos \thetaSeq{\lambda}\sin \left(\phiSeq{\mu}-\phiSeq{\nu}\right) \nonumber
\end{align}
which is easy to verify by realizing that all of these operators written in terms of the directions $\mathbf{n}_i$ are independent of the frame. 
For terms involving three directions, it is useful to note that $\mathbf{n}_\mu \cdot (\mathbf{n}_\nu \times \mathbf{n}_\lambda) = \mathbf{n}_\nu \cdot (\mathbf{n}_\lambda \times \mathbf{n}_\mu) = \mathbf{n}_\lambda \cdot (\mathbf{n}_\mu \times \mathbf{n}_\nu)$.
The lack of dependence on the Euler angles in the above expressions is consistent for two (related) reasons. The first is that all these expressions are scalars and therefore frame-independent. The second is that any one plaquette variable is always gauge-invariant and therefore cannot retain any information about the residual global gauge transformation. Similar to the electric Hamiltonian, it is important to mention that any one plaquette only depends on a finite number of physical link variables and furthermore, this number of variables is volume-independent. Additionally, much like in the case of the electric Hamiltonian, not all of the loop variables will appear in the Hamiltonian and the particular combination of loop variables and their specific prefactors will depend on the maximal-tree gauge-fixing convention chosen.

\section{Construction of Sequestered Mixed Basis}
\label{sec:mixed} 
In \cref{subsec:mixed_basis} we reviewed the construction in Ref.~\cite{DAndrea:2023qnr} of a mixed basis, which replaced the angular variables $\thetaOri{\kappa}$ and $\phiOri{\kappa}$ by the angular momentum quantum numbers $\LOri{\kappa}$ and $\MOri{\kappa}$. 
The overlap between states in the magnetic and mixed basis was given in \cref{eq:mixedoriginalOverlap} as
\begin{equation}
    \bra{\thetaOri{\kappa} \phiOri{\kappa}}\ket{\LOri{\kappa} \MOri{\kappa}}
      = \YLMSymbol_{\LOri{\kappa}\MOri{\kappa}}(\thetaOri{\kappa},\phiOri{\kappa}) \,,
\end{equation}
where we ignore the dependence on hyperspherical radial coordinates $\omega_\kappa$ for now. The quantum numbers $\LOri{\kappa}$ and $\MOri{\kappa}$ are the eigenvalues of the angular momentum operators forming the CSCO of the original mixed basis
\begin{align}
    \{\text{CSCO}\}_O = \left\{\LOpSq_1,\LOpComp{z}_1,\LOpSq_2,\LOpComp{z}_2,\dots \LOpSq_\NL,\LOpComp{z}_\NL \right\}
    \,.
\end{align}
This CSCO contains two operators for each loop $\kappa$, for a total of $2\NL$ operators.
A given basis state is therefore given by 
\begin{equation}
\ket{\LOri{1}, \MOri{1}, \ldots ,\LOri{\NL}, \MOri{\NL}} = \ket{\LOri{1}, m_1}\otimes \cdots \otimes \ket{\LOri{\NL}, \MOri{\NL}}
    \,,
\end{equation}
and the action of the operators in the CSCO on this state are
\begin{align}
    \LOpSq_\kappa \ket{\LOri{1}, \MOri{1}, \ldots ,\LOri{\NL}, m_{\NL}} &= \LOri{\kappa} (\LOri{\kappa}+1) \ket{\LOri{1}, \MOri{1}, \ldots ,\LOri{\NL}, m_{\NL}} \nonumber\\
    \LOpComp{z}_\kappa \ket{\LOri{1}, \MOri{1}, \ldots ,\LOri{\NL}, m_{\NL}} &= \MOri{\kappa} \ket{\LOri{1}, \MOri{1}, \ldots ,\LOri{\NL}, m_{\NL}}
    \,.
\end{align}

The sequestered mixed basis is defined through the eigenvalues of a different choice of CSCO, which has a few important differences.
Since the remaining gauge redundancy originates from the overall rotation of the system, we choose the total momentum operators $\LTotOpSq$, $\LTotOpComp{z}$ and $\LpTotOpComp{z}$ to be three of the $2\NL$ operators that make up this CSCO. 
We also choose to include the $2(\NL-2)$ operators $\LSeqOpSq_\mu$ and $\LSeqOpComp{z}_\mu$ with $3 \leq \mu \leq \NL$.
This leaves one final operator, which must depend on $\thetaDef$.

It turns out that there is not one unique choice for this operator, but an entire class of operators. A particularly convenient one, in terms of resource scaling, is
\begin{align}
\NSeqSq &= -\pdv[2]{\thetaDef} - \cot\thetaDef \pdv{\thetaDef} - \csc^2\thetaDef \left(\pdv{}{\gamma}-\sum_{\mu = 3}^\NL \pdv{}{\phiSeq{\mu}}\right)^2 \nonumber \\
&= -\pdv[2]{\thetaDef} - \cot\thetaDef \pdv{\thetaDef} + \csc^2 \thetaDef \LSigmaOpSq \,.
\end{align}
One can show through direct computation that this operator commutes with all the others operators in the CSCO. Furthermore, the operator $\NSeqSq$ has a few key properties. The first is its relation to the generating function of associate Legendre polynomials. Recall that associate Legendre polynomials obey the eigenstate equation
\begin{align}
\left(-\pdv[2]{\theta} - \cot\thetaDef \pdv{\theta} + m^2 \csc^2\theta\right) \ALPSymbol_n^m(\cos\theta) = n(n+1)\ALPSymbol_{n}^{m}(\cos\theta)  
\label{eq:ALPGen}
\end{align}
where $n, m$ are integers and $n\geq m \geq -n$ and $n \geq 0$. 
As long as $\sigma$, playing the role of $m$, is chosen to be the eigenvalue of $\LSigmaOpSq$, $\ALPSymbol_n^\sigma(\cos \thetaDef)$ will be an eigenfunction of the operator $\NSeqSq$, with eigenvalues $n(n+1)$.

This does beg the question of why this is the operator that we choose for the CSCO. One might wonder, for example, whether it is possible to use Legendre polynomials $\ALPSymbol_n$, defined as $\ALPSymbol_n(\cos \thetaDef)\equiv \ALPSymbol_{n}^{0}(\cos \thetaDef)$, as eigenfunctions of the sequestered basis Hilbert space. This choice ends up being a particularly poor one, as there are terms in the electric bilinears that are proportional to powers of $\csc\thetaDef$ and $\cot\thetaDef$. However, the overlap between two Legendre polynomials and such term, which is given by
\begin{align}
\int d\left(\cos \theta\right) \csc^2(\theta) \ALPSymbol_n(\cos \theta) \ALPSymbol_{n'}(\cos \theta) \quad \text{or} \quad \int d\left(\cos \theta\right) \cot \theta \csc \theta \ALPSymbol_n(\cos \theta) \ALPSymbol_{n'}(\cos \theta)
\end{align}
is divergent. One way to interpret this result is that the Legendre polynomials do not span the Hilbert space of this theory. Another choice is to work with $\ALPSymbol_{n}^{m}(\cos \thetaDef)$, with $m$ some non-zero, but fixed, integer value. As long as $m \geq 2$, the integrals
\begin{align}
\int d\left(\cos \theta\right) \csc^2(\theta) \ALPSymbol_{n}^{m}(\cos \theta) \ALPSymbol_{n'}^{m}(\cos \theta) \quad \text{or} \quad \int d\left(\cos \theta\right) \cot \theta \csc \theta \ALPSymbol_{n}^{m}(\cos \theta) \ALPSymbol_{n'}^{m}(\cos \theta)
\end{align}
are finite.  However, these integrals are also non-zero for many combinations of $(n, n')$, which is not favorable from the perspective of resource efficiency. Therefore, for these reasons, we choose $\NSeqSq$ to be the last operator needed to complete the CSCO. As a last comment, it turns out that this operator is the same as $\hat L_2^2$, defined by \cref{eq:LSeqOpdef}, but written in terms of sequestered basis variables. We choose to use $\NSeqSq$ as the name of this operator in order to allow for the possibility of using other operators for this last component of the CSCO.

The CSCO in the sequestered basis is therefore given by
\begin{align}
    \{\text{CSCO}\}_S &= \left\{\NSeqSq,\LSeqOpSq_3,\LSeqOpComp{z}_3, \dots, \LSeqOpSq_\NL,\LSeqOpComp{z}_\NL,\LTotOpSq, \LTotOpComp{z}, \LpTotOpComp{z}\right\}
\end{align}

A state in the sequestered mixed basis, ignoring the $\omega$ quantum numbers for now, is therefore characterized by integers $n$, $\{\LSeq{\mu} ,\MSeq{\mu}\}$, $\LTot,  \MTot,  \NTot$, with the basis states given by
\begin{align}
\label{eq:seq_mixed_basis_state}
    \ket{\Omega} = \ket{\NSeq} \otimes \left(\prod_{\mu=3}^{\NL} \ket{\LSeq{\mu} ,\MSeq{\mu}}\right) \otimes \ket{\LTot,  \MTot,  \NTot} \equiv \ket{\NSeq , \{\LSeq{\mu} ,\MSeq{\mu}\}; \LTot,  \MTot,  \NTot}
    \,.
\end{align}
The action of the operators on the basis states of the sequestered basis Hilbert space are 
\begin{gather}
\NSeqSq \ket{\Omega} = n(n+1) \ket{\Omega} \qquad  \LSeqOpSq_\mu \ket{\Omega} = \LSeq{\mu}(\LSeq{\mu}+1) \ket{\Omega} \qquad\LSeqOpComp{z}_\mu \ket{\Omega}= \MSeq{\mu} \ket{\Omega} \nonumber \\
\LTotOpSq \ket{\Omega}= \LTot(\LTot+1) \ket{\Omega} \qquad \LTotOpComp{z} \ket{\Omega} = \MTot \ket{\Omega}\qquad \LpTotOpComp{z} \ket{\Omega} = -\NTot \ket{\Omega}\,.
\end{gather}
where the negative sign in the last equation is due to the fact that $\LpTotOp$ satisfies anomalous commutation relations.
From these expressions one also finds 
\begin{align}
\LSigmaOpComp{z}\ket{\Omega} = -(N + \sum_\mu \MSeq{\mu})\ket{\Omega} \equiv - \sigma \ket{\Omega}
    \,.
\end{align}

The overlap with the sequestered basis in angular representation can be written as
\begin{align}
\braket{\thetaDef, \{\thetaSeq{\mu}, \phiSeq{\mu}\}; \alpha, \beta, \gamma}{\NSeq, \{\LSeq{\mu} \,\MSeq{\mu}\}; \LTot \MTot \NTot} = \ALPSymbol_n^\sigma(\cos \thetaDef) \left(\prod_{\mu = 3}^\NL\YLM{\mu}\right)\WD
\,,
\end{align}
where the Spherical Harmonics, the associated Legendre polynomials and the Wigner D functions, are discussed in more detail in \cref{app:normalizations}. Note that we use an unconventional normalization for both the associate Legendre polynomials and Wigner D functions.

The allowed values for the sequestered basis quantum numbers are as follows:
\begin{align}
\label{eq:intConstraints}
\LSeq{\mu} &\in \mathbb{N}^0 \qquad \MSeq{\mu} \in \left[-\LSeq{\mu}, \LSeq{\mu}\right] \nonumber \\
\LTot &\in \mathbb{N}^0 \qquad \MTot \in \left[-\LTot, \LTot\right],\quad \NTot \in \left[-\LTot, \LTot\right] \nonumber \\
\NSeq &\in \mathbb{N}^0, \geq |\sigma|
\end{align}
The constraint on $n$ arises from the property of associated Legendre polynomials (see \cref{eq:ASPZero}).
This constraint results in, for a system with three rods, the state $\ket{3, 2, 1, 2; -1, 1}$ is an allowed state, while $\ket{0, 2, 1, 2; -1, 1}$ is not, as $\sigma = 2$. The complexity of constructing this basis is addressed in \cref{sec:ResourceScaling}.

Reintroducing the radial coordinates, the Hilbert space of the mixed-sequestered basis is spanned by the states
\begin{align}
\ket{\{\omega_\kappa\}; \NSeq, \{\LSeq{\mu}, \MSeq{\mu}\}; \LTot \MTot \NTot}
\end{align}
with 
\begin{align}
&\braket{\{\omega'_\kappa\}, \thetaDef,\{\thetaSeq{\mu}, \phiSeq{\mu}\}; \alpha, \beta, \gamma }{\{\omega_\kappa\}; \NSeq, \{\LSeq{\mu}, \MSeq{\mu}\}; \LTot \MTot \NTot} \nonumber \\ 
&\qquad\qquad\qquad=\prod_{\kappa}\left(\frac{\delta(\omega_\kappa-\omega'_\kappa)}{2\sin \frac{\omega_\kappa}{2}}\right)\ALPSymbol_n^\sigma(\cos \thetaDef) \left(\prod_{\mu = 3}^\NL\YLM{\mu}\right)\WD \, .
\label{eq:SMBOverlap}
\end{align}
This sequestered basis makes it easy to fix the last remaining gauge redundancy by choosing the appropriate values for $L$, $M$ and $N$. It is important to note that there is no operator in the Hamiltonian that changes the quantum numbers $\LTot$ and $\MTot$. However, the quantum number $\NTot$ is allowed to change, though it is bounded by $\LTot \geq |\NTot|$. 
In particular, in a pure gauge theory, gauge fixing would simply set $L = M = N = 0$. As a last comment, one might worry that using associate Legendre polynomials as eigenfunctions might lead to a set of basis states that are not complete and orthogonal. While it is true that associate Legendre polynomials are not generally mutually orthogonal, this basis is complete and orthonormal due to $\sigma$ being related to $\MSeq{\mu}$ and $\NTot$. This is discussed in more detail in \cref{sec:MBEig}.

The Hamiltonian is given by the same expressions as those presented in \cref{sec:seq_basis,sec:EEResults}, but the angular information, given in terms of dot and cross products involving $\nSeqBody{\mu}$, $\LSeqOp_\mu$, $\LSeqOpComp{z}_\sigma$, $\LTotOp$, $\LpTotOp$ as well as derivatives $\pdv{\thetaDef}$ now needs to be written in the new basis states. Additionally, due to the factors of $\csc (\omega_\kappa/2)$ in \cref{eq:SMBOverlap}, the coefficients in front of the first and second order derivatives with respect to $\omega_\kappa$ are altered. The change in normalization with respect to $\omega_\kappa$ can easily be done via the substitution
\begin{align}
-\pdv[2]{}{\omega}-\cotw{}\pdv{}{\omega}&\Longrightarrow -\pdv[2]{}{\omega} - \frac{1}{4} 
\end{align}
for bilinears that only include one rod and
\begin{align}
-\CA_{\delta \delta} \pdv{}{\omega_\mu}{\omega_\nu}&-\left(\CA_{\delta ;\mu}\cotw{\nu}+\CA_{0;\mu}\right)\pdv{}{\omega_\mu}-\left(\CA_{\delta ;\nu}\cotw{\mu}+\CA_{0;\nu}\right)\pdv{}{\omega_\nu} \Longrightarrow - \CA_{\delta \delta}\pdv{}{\omega_\mu}{\omega_\nu} \nonumber \\
& - \frac{1}{2}\left(2 \CA_{0;\nu}+ \left(2 \CA_{\delta;\nu}-\CA_{\delta \delta}\right)\cotw{\mu}\right)\pdv{}{\omega_\nu}- \frac{1}{2}\left(2 \CA_{0;\mu}+ \left(2 \CA_{\delta;\mu}-\CA_{\delta \delta}\right)\cotw{\nu}\right)\pdv{}{\omega_\mu} \nonumber \\
&+\frac{1}{4}\left(2 \CA_{0\mu}\cotw{\mu}+2 \CA_{0\nu}\cotw{\nu}+\left(2\left(\CA_{\delta;\mu}+\CA_{\delta;\nu}\right)-\CA_{\delta \delta}\right)\cotw{\mu}\cotw{\nu}\right) 
\end{align}

Determining the behavior of the other operators on the new basis states is a bit more involved to determine. However, it can be done. For example, the action of the scalar product of the direction of $\nSeqBody{\mu}$ and the operator $\LSeqOp_{\nu}$ on the state $\ket{\Omega}$ can be written as
\begin{align}
\label{eq:nmunnu}
\nSeqBody{\mu}\cdot \LSeqOp_{\nu}\ket{\Omega} =&\frac{1}{2}\sum_{\delta_\ell=\pm}\sum_{\delta_m =\pm}\bigg[\delta_{\LSeq{\mu}}\delta_m \OpCoeff{\delta_\ell}{-\delta_m}{\LSeq{\mu}}{\MSeq{\mu}}\OpCoeff{0}{\delta_m}{\LSeq{\nu}}{\MSeq{\nu}}\ket{\LSeq{\mu}+\delta_\ell, \MSeq{\mu}-\delta_m}\ket{\LSeq{\nu}, \MSeq{\nu}+\delta_m} \nonumber \\
&+ \OpCoeff{\delta_\ell}{0}{\LSeq{\mu}}{\MSeq{\mu}}\OpCoeff{0}{0}{\LSeq{\nu}}{\MSeq{\nu}}\ket{\LSeq{\mu}+\delta_\ell, \MSeq{\mu}}\ket{\LSeq{\nu}, \MSeq{\nu}}\bigg]\otimes \ket{\NSeq}\otimes\left(\prod_{\rho \neq \mu, \nu} \ket{\LSeq{\rho}, \MSeq{\rho}}\right)\otimes \ket{\LTot,\MTot, \NTot}
\end{align}
where we have used the shorthand notation
\begin{align}
\ket{\ell + \delta_\ell,\, m} &\equiv \begin{cases}
\ket{\ell+1,\, m} \qquad &\qquad \delta_\ell = + \\
\ket{\ell-1,\, m } \qquad &\qquad
 \delta_\ell = - \\
\end{cases}
\end{align}
and other related ones. 
The expression for the coefficients $\OpCoeff{\delta_\ell}{\delta_m}{\LSeq{}}{\MSeq{}}$ are given in \cref{app:RecursionRel}, and are closely related to the factors that arise in the action of raising and lowering operators in \sutwo.
These coefficients vanish if the values of $\ell$ or $m$ are changed to fall outside of the allowed range $\abs{m} \leq \ell$.

An important aspect of \cref{eq:nmunnu} is that it implies that this operator raises and lowers the quantum numbers $\LSeq{\mu}$, $\MSeq{\mu}$, $\LSeq{\nu}$ and $\MSeq{\nu}$ by one unit up or down.
In fact, any operator that only involves $\nSeqBody{\mu}$ and $\LSeqOp_\mu$ will depend on at most two directions $\mu$ and $\nu$, and only affects the quantum numbers $\LSeq{\mu}$, $\MSeq{\mu}$, $\LSeq{\nu}$ and $\MSeq{\nu}$, by at most one unit.
Also note, that this operator changes the values of $\MSeq{\mu}$ and $\MSeq{\nu}$ such that their sum, and therefore the value of $\sigma$, remains unchanged. 
Since $\sigma$ is associated with rods 1 and 2, it makes sense that its value remains unchanged by this operator.

A more complicated operator is 
\begin{align}
\label{eq:eta2dotetamu}
\nSeqBody{2}\cdot \nSeqBody{\mu}\ket{\Omega}=&\frac{1}{2}\sum_{\substack{\delta_n =\pm\\\delta \LSeq{\mu}=\pm}}\sum_{\delta m = \pm}\left(\OpCoeff{\delta_n}{0}{\NSeq}{\sigma}\OpCoeff{\delta_\LSeq{\mu}}{0}{\LSeq{\mu}}{\MSeq{\mu}}\ket{\LSeq{\mu}+\delta_\LSeq{\mu},\MSeq{\mu}}+\delta_n \delta_\LSeq{\mu}\OpCoeff{\delta_n}{\delta_m}{\NSeq}{\sigma}\OpCoeff{\delta_\LSeq{\mu}}{\delta_m}{\LSeq{\mu}}{\MSeq{\mu}}\ket{\LSeq{\mu}+\delta_\LSeq{\mu}, \MSeq{\mu}+\delta_m}\right)\nonumber \\
&\otimes \ket{\NSeq+\delta_n}\otimes\left(\prod_{\rho \neq \mu} \ket{\LSeq{\rho}, \MSeq{\rho}}\right)\otimes \ket{\LTot,\MTot, \NTot}
\end{align}
As expected, this operator changes the quantum numbers $n$, $\LSeq{\mu}$ and $\MSeq{\mu}$, but again by at most one unit. 
Since this operator involves the direction $\nSeqBody{2}$ of rod 2, and only a single direction $\nSeqBody{\mu}$, it is expected that the value of $\sigma$ can be changed by this operator. 
The coefficient $\OpCoeff{\delta_n}{0}{\NSeq}{\sigma}$ ensures that the action of this operator annihilates the system if it leads to quantum numbers violating $\abs{\sigma} \leq n$.

In \cref{app:RecursionRel} all angular operators required for the terms in the electric and magnetic Hamiltonian are given in the mixed basis.
While some of the expressions are somewhat lengthy, they all have a similar form to those presented above. 
In particular, all of them can change quantum numbers by at most a single unit, and all have coefficients appropriate to ensure that their action only gives rise to states that satisfy the constraints given in \cref{eq:intConstraints}.
While we refer the reader to \cref{app:RecursionRel} for the explicit expressions, we summarize the operators in \cref{tab:1,tab:2,tab:3}, where we show which quantum numbers are changed by a given operator, and how many terms are required for each of them.

As an example of how to read these tables, the operator $\nSeqBody{\mu}\cdot \LSeqOp_{\nu}$ in Eq.~\eqref{eq:nmunnu} appears in the operator $\boldsymbol{\CE}_{\mu\zeta_\mu} \cdot \boldsymbol{\CE}_{\nu\zeta_\nu}$, which is summarized in Table~\ref{tab:3}. Using the notation of Table~\ref{tab:3}, the operator $\nSeqBody{\mu}\cdot \LSeqOp_{\nu}$ can changes quantum numbers in two different ways: either as a change in $\Delta \LSeq{\mu} = \pm 1$ OR a change in $\Delta \ell_\nu = \pm 1$, $\Delta m_\mu = \pm 1$ and $\Delta m_\nu = \pm 1$, as long as $\sigma$ is conserved.
In a similar way, the operator $\nSeqBody{2}\cdot \nSeqBody{\mu}$ in Eq.~\eqref{eq:eta2dotetamu} appears in Table~\ref{tab:3} and can change the quantum numbers in two different ways: either as a change in $\Delta n = \pm 1$ and $\Delta \LSeq{\mu} = \pm 1$ OR a change in $\Delta n = \pm 1, \Delta \LSeq{\mu} = \pm 1$ and  $\Delta \MSeq{\mu} = \pm 1$.

\begin{table}[h!]
\begin{tabular}{|c|c|c|c|c|}
\hline
\text{Bilinear}&\text{Changing quantum numbers}&$\sigma$ change&\text{Total (four rods)}& \text{Total (N rods) }\\\hline\hline
$\boldsymbol{\CE}_{1\zeta} \cdot \boldsymbol{\CE}_{1\zeta'}$&  no change & & 1 & 1 \\
                    &  $\Delta N = \pm 1$ & $\Delta\sigma = \pm 1$ & 2 & 2\\
                    &  $\Delta m_\mu^{[i]} = \pm 1$ & $\Delta\sigma = \pm 1$ & 4 & $2\NL - 4$\\
                    &  $\Delta N = \pm 1$; $\Delta m_\mu^{[i]} = \pm 1$ & $\Delta\sigma = 0$ & 4 & $2\NL - 4$\\
                    &  $\Delta m_\mu^{[i]} = \pm 1$, $\Delta m_\nu^{[i]} = \pm 1$, & $\Delta\sigma = 0$ & 2 & $(\NL-2)(\NL-3)$ \\ \hline
                    &\textbf{total:}& & 13 & $\NL(\NL-1) +1$\\ \hline\hline
$\boldsymbol{\CE}_{2\zeta} \cdot \boldsymbol{\CE}_{2\zeta'}$&  no change & & 1 & 1\\ \hline
                    &\textbf{total:}& & 1 & 1\\ \hline\hline
$\boldsymbol{\CE}_{\mu\zeta} \cdot \boldsymbol{\CE}_{\mu\zeta'}$& no change & & 1 & 1 \\ \hline
                    &\textbf{total:}& & 1  & 1\\ \hline
\end{tabular}
\caption{Bilinears involving a single rod. Note that the superscript [i] on $\Delta m_\mu^{[i]}$ and $\Delta m_\nu^{[i]}$ indicates that this an `indefinite' index and therefore these terms will come with a sum over $\mu, \nu = [3, \NL]$. 
When $\mu$ appears in the bilinear subscripts, just as in $\boldsymbol{\CE}_{\mu\zeta} \cdot \boldsymbol{\CE}_{\mu\zeta'}$, that index is not summed over. \label{tab:1}}
\end{table}
\begin{table}[h!]
\begin{tabular}{|c|c|c|c|c|}
\hline
\text{Bilinear}&\text{Changing quantum numbers}&$\sigma$ change&\text{Total (four rods)}& \text{Total (N rods) }\\\hline\hline
$\boldsymbol{\CE}_{1\zeta_1} \cdot \boldsymbol{\CE}_{2 \zeta_2}$& no change & & 1 & 1 \\
                    &  $n$ & $\Delta\sigma = \pm 0$ & 2 & 2\\
                    &  $\Delta N = \pm 1$ & $\Delta\sigma = \pm 1$ & 2 & 2 \\
                    &  $\Delta m_\mu^{[i]} = \pm 1$ & $\Delta\sigma = \pm 1$ & 4 & $2 \NL-4$ \\
                    &  $\Delta n = \pm 1$, $\Delta N = \pm 1$ & $\Delta\sigma = \pm 1$ & 4 & 4 \\
                    &  $\Delta n = \pm 1$, $\Delta m_\mu^{[i]} = \pm 1$ & $\Delta\sigma = \pm 1$ & 8 & $4 \NL - 8$ \\ \hline
                    &\textbf{total:}& & 21 & $3 (2 \NL- 1)$\\ \hline\hline
$\boldsymbol{\CE}_{1\zeta_1} \cdot \boldsymbol{\CE}_{\mu\zeta_\mu}$&  no change & & 1 & 1 \\
                    & $\Delta \ell_\mu = \pm 1$ & $\Delta\sigma = 0$ & 2 & 2\\
                    & $\Delta m_\mu = \pm 1$ & $\Delta\sigma = \pm 1$ & 2 & 2 \\
                    & $\Delta \ell_\mu = \pm 1$, $\Delta m_\mu = \pm 1$ & $\Delta\sigma = \pm 1$ & 4 & 4 \\
                    & $\Delta m_\mu = \pm 1$, $\Delta N = \pm 1$ & $\Delta \sigma = 0$ & 2 & 2 \\
                    & $\Delta m_\mu = \pm 1$, $\Delta m_\nu^{[i]} = \pm 1$ & $\Delta \sigma = 0$ & 2 & $2(\NL - 3)$ \\
                    & $\Delta \ell_\mu = \pm 1$, $\Delta m_\mu = \pm 1$, $\Delta N = \pm 1$ & $\Delta \sigma = 0$ & 4 & 4 \\
                    & $\Delta \ell_\mu = \pm 1$, $\Delta m_\mu = \pm 1$, $\Delta m_\nu^{[i]} = \pm 1$ & $\Delta \sigma = 0$ & 4 & $4(\NL -3)$ \\ \hline
                    &\textbf{total:}& & 21 & $3 (2 \NL- 1)$\\ \hline 
\end{tabular}
\caption{Bilinears involving rod 1 and any other rod. Note that the superscript [i] on $\Delta m_\nu^{[i]}$ indicates that this an `indefinite' index and therefore these terms will come with a sum over $\nu = [3, \NL]$. When $\mu$ appears in the bilinear subscripts, just as in $\boldsymbol{\CE}_{\mu\zeta} \cdot \boldsymbol{\CE}_{\mu\zeta'}$, that index is not summed over. So for example, the terms that change $\LSeq{\mu}$ and $\MSeq{\mu}$ are not summed over\label{tab:2}.}
\end{table}
\begin{table}
\begin{tabular}{|c|c|c|c|c|}
\hline\text{Bilinear}&\text{Changing quantum numbers}&$\sigma$ change&\text{Total (four rods)}& \text{Total (N rods) }\\\hline\hline
$\boldsymbol{\CE}_{2\zeta_2} \cdot \boldsymbol{\CE}_{\mu\zeta_\mu}$ & no change & & 1 & 1 \\
                    & $\Delta n = \pm 1$ & $\Delta \sigma = 0$ & 2 & 2 \\
                    & $\Delta \ell_\mu = \pm 1$ & $\Delta \sigma = 0$ & 2 & 2\\
                    & $\Delta m_\mu = \pm 1$ & $\Delta \sigma = \pm 1$ & 2 & 2 \\
                    & $\ell_\mu = \pm 1$, $\Delta m_\mu = \pm 1$ & $\Delta \sigma = \pm 1$ & 4 & 4 \\
                    & $\Delta n = \pm 1$, $\Delta \ell_\mu = \pm 1$ & $\Delta \sigma = 0$ & 4 & 4 \\
                    & $\Delta n = \pm 1$, $\Delta m_\mu = \pm 1$ & $\Delta \sigma = \pm 1$ & 4 & 4 \\
                    & $\Delta n = \pm 1$, $\Delta \ell_\mu = \pm 1$ $\Delta m_\mu = \pm 1 $ & $\Delta \sigma = \pm 1$ & 8 & 8\\ \hline
                    & \textbf{total:} & & 27 & 27 \\ \hline\hline
$\boldsymbol{\CE}_{\mu\zeta_\mu} \cdot \boldsymbol{\CE}_{\nu\zeta_\nu}$ & \text{no change} & & 1 & 1 \\
                    & $\Delta \ell_\mu = \pm 1$ & $\Delta \sigma = 0$ & 2 & 2 \\
                    & $\Delta \ell_\nu = \pm 1$ & $\Delta \sigma = 0$ & 2 & 2 \\
                    & $\Delta \ell_\mu = \pm 1$, $\Delta \ell_\nu = \pm 1$ & $\Delta \sigma = 0$ & 4 & 4 \\
                    & $\Delta m_\mu = \pm 1$, $\Delta m_\nu = \pm 1$ & $\Delta \sigma = 0$ & 2 & 2 \\
                    & $\Delta \ell_\mu = \pm 1$, $\Delta m_\mu = \pm 1$, $\Delta m_\nu = \pm 1$ & $\Delta \sigma = 0$ & 4 & 4 \\ 
                    & $\Delta \ell_\nu = \pm 1$, $\Delta m_\mu = \pm 1$, $\Delta m_\nu = \pm 1$ & $\Delta \sigma = 0$ & 4 & 4 \\
                    & $\Delta \ell_\mu = \pm 1$, $\Delta \ell_\nu = \pm 1$, $\Delta m_\mu = \pm 1$, $\Delta m_\nu = \pm 1$ & $\Delta \sigma = 0$ & 8 & 8 \\ \hline
                    &\textbf{total:} & & 27 & 27\\ \hline
\end{tabular}
\caption{Bilinears involving two different rods, but not rod 1. Note that in these bilinears, there is no sum over the $\NL-2$ rods that are not Rod 1 nor Rod 2. \label{tab:3}}
\end{table}

\section{Quantum simulation resource scaling}
\label{sec:ResourceScaling}

In this section, we study the asymptotic gate complexity of simulating the time evolution of the sequestered mixed-basis Hamiltonian as a function of the number of lattice sites.
Using simple counting arguments, we show that time evolution can be simulated, using either product formulas or quantum signal processing, for a gate cost that is polynomial in the number of lattice sites.
Before discussing the details, we stress that the purpose of this section is to demonstrate that, despite enforcing the final global gauge-symmetry, the gate cost does not scale exponentially with the volume, and not to come up with the most efficient implementation possible.
As already discussed in \cref{sec:mixed}, the number of lattice sites that a given operator acts on is volume independent, which implies that determining the volume scaling of the gate cost can be done independently of how the gate count scales with the local bosonic truncation.
For this reason, we leave a detailed study of how the gate cost scales with the number of qubits used to represent each $\omega_\kappa$ and the cutoffs $\ell_\text{max}$ and $n_\text{max}$ for future work.

To simplify the notation required to understand the volume scaling, we write 
\begin{equation}
\label{eq:H_for_counting}
    H_E = \sum_{j=0}^{N_E-1} H^{(E)}_j, \qquad H_B = \sum_{j=0}^{N_B-1} H^{(B)}_j,
\end{equation}
where $H^{(E)}_j (H^{(B)}_j)$ represent operators in $H_E (H_B)$ that act on a finite number of rods, and $N_E (N_B)$ is number of summands in $H_E (H_B)$.
While the decomposition in \cref{eq:H_for_counting} is not unique, the asymptotic scaling of $N_E$ and $N_B$ will be the same, regardless of how one groups the terms.
Furthermore, while different maximal-trees will result in slightly different Hamiltonians, any maximal-tree will contain all electric bilinears of the form $\boldsymbol{\CE}_{\kappa \zeta} \cdot \boldsymbol{\CE}_{\kappa \zeta'} \forall \kappa \in \NL$.
As discussed in more detail below, this observation implies that, while the overall pre-factor of the gate complexity for different maximal-tree choices will generally differ, the asymptotic volume scaling is independent of the maximal-tree used.

The magnetic Hamiltonian is a sum of traces of products of plaquette operators, the number of which increases linearly with the number of sites, i.e., $N_B = \CO(\NL)$.
Turning to the electric Hamiltonian, the most complicated term is of the form $\boldsymbol{\CE}_{1 \zeta} \cdot \boldsymbol{\CE}_{1 \zeta'}$, which, as seen in \cref{tab:1}, is a sum of $\CO(\NL^2)$ terms. 
Relative to $\boldsymbol{\CE}_{1 \zeta} \cdot \boldsymbol{\CE}_{1 \zeta'}$, all other bilinears in $H_E$ contribute a subleading number of terms, which implies that the number of terms in the electric Hamiltonian scales as $N_E = \CO(\NL^2)$.

The cost of simulating this theory using product formulas is the cost of a single Trotter step times the number of Trotter steps needed to simulate time evolution for time $t$ to error $\epsilon$.
A single Trotter step is composed of stages, and each stage generally requires exponentiating each term in the Hamiltonian.
For a $p^{\rm th}$ order product formula, the number of stages generally scales exponentially with $p$~\cite{Childs:2019hts}; for this reason one often chooses $p = \CO(1)$, which, combined with the fact that the Hamiltonian is a sum of $\CO(\NL^2)$ terms, implies the gate cost of a single Trotter step is $\CO(\NL^2)$.
The number of Trotter steps $N_\text{steps}$ required for a $p^{\rm th}$ order PF (often referred to as the `Trotter number') is given by~\cite{Childs:2019hts}
\begin{equation}
    N_\text{steps} = \CO\left(\frac{\tilde{\alpha}^{\frac{1}{p}}t^{1+\frac{1}{p}}}{\epsilon^{\frac{1}{p}}} \right),
\end{equation}
where the parameter $\tilde{\alpha}$ is a measure of the non-commutivity of terms in the Hamiltonian, and is given by
\begin{equation}
    \tilde{\alpha} = \sum_{\gamma_1, \gamma_2, \dots, \gamma_{p+1}=1}^\Gamma \| [H_{\gamma_{p+1}}, \dots, [H_{\gamma_2}, H_{\gamma_1}]] \|, \qquad H_{\gamma_\kappa} \in \{H_j^{(E)}, H_j^{(B)} \},
\end{equation}
where $\Gamma = N_E + N_B = \CO(\NL^2)$ is the total number of terms in the full Hamiltonian.
Assuming all commutators in $\tilde{\alpha}$ are non-zero for simplicity, the number of commutators in the sum over all $\gamma_j$'s is $\CO(\NL^{2(p+1)})$.
This implies that the volume scaling of the parameter $\tilde{\alpha}$ is bounded by $\tilde{\alpha} = \CO(\NL^{2(p+1)})$.
Multiplying the number of Trotter steps by the cost of a single step (recalling that $p = \CO(1)$) leads to a total gate cost
\begin{equation}
\label{eq:cost_pf}
    \text{Gates}_{\rm PF} = \CO\left(\frac{\NL^{4+\frac{2}{p}} t^{1+\frac{1}{p}}}{\epsilon^{\frac{1}{p}}} \right).
\end{equation}
The gate cost of simulating this theory using product formulas is therefore polynomial in the number of lattice sites.

As an alternative to product formulas, the time-evolution operator can also be implemented using near-optimal techniques based on Quantum Signal Processing~\cite{Low:2016sck,Low:2016znh,Motlagh:2023oqc,Kikuchi:2023qbb}.
Following the discussion of Ref.~\cite{Hariprakash:2023tla}, one can show
\begin{equation}
    \text{Gates}_\text{QSP} = \CO\left(\NL^2 \log \NL \left(\NL^2 t + \log\left( \frac{1}{\epsilon} \right) \right) \right),
\end{equation}
which is not only polynomial in the number of lattice sites, but has better asymptotic scaling than using PFs given in \cref{eq:cost_pf}.

We now comment on the classical cost of calculating the matrix elements of the Hamiltonian.
This procedure is complicated by the fact that the parameter $\sigma$ is not a true dynamical quantum number, but rather a linear combination of other quantum numbers that must be calculated for each matrix element.
In particular, because $\sigma$ depends on the quantum numbers of all rods (recall $\sigma = \NTot + \sum_\mu \MSeq{\mu}$), classically calculating $\sigma$ for each basis state requires resources exponential in the volume.
This problem, however, could in principle be overcome by pre-computing $\sigma$ for a given initial state, storing its value in an ancillary register, and then querying said register throughout the simulation when needed.
We leave further exploration for future work.

We conclude by restating that the point of this simple volume scaling analysis was to demonstrate that one can simulate this fully gauge-fixed formulation of SU(2) lattice gauge theory using a number of gates that scales polynomially with the number of lattice sites.
It is likely that the asymptotic scaling we quote can be improved through a more careful treatment, which we leave for future work.

\section{Conclusions}
\label{sec:Conclusions}

In this work, we demonstrated how to construct a fully gauge-fixed Hamiltonian for a pure \sutwo lattice gauge theory.
Extending on the formulation in Ref.~\cite{DAndrea:2023qnr}, we work in the maximal tree gauge, where all local gauge symmetries are fixed and only a global gauge symmetry remains.
To fix the final global gauge symmetry, we use the intuition gained via the simple geometric picture of \sutwo as a set of rotating rods, which follows from the diffeomorphism between \sutwo and ${\rm S}^3$.
In this way, one can define a basis for the Hilbert space using body-frame coordinates, which can be related to the usual lab-frame coordinates through a rotation parameterized by the three Euler angles. 
The Hilbert space expressed in this `sequestered' basis is then cleanly partitioned into sectors with different total charge specified by the Euler angle quantum numbers.

In analogy to the work in Ref.~\cite{DAndrea:2023qnr}, we then convert the sequestered basis from the magnetic basis (axis-angle coordinates) to a mixed-basis, where the angular quantum numbers are expressed using electric basis quantum numbers.
To do so, we first identify a complete set of commuting observables (CSCO) associated with the angular quantum numbers.
By expanding the wavefunction in terms of the eigenfunctions of the CSCO, the states in the angular Hilbert space are now specified by integer eigenvalues.
In this mixed basis, gauge-fixing to a particular total charge sector becomes trivial, and is done by simply choosing the definite total angular momentum quantum numbers in the simulated Hilbert space.
We then showed that, despite the global nature of the final gauge-fixing procedure, the degree of coupling of the gauge-fixed Hamiltonian is independent of the volume. 
Using this fact, we demonstrated that this theory can be simulated, using either Product Formulas or Quantum Signal Processing, with a number of gates scaling polynomially in the volume.

In addition to the simplicity of gauge-fixing to a particular total charge sector, the formulation in this work inherits the properties from the formulation developed in Ref.~\cite{DAndrea:2023qnr}, in that it is simultaneously systematically improvable, efficient at all values of the gauge coupling, and gauge-invariant.
Conversely, all current digitizations of Hamiltonians written in the temporal gauge, \ie, performing the minimal gauge-fixing necessary to write a Hamiltonian, must sacrifice one or more of these properties.
Electric basis digitizations, while gauge-invariant and systematically improvable, require increasing the electric field cutoff as the lattice spacing is decreased. 
Gauge-invariant magnetic basis formulations in the temporal gauge can be achieved using discrete subgroups of the continuous group.
However, because there exist only finitely many discrete subgroups of a given non-Abelian group, there is a theoretical limit to how small a lattice spacing can be reliably studied~\cite{Gustafson:2023kvd}.

Our work naturally leads to a number of interesting and also necessary extensions.
Perhaps the most pressing direction is the extension to include fermions, which may require a modification of the maximal-tree gauge fixing procedure.
Once a gauge-fixed theory with fermions has been developed, it will be interesting to perform a study comparing the cost of simulating it to other formulations of \sutwo LGT, some examples being Loop-String-Hadron~\cite{Raychowdhury:2018osk, Raychowdhury:2019iki, Davoudi:2022xmb} and discrete subgroup~\cite{Gustafson:2022xdt, Gustafson:2023kvd, Assi:2024pdn, Lamm:2024jnl, Muarari:2024dqx} formulations.
Another crucial step, necessary for the study of Quantum Chromodynamics, is extending this formalism to \suthree gauge theories. The change from \sutwo to \suthree has proven to be a hurdle for many Hamiltonian formulations of LGTs.

We conclude with a discussion about the role of gauge-fixing in quantum simulations of LGTs, which extends beyond the exploration of broader classes of digitizations.
Because gauge-fixing generally reduces the size of the Hilbert space, it also reduces the number of qubits needed to simulate the theory on a quantum device.
Additionally, complete gauge-fixing simplifies the problem of preparing states of definite total charge, as any state in the Hilbert space by construction has the desired total charge. 
While gauge-fixing leads to several desirable properties with respect to quantum simulation, it also comes with downsides compared to formulations where the gauge-redundancies are not removed. 
Firstly, because gauge-fixing enforces constraints among different lattice sites, it generally increases the non-locality of operators in the theory, either in the physical distance between coupled lattice sites~\cite{DAndrea:2023qnr} or in the number of lattice sites a single operator acts on~\cite{Bauer:2021gek, Grabowska:2022uos, Kane:2022ejm}.
While we showed that simulating this fully gauge-fixed Hamiltonian can be done with a gate cost polynomial in the volume, the volume scaling is generally worse compared to, \eg, the standard Kogut-Susskind Hamiltonian, see, \eg, Ref.~\cite{Rhodes:2024zbr}.
Additionally, it was recently shown that the gauge-redundancy of lattice gauge theories can be exploited to perform error correction~\cite{Carena:2024dzu}.
Detailed studies comparing these various trade-offs are therefore imperative to determine the degree of gauge-fixing that will result in the fasted path to practical quantum simulations.

\section{Acknowledgements}
The authors would like to thank Ivan Burbano for valuable discussions. 
DMG is supported in part by the U.S. Department of Energy, Office of Science, Office of Nuclear Physics, InQubator for Quantum Simulation (IQuS) (\url{https:// iqus.uw.edu}) under Award Number DOE (NP) Award DE-SC0020970 via the program on Quantum Horizons: QIS Research and Innovation for Nuclear Science. 
DMG is supported, in part, through the Department of Physics\footnote{\url{https://phys.washington.edu}}
and the College of Arts and Sciences\footnote{\url{https://www.artsci.washington.edu}} at the University of Washington. 
CFK is supported in part by the Department of Physics, Maryland Center for Fundamental Physics, and the College of Computer, Mathematical, and Natural Sciences at the University of Maryland, College Park.
This material is based upon work supported by the U.S. Department of
Energy, Office of Science, Office of Advanced Scientific Computing Research, Department of
Energy Computational Science Graduate Fellowship under Award Number DE-SC0020347. 
CWB was supported by the DOE, Office of Science under contract DE-AC02-05CH11231, partially through Quantum Information Science Enabled Discovery (QuantISED) for High Energy Physics (KA2401032)

\bibliographystyle{apsrev4-1}
\bibliography{references}

\begin{thebibliography}{132}%
\makeatletter
\providecommand \@ifxundefined [1]{%
 \@ifx{#1\undefined}
}%
\providecommand \@ifnum [1]{%
 \ifnum #1\expandafter \@firstoftwo
 \else \expandafter \@secondoftwo
 \fi
}%
\providecommand \@ifx [1]{%
 \ifx #1\expandafter \@firstoftwo
 \else \expandafter \@secondoftwo
 \fi
}%
\providecommand \natexlab [1]{#1}%
\providecommand \enquote  [1]{``#1''}%
\providecommand \bibnamefont  [1]{#1}%
\providecommand \bibfnamefont [1]{#1}%
\providecommand \citenamefont [1]{#1}%
\providecommand \href@noop [0]{\@secondoftwo}%
\providecommand \href [0]{\begingroup \@sanitize@url \@href}%
\providecommand \@href[1]{\@@startlink{#1}\@@href}%
\providecommand \@@href[1]{\endgroup#1\@@endlink}%
\providecommand \@sanitize@url [0]{\catcode `\\12\catcode `\$12\catcode `\&12\catcode `\#12\catcode `\^12\catcode `\_12\catcode `\%12\relax}%
\providecommand \@@startlink[1]{}%
\providecommand \@@endlink[0]{}%
\providecommand \url  [0]{\begingroup\@sanitize@url \@url }%
\providecommand \@url [1]{\endgroup\@href {#1}{\urlprefix }}%
\providecommand \urlprefix  [0]{URL }%
\providecommand \Eprint [0]{\href }%
\providecommand \doibase [0]{http://dx.doi.org/}%
\providecommand \selectlanguage [0]{\@gobble}%
\providecommand \bibinfo  [0]{\@secondoftwo}%
\providecommand \bibfield  [0]{\@secondoftwo}%
\providecommand \translation [1]{[#1]}%
\providecommand \BibitemOpen [0]{}%
\providecommand \bibitemStop [0]{}%
\providecommand \bibitemNoStop [0]{.\EOS\space}%
\providecommand \EOS [0]{\spacefactor3000\relax}%
\providecommand \BibitemShut  [1]{\csname bibitem#1\endcsname}%
\let\auto@bib@innerbib\@empty
\bibitem [{\citenamefont {Jordan}\ \emph {et~al.}(2012)\citenamefont {Jordan}, \citenamefont {Lee},\ and\ \citenamefont {Preskill}}]{Jordan:2012xnu}%
  \BibitemOpen
  \bibfield  {author} {\bibinfo {author} {\bibfnamefont {S.~P.}\ \bibnamefont {Jordan}}, \bibinfo {author} {\bibfnamefont {K.~S.~M.}\ \bibnamefont {Lee}}, \ and\ \bibinfo {author} {\bibfnamefont {J.}~\bibnamefont {Preskill}},\ }\href {\doibase 10.1126/science.1217069} {\bibfield  {journal} {\bibinfo  {journal} {Science}\ }\textbf {\bibinfo {volume} {336}},\ \bibinfo {pages} {1130} (\bibinfo {year} {2012})},\ \Eprint {http://arxiv.org/abs/1111.3633} {arXiv:1111.3633 [quant-ph]} \BibitemShut {NoStop}%
\bibitem [{\citenamefont {Bauer}\ \emph {et~al.}(2023{\natexlab{a}})\citenamefont {Bauer} \emph {et~al.}}]{Bauer:2022hpo}%
  \BibitemOpen
  \bibfield  {author} {\bibinfo {author} {\bibfnamefont {C.~W.}\ \bibnamefont {Bauer}} \emph {et~al.},\ }\href {\doibase 10.1103/PRXQuantum.4.027001} {\bibfield  {journal} {\bibinfo  {journal} {PRX Quantum}\ }\textbf {\bibinfo {volume} {4}},\ \bibinfo {pages} {027001} (\bibinfo {year} {2023}{\natexlab{a}})},\ \Eprint {http://arxiv.org/abs/2204.03381} {arXiv:2204.03381 [quant-ph]} \BibitemShut {NoStop}%
\bibitem [{\citenamefont {Davoudi}\ \emph {et~al.}(2022)\citenamefont {Davoudi} \emph {et~al.}}]{Davoudi:2022bnl}%
  \BibitemOpen
  \bibfield  {author} {\bibinfo {author} {\bibfnamefont {Z.}~\bibnamefont {Davoudi}} \emph {et~al.},\ }in\ \href@noop {} {\emph {\bibinfo {booktitle} {{Snowmass 2021}}}}\ (\bibinfo {year} {2022})\ \Eprint {http://arxiv.org/abs/2209.10758} {arXiv:2209.10758 [hep-lat]} \BibitemShut {NoStop}%
\bibitem [{\citenamefont {Di~Meglio}\ \emph {et~al.}(2024)\citenamefont {Di~Meglio} \emph {et~al.}}]{DiMeglio:2023nsa}%
  \BibitemOpen
  \bibfield  {author} {\bibinfo {author} {\bibfnamefont {A.}~\bibnamefont {Di~Meglio}} \emph {et~al.},\ }\href {\doibase 10.1103/PRXQuantum.5.037001} {\bibfield  {journal} {\bibinfo  {journal} {PRX Quantum}\ }\textbf {\bibinfo {volume} {5}},\ \bibinfo {pages} {037001} (\bibinfo {year} {2024})},\ \Eprint {http://arxiv.org/abs/2307.03236} {arXiv:2307.03236 [quant-ph]} \BibitemShut {NoStop}%
\bibitem [{\citenamefont {Bauer}\ \emph {et~al.}(2023{\natexlab{b}})\citenamefont {Bauer}, \citenamefont {Davoudi}, \citenamefont {Klco},\ and\ \citenamefont {Savage}}]{Bauer:2023qgm}%
  \BibitemOpen
  \bibfield  {author} {\bibinfo {author} {\bibfnamefont {C.~W.}\ \bibnamefont {Bauer}}, \bibinfo {author} {\bibfnamefont {Z.}~\bibnamefont {Davoudi}}, \bibinfo {author} {\bibfnamefont {N.}~\bibnamefont {Klco}}, \ and\ \bibinfo {author} {\bibfnamefont {M.~J.}\ \bibnamefont {Savage}},\ }\href {\doibase 10.1038/s42254-023-00599-8} {\bibfield  {journal} {\bibinfo  {journal} {Nature Rev. Phys.}\ }\textbf {\bibinfo {volume} {5}},\ \bibinfo {pages} {420} (\bibinfo {year} {2023}{\natexlab{b}})},\ \Eprint {http://arxiv.org/abs/2404.06298} {arXiv:2404.06298 [hep-ph]} \BibitemShut {NoStop}%
\bibitem [{\citenamefont {Watson}\ \emph {et~al.}(2023)\citenamefont {Watson}, \citenamefont {Bringewatt}, \citenamefont {Shaw}, \citenamefont {Childs}, \citenamefont {Gorshkov},\ and\ \citenamefont {Davoudi}}]{Watson:2023oov}%
  \BibitemOpen
  \bibfield  {author} {\bibinfo {author} {\bibfnamefont {J.~D.}\ \bibnamefont {Watson}}, \bibinfo {author} {\bibfnamefont {J.}~\bibnamefont {Bringewatt}}, \bibinfo {author} {\bibfnamefont {A.~F.}\ \bibnamefont {Shaw}}, \bibinfo {author} {\bibfnamefont {A.~M.}\ \bibnamefont {Childs}}, \bibinfo {author} {\bibfnamefont {A.~V.}\ \bibnamefont {Gorshkov}}, \ and\ \bibinfo {author} {\bibfnamefont {Z.}~\bibnamefont {Davoudi}},\ }\href@noop {} {\  (\bibinfo {year} {2023})},\ \Eprint {http://arxiv.org/abs/2312.05344} {arXiv:2312.05344 [quant-ph]} \BibitemShut {NoStop}%
\bibitem [{\citenamefont {Byrnes}\ and\ \citenamefont {Yamamoto}(2006)}]{Byrnes:2005qx}%
  \BibitemOpen
  \bibfield  {author} {\bibinfo {author} {\bibfnamefont {T.}~\bibnamefont {Byrnes}}\ and\ \bibinfo {author} {\bibfnamefont {Y.}~\bibnamefont {Yamamoto}},\ }\href {\doibase 10.1103/PhysRevA.73.022328} {\bibfield  {journal} {\bibinfo  {journal} {Phys. Rev. A}\ }\textbf {\bibinfo {volume} {73}},\ \bibinfo {pages} {022328} (\bibinfo {year} {2006})},\ \Eprint {http://arxiv.org/abs/quant-ph/0510027} {arXiv:quant-ph/0510027} \BibitemShut {NoStop}%
\bibitem [{\citenamefont {Mildenberger}\ \emph {et~al.}(2022)\citenamefont {Mildenberger}, \citenamefont {Mruczkiewicz}, \citenamefont {Halimeh}, \citenamefont {Jiang},\ and\ \citenamefont {Hauke}}]{Mildenberger:2022jqr}%
  \BibitemOpen
  \bibfield  {author} {\bibinfo {author} {\bibfnamefont {J.}~\bibnamefont {Mildenberger}}, \bibinfo {author} {\bibfnamefont {W.}~\bibnamefont {Mruczkiewicz}}, \bibinfo {author} {\bibfnamefont {J.~C.}\ \bibnamefont {Halimeh}}, \bibinfo {author} {\bibfnamefont {Z.}~\bibnamefont {Jiang}}, \ and\ \bibinfo {author} {\bibfnamefont {P.}~\bibnamefont {Hauke}},\ }\href@noop {} {\enquote {\bibinfo {title} {{Probing confinement in a $\mathbb{Z}_2$ lattice gauge theory on a quantum computer}},}\ } (\bibinfo {year} {2022}),\ \Eprint {http://arxiv.org/abs/2203.08905} {arXiv:2203.08905 [quant-ph]} \BibitemShut {NoStop}%
\bibitem [{\citenamefont {Pardo}\ \emph {et~al.}(2023)\citenamefont {Pardo}, \citenamefont {Greenberg}, \citenamefont {Fortinsky}, \citenamefont {Katz},\ and\ \citenamefont {Zohar}}]{Pardo:2022hrp}%
  \BibitemOpen
  \bibfield  {author} {\bibinfo {author} {\bibfnamefont {G.}~\bibnamefont {Pardo}}, \bibinfo {author} {\bibfnamefont {T.}~\bibnamefont {Greenberg}}, \bibinfo {author} {\bibfnamefont {A.}~\bibnamefont {Fortinsky}}, \bibinfo {author} {\bibfnamefont {N.}~\bibnamefont {Katz}}, \ and\ \bibinfo {author} {\bibfnamefont {E.}~\bibnamefont {Zohar}},\ }\href {\doibase 10.1103/PhysRevResearch.5.023077} {\bibfield  {journal} {\bibinfo  {journal} {Phys. Rev. Res.}\ }\textbf {\bibinfo {volume} {5}},\ \bibinfo {pages} {023077} (\bibinfo {year} {2023})},\ \Eprint {http://arxiv.org/abs/2206.00685} {arXiv:2206.00685 [quant-ph]} \BibitemShut {NoStop}%
\bibitem [{\citenamefont {Banerjee}\ \emph {et~al.}(2012)\citenamefont {Banerjee}, \citenamefont {Dalmonte}, \citenamefont {Muller}, \citenamefont {Rico}, \citenamefont {Stebler}, \citenamefont {Wiese},\ and\ \citenamefont {Zoller}}]{Banerjee:2012pg}%
  \BibitemOpen
  \bibfield  {author} {\bibinfo {author} {\bibfnamefont {D.}~\bibnamefont {Banerjee}}, \bibinfo {author} {\bibfnamefont {M.}~\bibnamefont {Dalmonte}}, \bibinfo {author} {\bibfnamefont {M.}~\bibnamefont {Muller}}, \bibinfo {author} {\bibfnamefont {E.}~\bibnamefont {Rico}}, \bibinfo {author} {\bibfnamefont {P.}~\bibnamefont {Stebler}}, \bibinfo {author} {\bibfnamefont {U.~J.}\ \bibnamefont {Wiese}}, \ and\ \bibinfo {author} {\bibfnamefont {P.}~\bibnamefont {Zoller}},\ }\href {\doibase 10.1103/PhysRevLett.109.175302} {\bibfield  {journal} {\bibinfo  {journal} {Phys. Rev. Lett.}\ }\textbf {\bibinfo {volume} {109}},\ \bibinfo {pages} {175302} (\bibinfo {year} {2012})},\ \Eprint {http://arxiv.org/abs/1205.6366} {arXiv:1205.6366 [cond-mat.quant-gas]} \BibitemShut {NoStop}%
\bibitem [{\citenamefont {Hauke}\ \emph {et~al.}(2013)\citenamefont {Hauke}, \citenamefont {Marcos}, \citenamefont {Dalmonte},\ and\ \citenamefont {Zoller}}]{Hauke:2013jga}%
  \BibitemOpen
  \bibfield  {author} {\bibinfo {author} {\bibfnamefont {P.}~\bibnamefont {Hauke}}, \bibinfo {author} {\bibfnamefont {D.}~\bibnamefont {Marcos}}, \bibinfo {author} {\bibfnamefont {M.}~\bibnamefont {Dalmonte}}, \ and\ \bibinfo {author} {\bibfnamefont {P.}~\bibnamefont {Zoller}},\ }\href {\doibase 10.1103/PhysRevX.3.041018} {\bibfield  {journal} {\bibinfo  {journal} {Phys. Rev. X}\ }\textbf {\bibinfo {volume} {3}},\ \bibinfo {pages} {041018} (\bibinfo {year} {2013})},\ \Eprint {http://arxiv.org/abs/1306.2162} {arXiv:1306.2162 [cond-mat.quant-gas]} \BibitemShut {NoStop}%
\bibitem [{\citenamefont {Zohar}\ \emph {et~al.}(2013{\natexlab{a}})\citenamefont {Zohar}, \citenamefont {Cirac},\ and\ \citenamefont {Reznik}}]{Zohar:2013zla}%
  \BibitemOpen
  \bibfield  {author} {\bibinfo {author} {\bibfnamefont {E.}~\bibnamefont {Zohar}}, \bibinfo {author} {\bibfnamefont {J.~I.}\ \bibnamefont {Cirac}}, \ and\ \bibinfo {author} {\bibfnamefont {B.}~\bibnamefont {Reznik}},\ }\href {\doibase 10.1103/PhysRevA.88.023617} {\bibfield  {journal} {\bibinfo  {journal} {Phys. Rev. A}\ }\textbf {\bibinfo {volume} {88}},\ \bibinfo {pages} {023617} (\bibinfo {year} {2013}{\natexlab{a}})},\ \Eprint {http://arxiv.org/abs/1303.5040} {arXiv:1303.5040 [quant-ph]} \BibitemShut {NoStop}%
\bibitem [{\citenamefont {K\"uhn}\ \emph {et~al.}(2014)\citenamefont {K\"uhn}, \citenamefont {Cirac},\ and\ \citenamefont {Ba\~nuls}}]{Kuhn:2014rha}%
  \BibitemOpen
  \bibfield  {author} {\bibinfo {author} {\bibfnamefont {S.}~\bibnamefont {K\"uhn}}, \bibinfo {author} {\bibfnamefont {J.~I.}\ \bibnamefont {Cirac}}, \ and\ \bibinfo {author} {\bibfnamefont {M.-C.}\ \bibnamefont {Ba\~nuls}},\ }\href {\doibase 10.1103/PhysRevA.90.042305} {\bibfield  {journal} {\bibinfo  {journal} {Phys. Rev. A}\ }\textbf {\bibinfo {volume} {90}},\ \bibinfo {pages} {042305} (\bibinfo {year} {2014})},\ \Eprint {http://arxiv.org/abs/1407.4995} {arXiv:1407.4995 [quant-ph]} \BibitemShut {NoStop}%
\bibitem [{\citenamefont {Kasper}\ \emph {et~al.}(2016)\citenamefont {Kasper}, \citenamefont {Hebenstreit}, \citenamefont {Oberthaler},\ and\ \citenamefont {Berges}}]{Kasper:2015cca}%
  \BibitemOpen
  \bibfield  {author} {\bibinfo {author} {\bibfnamefont {V.}~\bibnamefont {Kasper}}, \bibinfo {author} {\bibfnamefont {F.}~\bibnamefont {Hebenstreit}}, \bibinfo {author} {\bibfnamefont {M.}~\bibnamefont {Oberthaler}}, \ and\ \bibinfo {author} {\bibfnamefont {J.}~\bibnamefont {Berges}},\ }\href {\doibase 10.1016/j.physletb.2016.07.036} {\bibfield  {journal} {\bibinfo  {journal} {Phys. Lett. B}\ }\textbf {\bibinfo {volume} {760}},\ \bibinfo {pages} {742} (\bibinfo {year} {2016})},\ \Eprint {http://arxiv.org/abs/1506.01238} {arXiv:1506.01238 [cond-mat.quant-gas]} \BibitemShut {NoStop}%
\bibitem [{\citenamefont {Zohar}\ \emph {et~al.}(2016)\citenamefont {Zohar}, \citenamefont {Cirac},\ and\ \citenamefont {Reznik}}]{Zohar:2015hwa}%
  \BibitemOpen
  \bibfield  {author} {\bibinfo {author} {\bibfnamefont {E.}~\bibnamefont {Zohar}}, \bibinfo {author} {\bibfnamefont {J.~I.}\ \bibnamefont {Cirac}}, \ and\ \bibinfo {author} {\bibfnamefont {B.}~\bibnamefont {Reznik}},\ }\href {\doibase 10.1088/0034-4885/79/1/014401} {\bibfield  {journal} {\bibinfo  {journal} {Rept. Prog. Phys.}\ }\textbf {\bibinfo {volume} {79}},\ \bibinfo {pages} {014401} (\bibinfo {year} {2016})},\ \Eprint {http://arxiv.org/abs/1503.02312} {arXiv:1503.02312 [quant-ph]} \BibitemShut {NoStop}%
\bibitem [{\citenamefont {{Martinez}}\ \emph {et~al.}(2016)\citenamefont {{Martinez}}, \citenamefont {{Muschik}}, \citenamefont {{Schindler}}, \citenamefont {{Nigg}}, \citenamefont {{Erhard}}, \citenamefont {{Heyl}}, \citenamefont {{Hauke}}, \citenamefont {{Dalmonte}}, \citenamefont {{Monz}}, \citenamefont {{Zoller}},\ and\ \citenamefont {{Blatt}}}]{Martinez:2016yna}%
  \BibitemOpen
  \bibfield  {author} {\bibinfo {author} {\bibfnamefont {E.~A.}\ \bibnamefont {{Martinez}}}, \bibinfo {author} {\bibfnamefont {C.~A.}\ \bibnamefont {{Muschik}}}, \bibinfo {author} {\bibfnamefont {P.}~\bibnamefont {{Schindler}}}, \bibinfo {author} {\bibfnamefont {D.}~\bibnamefont {{Nigg}}}, \bibinfo {author} {\bibfnamefont {A.}~\bibnamefont {{Erhard}}}, \bibinfo {author} {\bibfnamefont {M.}~\bibnamefont {{Heyl}}}, \bibinfo {author} {\bibfnamefont {P.}~\bibnamefont {{Hauke}}}, \bibinfo {author} {\bibfnamefont {M.}~\bibnamefont {{Dalmonte}}}, \bibinfo {author} {\bibfnamefont {T.}~\bibnamefont {{Monz}}}, \bibinfo {author} {\bibfnamefont {P.}~\bibnamefont {{Zoller}}}, \ and\ \bibinfo {author} {\bibfnamefont {R.}~\bibnamefont {{Blatt}}},\ }\href {\doibase 10.1038/nature18318} {\bibfield  {journal} {\bibinfo  {journal} {Nature}\ }\textbf {\bibinfo {volume} {534}},\ \bibinfo {pages} {516} (\bibinfo {year} {2016})},\ \Eprint {http://arxiv.org/abs/1605.04570} {arXiv:1605.04570 [quant-ph]} \BibitemShut {NoStop}%
\bibitem [{\citenamefont {Yang}\ \emph {et~al.}(2016)\citenamefont {Yang}, \citenamefont {Giri}, \citenamefont {Johanning}, \citenamefont {Wunderlich}, \citenamefont {Zoller},\ and\ \citenamefont {Hauke}}]{Yang:2016hjn}%
  \BibitemOpen
  \bibfield  {author} {\bibinfo {author} {\bibfnamefont {D.}~\bibnamefont {Yang}}, \bibinfo {author} {\bibfnamefont {G.~S.}\ \bibnamefont {Giri}}, \bibinfo {author} {\bibfnamefont {M.}~\bibnamefont {Johanning}}, \bibinfo {author} {\bibfnamefont {C.}~\bibnamefont {Wunderlich}}, \bibinfo {author} {\bibfnamefont {P.}~\bibnamefont {Zoller}}, \ and\ \bibinfo {author} {\bibfnamefont {P.}~\bibnamefont {Hauke}},\ }\href {\doibase 10.1103/PhysRevA.94.052321} {\bibfield  {journal} {\bibinfo  {journal} {Phys. Rev. A}\ }\textbf {\bibinfo {volume} {94}},\ \bibinfo {pages} {052321} (\bibinfo {year} {2016})},\ \Eprint {http://arxiv.org/abs/1604.03124} {arXiv:1604.03124 [quant-ph]} \BibitemShut {NoStop}%
\bibitem [{\citenamefont {Kokail}\ \emph {et~al.}(2019)\citenamefont {Kokail} \emph {et~al.}}]{Kokail:2018eiw}%
  \BibitemOpen
  \bibfield  {author} {\bibinfo {author} {\bibfnamefont {C.}~\bibnamefont {Kokail}} \emph {et~al.},\ }\href {\doibase 10.1038/s41586-019-1177-4} {\bibfield  {journal} {\bibinfo  {journal} {Nature}\ }\textbf {\bibinfo {volume} {569}},\ \bibinfo {pages} {355} (\bibinfo {year} {2019})},\ \Eprint {http://arxiv.org/abs/1810.03421} {arXiv:1810.03421 [quant-ph]} \BibitemShut {NoStop}%
\bibitem [{\citenamefont {Klco}\ \emph {et~al.}(2018)\citenamefont {Klco}, \citenamefont {Dumitrescu}, \citenamefont {McCaskey}, \citenamefont {Morris}, \citenamefont {Pooser}, \citenamefont {Sanz}, \citenamefont {Solano}, \citenamefont {Lougovski},\ and\ \citenamefont {Savage}}]{Klco:2018kyo}%
  \BibitemOpen
  \bibfield  {author} {\bibinfo {author} {\bibfnamefont {N.}~\bibnamefont {Klco}}, \bibinfo {author} {\bibfnamefont {E.~F.}\ \bibnamefont {Dumitrescu}}, \bibinfo {author} {\bibfnamefont {A.~J.}\ \bibnamefont {McCaskey}}, \bibinfo {author} {\bibfnamefont {T.~D.}\ \bibnamefont {Morris}}, \bibinfo {author} {\bibfnamefont {R.~C.}\ \bibnamefont {Pooser}}, \bibinfo {author} {\bibfnamefont {M.}~\bibnamefont {Sanz}}, \bibinfo {author} {\bibfnamefont {E.}~\bibnamefont {Solano}}, \bibinfo {author} {\bibfnamefont {P.}~\bibnamefont {Lougovski}}, \ and\ \bibinfo {author} {\bibfnamefont {M.~J.}\ \bibnamefont {Savage}},\ }\href {\doibase 10.1103/PhysRevA.98.032331} {\bibfield  {journal} {\bibinfo  {journal} {Phys. Rev. A}\ }\textbf {\bibinfo {volume} {98}},\ \bibinfo {pages} {032331} (\bibinfo {year} {2018})},\ \Eprint {http://arxiv.org/abs/1803.03326} {arXiv:1803.03326 [quant-ph]} \BibitemShut {NoStop}%
\bibitem [{\citenamefont {Lu}\ \emph {et~al.}(2019)\citenamefont {Lu} \emph {et~al.}}]{Lu:2018pjk}%
  \BibitemOpen
  \bibfield  {author} {\bibinfo {author} {\bibfnamefont {H.-H.}\ \bibnamefont {Lu}} \emph {et~al.},\ }\href {\doibase 10.1103/PhysRevA.100.012320} {\bibfield  {journal} {\bibinfo  {journal} {Phys. Rev. A}\ }\textbf {\bibinfo {volume} {100}},\ \bibinfo {pages} {012320} (\bibinfo {year} {2019})},\ \Eprint {http://arxiv.org/abs/1810.03959} {arXiv:1810.03959 [quant-ph]} \BibitemShut {NoStop}%
\bibitem [{\citenamefont {Kaplan}\ and\ \citenamefont {Stryker}(2020)}]{Kaplan:2018vnj}%
  \BibitemOpen
  \bibfield  {author} {\bibinfo {author} {\bibfnamefont {D.~B.}\ \bibnamefont {Kaplan}}\ and\ \bibinfo {author} {\bibfnamefont {J.~R.}\ \bibnamefont {Stryker}},\ }\href {\doibase 10.1103/PhysRevD.102.094515} {\bibfield  {journal} {\bibinfo  {journal} {Phys. Rev. D}\ }\textbf {\bibinfo {volume} {102}},\ \bibinfo {pages} {094515} (\bibinfo {year} {2020})},\ \Eprint {http://arxiv.org/abs/1806.08797} {arXiv:1806.08797 [hep-lat]} \BibitemShut {NoStop}%
\bibitem [{\citenamefont {Mil}\ \emph {et~al.}(2020)\citenamefont {Mil}, \citenamefont {Zache}, \citenamefont {Hegde}, \citenamefont {Xia}, \citenamefont {Bhatt}, \citenamefont {Oberthaler}, \citenamefont {Hauke}, \citenamefont {Berges},\ and\ \citenamefont {Jendrzejewski}}]{Mil:2019pbt}%
  \BibitemOpen
  \bibfield  {author} {\bibinfo {author} {\bibfnamefont {A.}~\bibnamefont {Mil}}, \bibinfo {author} {\bibfnamefont {T.~V.}\ \bibnamefont {Zache}}, \bibinfo {author} {\bibfnamefont {A.}~\bibnamefont {Hegde}}, \bibinfo {author} {\bibfnamefont {A.}~\bibnamefont {Xia}}, \bibinfo {author} {\bibfnamefont {R.~P.}\ \bibnamefont {Bhatt}}, \bibinfo {author} {\bibfnamefont {M.~K.}\ \bibnamefont {Oberthaler}}, \bibinfo {author} {\bibfnamefont {P.}~\bibnamefont {Hauke}}, \bibinfo {author} {\bibfnamefont {J.}~\bibnamefont {Berges}}, \ and\ \bibinfo {author} {\bibfnamefont {F.}~\bibnamefont {Jendrzejewski}},\ }\href {\doibase 10.1126/science.aaz5312} {\bibfield  {journal} {\bibinfo  {journal} {Science}\ }\textbf {\bibinfo {volume} {367}},\ \bibinfo {pages} {1128} (\bibinfo {year} {2020})},\ \Eprint {http://arxiv.org/abs/1909.07641} {arXiv:1909.07641 [cond-mat.quant-gas]} \BibitemShut {NoStop}%
\bibitem [{\citenamefont {Davoudi}\ \emph {et~al.}(2020)\citenamefont {Davoudi}, \citenamefont {Hafezi}, \citenamefont {Monroe}, \citenamefont {Pagano}, \citenamefont {Seif},\ and\ \citenamefont {Shaw}}]{Davoudi:2019bhy}%
  \BibitemOpen
  \bibfield  {author} {\bibinfo {author} {\bibfnamefont {Z.}~\bibnamefont {Davoudi}}, \bibinfo {author} {\bibfnamefont {M.}~\bibnamefont {Hafezi}}, \bibinfo {author} {\bibfnamefont {C.}~\bibnamefont {Monroe}}, \bibinfo {author} {\bibfnamefont {G.}~\bibnamefont {Pagano}}, \bibinfo {author} {\bibfnamefont {A.}~\bibnamefont {Seif}}, \ and\ \bibinfo {author} {\bibfnamefont {A.}~\bibnamefont {Shaw}},\ }\href {\doibase 10.1103/PhysRevResearch.2.023015} {\bibfield  {journal} {\bibinfo  {journal} {Phys. Rev. Res.}\ }\textbf {\bibinfo {volume} {2}},\ \bibinfo {pages} {023015} (\bibinfo {year} {2020})},\ \Eprint {http://arxiv.org/abs/1908.03210} {arXiv:1908.03210 [quant-ph]} \BibitemShut {NoStop}%
\bibitem [{\citenamefont {Surace}\ \emph {et~al.}(2020)\citenamefont {Surace}, \citenamefont {Mazza}, \citenamefont {Giudici}, \citenamefont {Lerose}, \citenamefont {Gambassi},\ and\ \citenamefont {Dalmonte}}]{Surace:2019dtp}%
  \BibitemOpen
  \bibfield  {author} {\bibinfo {author} {\bibfnamefont {F.~M.}\ \bibnamefont {Surace}}, \bibinfo {author} {\bibfnamefont {P.~P.}\ \bibnamefont {Mazza}}, \bibinfo {author} {\bibfnamefont {G.}~\bibnamefont {Giudici}}, \bibinfo {author} {\bibfnamefont {A.}~\bibnamefont {Lerose}}, \bibinfo {author} {\bibfnamefont {A.}~\bibnamefont {Gambassi}}, \ and\ \bibinfo {author} {\bibfnamefont {M.}~\bibnamefont {Dalmonte}},\ }\href {\doibase 10.1103/PhysRevX.10.021041} {\bibfield  {journal} {\bibinfo  {journal} {Phys. Rev. X}\ }\textbf {\bibinfo {volume} {10}},\ \bibinfo {pages} {021041} (\bibinfo {year} {2020})},\ \Eprint {http://arxiv.org/abs/1902.09551} {arXiv:1902.09551 [cond-mat.quant-gas]} \BibitemShut {NoStop}%
\bibitem [{\citenamefont {Haase}\ \emph {et~al.}(2021)\citenamefont {Haase}, \citenamefont {Dellantonio}, \citenamefont {Celi}, \citenamefont {Paulson}, \citenamefont {Kan}, \citenamefont {Jansen},\ and\ \citenamefont {Muschik}}]{Haase:2020kaj}%
  \BibitemOpen
  \bibfield  {author} {\bibinfo {author} {\bibfnamefont {J.~F.}\ \bibnamefont {Haase}}, \bibinfo {author} {\bibfnamefont {L.}~\bibnamefont {Dellantonio}}, \bibinfo {author} {\bibfnamefont {A.}~\bibnamefont {Celi}}, \bibinfo {author} {\bibfnamefont {D.}~\bibnamefont {Paulson}}, \bibinfo {author} {\bibfnamefont {A.}~\bibnamefont {Kan}}, \bibinfo {author} {\bibfnamefont {K.}~\bibnamefont {Jansen}}, \ and\ \bibinfo {author} {\bibfnamefont {C.~A.}\ \bibnamefont {Muschik}},\ }\href {\doibase 10.22331/q-2021-02-04-393} {\bibfield  {journal} {\bibinfo  {journal} {Quantum}\ }\textbf {\bibinfo {volume} {5}},\ \bibinfo {pages} {393} (\bibinfo {year} {2021})},\ \Eprint {http://arxiv.org/abs/2006.14160} {arXiv:2006.14160 [quant-ph]} \BibitemShut {NoStop}%
\bibitem [{\citenamefont {Luo}\ \emph {et~al.}(2020)\citenamefont {Luo}, \citenamefont {Shen}, \citenamefont {Highman}, \citenamefont {Clark}, \citenamefont {DeMarco}, \citenamefont {El-Khadra},\ and\ \citenamefont {Gadway}}]{Luo:2019vmi}%
  \BibitemOpen
  \bibfield  {author} {\bibinfo {author} {\bibfnamefont {D.}~\bibnamefont {Luo}}, \bibinfo {author} {\bibfnamefont {J.}~\bibnamefont {Shen}}, \bibinfo {author} {\bibfnamefont {M.}~\bibnamefont {Highman}}, \bibinfo {author} {\bibfnamefont {B.~K.}\ \bibnamefont {Clark}}, \bibinfo {author} {\bibfnamefont {B.}~\bibnamefont {DeMarco}}, \bibinfo {author} {\bibfnamefont {A.~X.}\ \bibnamefont {El-Khadra}}, \ and\ \bibinfo {author} {\bibfnamefont {B.}~\bibnamefont {Gadway}},\ }\href {\doibase 10.1103/PhysRevA.102.032617} {\bibfield  {journal} {\bibinfo  {journal} {Phys. Rev. A}\ }\textbf {\bibinfo {volume} {102}},\ \bibinfo {pages} {032617} (\bibinfo {year} {2020})},\ \Eprint {http://arxiv.org/abs/1912.11488} {arXiv:1912.11488 [quant-ph]} \BibitemShut {NoStop}%
\bibitem [{\citenamefont {Shaw}\ \emph {et~al.}(2020)\citenamefont {Shaw}, \citenamefont {Lougovski}, \citenamefont {Stryker},\ and\ \citenamefont {Wiebe}}]{Shaw:2020udc}%
  \BibitemOpen
  \bibfield  {author} {\bibinfo {author} {\bibfnamefont {A.~F.}\ \bibnamefont {Shaw}}, \bibinfo {author} {\bibfnamefont {P.}~\bibnamefont {Lougovski}}, \bibinfo {author} {\bibfnamefont {J.~R.}\ \bibnamefont {Stryker}}, \ and\ \bibinfo {author} {\bibfnamefont {N.}~\bibnamefont {Wiebe}},\ }\href {\doibase 10.22331/q-2020-08-10-306} {\bibfield  {journal} {\bibinfo  {journal} {Quantum}\ }\textbf {\bibinfo {volume} {4}},\ \bibinfo {pages} {306} (\bibinfo {year} {2020})},\ \Eprint {http://arxiv.org/abs/2002.11146} {arXiv:2002.11146 [quant-ph]} \BibitemShut {NoStop}%
\bibitem [{\citenamefont {Yang}\ \emph {et~al.}(2020)\citenamefont {Yang}, \citenamefont {Sun}, \citenamefont {Ott}, \citenamefont {Wang}, \citenamefont {Zache}, \citenamefont {Halimeh}, \citenamefont {Yuan}, \citenamefont {Hauke},\ and\ \citenamefont {Pan}}]{Yang:2020yer}%
  \BibitemOpen
  \bibfield  {author} {\bibinfo {author} {\bibfnamefont {B.}~\bibnamefont {Yang}}, \bibinfo {author} {\bibfnamefont {H.}~\bibnamefont {Sun}}, \bibinfo {author} {\bibfnamefont {R.}~\bibnamefont {Ott}}, \bibinfo {author} {\bibfnamefont {H.-Y.}\ \bibnamefont {Wang}}, \bibinfo {author} {\bibfnamefont {T.~V.}\ \bibnamefont {Zache}}, \bibinfo {author} {\bibfnamefont {J.~C.}\ \bibnamefont {Halimeh}}, \bibinfo {author} {\bibfnamefont {Z.-S.}\ \bibnamefont {Yuan}}, \bibinfo {author} {\bibfnamefont {P.}~\bibnamefont {Hauke}}, \ and\ \bibinfo {author} {\bibfnamefont {J.-W.}\ \bibnamefont {Pan}},\ }\href {\doibase 10.1038/s41586-020-2910-8} {\bibfield  {journal} {\bibinfo  {journal} {Nature}\ }\textbf {\bibinfo {volume} {587}},\ \bibinfo {pages} {392} (\bibinfo {year} {2020})},\ \Eprint {http://arxiv.org/abs/2003.08945} {arXiv:2003.08945 [cond-mat.quant-gas]} \BibitemShut {NoStop}%
\bibitem [{\citenamefont {Ott}\ \emph {et~al.}(2021)\citenamefont {Ott}, \citenamefont {Zache}, \citenamefont {Jendrzejewski},\ and\ \citenamefont {Berges}}]{Ott:2020ycj}%
  \BibitemOpen
  \bibfield  {author} {\bibinfo {author} {\bibfnamefont {R.}~\bibnamefont {Ott}}, \bibinfo {author} {\bibfnamefont {T.~V.}\ \bibnamefont {Zache}}, \bibinfo {author} {\bibfnamefont {F.}~\bibnamefont {Jendrzejewski}}, \ and\ \bibinfo {author} {\bibfnamefont {J.}~\bibnamefont {Berges}},\ }\href {\doibase 10.1103/PhysRevLett.127.130504} {\bibfield  {journal} {\bibinfo  {journal} {Phys. Rev. Lett.}\ }\textbf {\bibinfo {volume} {127}},\ \bibinfo {pages} {130504} (\bibinfo {year} {2021})},\ \Eprint {http://arxiv.org/abs/2012.10432} {arXiv:2012.10432 [cond-mat.quant-gas]} \BibitemShut {NoStop}%
\bibitem [{\citenamefont {Paulson}\ \emph {et~al.}(2021)\citenamefont {Paulson} \emph {et~al.}}]{Paulson:2020zjd}%
  \BibitemOpen
  \bibfield  {author} {\bibinfo {author} {\bibfnamefont {D.}~\bibnamefont {Paulson}} \emph {et~al.},\ }\href {\doibase 10.1103/PRXQuantum.2.030334} {\bibfield  {journal} {\bibinfo  {journal} {PRX Quantum}\ }\textbf {\bibinfo {volume} {2}},\ \bibinfo {pages} {030334} (\bibinfo {year} {2021})},\ \Eprint {http://arxiv.org/abs/2008.09252} {arXiv:2008.09252 [quant-ph]} \BibitemShut {NoStop}%
\bibitem [{\citenamefont {Nguyen}\ \emph {et~al.}(2022)\citenamefont {Nguyen}, \citenamefont {Tran}, \citenamefont {Zhu}, \citenamefont {Green}, \citenamefont {Alderete}, \citenamefont {Davoudi},\ and\ \citenamefont {Linke}}]{Nguyen:2021hyk}%
  \BibitemOpen
  \bibfield  {author} {\bibinfo {author} {\bibfnamefont {N.~H.}\ \bibnamefont {Nguyen}}, \bibinfo {author} {\bibfnamefont {M.~C.}\ \bibnamefont {Tran}}, \bibinfo {author} {\bibfnamefont {Y.}~\bibnamefont {Zhu}}, \bibinfo {author} {\bibfnamefont {A.~M.}\ \bibnamefont {Green}}, \bibinfo {author} {\bibfnamefont {C.~H.}\ \bibnamefont {Alderete}}, \bibinfo {author} {\bibfnamefont {Z.}~\bibnamefont {Davoudi}}, \ and\ \bibinfo {author} {\bibfnamefont {N.~M.}\ \bibnamefont {Linke}},\ }\href {\doibase 10.1103/PRXQuantum.3.020324} {\bibfield  {journal} {\bibinfo  {journal} {PRX Quantum}\ }\textbf {\bibinfo {volume} {3}},\ \bibinfo {pages} {020324} (\bibinfo {year} {2022})},\ \Eprint {http://arxiv.org/abs/2112.14262} {arXiv:2112.14262 [quant-ph]} \BibitemShut {NoStop}%
\bibitem [{\citenamefont {Zhou}\ \emph {et~al.}(2022)\citenamefont {Zhou}, \citenamefont {Su}, \citenamefont {Halimeh}, \citenamefont {Ott}, \citenamefont {Sun}, \citenamefont {Hauke}, \citenamefont {Yang}, \citenamefont {Yuan}, \citenamefont {Berges},\ and\ \citenamefont {Pan}}]{Zhou:2021kdl}%
  \BibitemOpen
  \bibfield  {author} {\bibinfo {author} {\bibfnamefont {Z.-Y.}\ \bibnamefont {Zhou}}, \bibinfo {author} {\bibfnamefont {G.-X.}\ \bibnamefont {Su}}, \bibinfo {author} {\bibfnamefont {J.~C.}\ \bibnamefont {Halimeh}}, \bibinfo {author} {\bibfnamefont {R.}~\bibnamefont {Ott}}, \bibinfo {author} {\bibfnamefont {H.}~\bibnamefont {Sun}}, \bibinfo {author} {\bibfnamefont {P.}~\bibnamefont {Hauke}}, \bibinfo {author} {\bibfnamefont {B.}~\bibnamefont {Yang}}, \bibinfo {author} {\bibfnamefont {Z.-S.}\ \bibnamefont {Yuan}}, \bibinfo {author} {\bibfnamefont {J.}~\bibnamefont {Berges}}, \ and\ \bibinfo {author} {\bibfnamefont {J.-W.}\ \bibnamefont {Pan}},\ }\href {\doibase 10.1126/science.abl6277} {\bibfield  {journal} {\bibinfo  {journal} {Science}\ }\textbf {\bibinfo {volume} {377}},\ \bibinfo {pages} {311} (\bibinfo {year} {2022})},\ \Eprint {http://arxiv.org/abs/2107.13563} {arXiv:2107.13563 [cond-mat.quant-gas]} \BibitemShut {NoStop}%
\bibitem [{\citenamefont {Riechert}\ \emph {et~al.}(2022)\citenamefont {Riechert}, \citenamefont {Halimeh}, \citenamefont {Kasper}, \citenamefont {Bretheau}, \citenamefont {Zohar}, \citenamefont {Hauke},\ and\ \citenamefont {Jendrzejewski}}]{Riechert:2021ink}%
  \BibitemOpen
  \bibfield  {author} {\bibinfo {author} {\bibfnamefont {H.}~\bibnamefont {Riechert}}, \bibinfo {author} {\bibfnamefont {J.~C.}\ \bibnamefont {Halimeh}}, \bibinfo {author} {\bibfnamefont {V.}~\bibnamefont {Kasper}}, \bibinfo {author} {\bibfnamefont {L.}~\bibnamefont {Bretheau}}, \bibinfo {author} {\bibfnamefont {E.}~\bibnamefont {Zohar}}, \bibinfo {author} {\bibfnamefont {P.}~\bibnamefont {Hauke}}, \ and\ \bibinfo {author} {\bibfnamefont {F.}~\bibnamefont {Jendrzejewski}},\ }\href {\doibase 10.1103/PhysRevB.105.205141} {\bibfield  {journal} {\bibinfo  {journal} {Phys. Rev. B}\ }\textbf {\bibinfo {volume} {105}},\ \bibinfo {pages} {205141} (\bibinfo {year} {2022})},\ \Eprint {http://arxiv.org/abs/2108.01086} {arXiv:2108.01086 [cond-mat.mes-hall]} \BibitemShut {NoStop}%
\bibitem [{\citenamefont {Bauer}\ and\ \citenamefont {Grabowska}(2023)}]{Bauer:2021gek}%
  \BibitemOpen
  \bibfield  {author} {\bibinfo {author} {\bibfnamefont {C.~W.}\ \bibnamefont {Bauer}}\ and\ \bibinfo {author} {\bibfnamefont {D.~M.}\ \bibnamefont {Grabowska}},\ }\href {\doibase 10.1103/PhysRevD.107.L031503} {\bibfield  {journal} {\bibinfo  {journal} {Phys. Rev. D}\ }\textbf {\bibinfo {volume} {107}},\ \bibinfo {pages} {L031503} (\bibinfo {year} {2023})},\ \Eprint {http://arxiv.org/abs/2111.08015} {arXiv:2111.08015 [hep-ph]} \BibitemShut {NoStop}%
\bibitem [{\citenamefont {Kane}\ \emph {et~al.}(2022)\citenamefont {Kane}, \citenamefont {Grabowska}, \citenamefont {Nachman},\ and\ \citenamefont {Bauer}}]{Kane:2022ejm}%
  \BibitemOpen
  \bibfield  {author} {\bibinfo {author} {\bibfnamefont {C.}~\bibnamefont {Kane}}, \bibinfo {author} {\bibfnamefont {D.~M.}\ \bibnamefont {Grabowska}}, \bibinfo {author} {\bibfnamefont {B.}~\bibnamefont {Nachman}}, \ and\ \bibinfo {author} {\bibfnamefont {C.~W.}\ \bibnamefont {Bauer}},\ }\href@noop {} {\  (\bibinfo {year} {2022})},\ \Eprint {http://arxiv.org/abs/2211.10497} {arXiv:2211.10497 [quant-ph]} \BibitemShut {NoStop}%
\bibitem [{\citenamefont {Grabowska}\ \emph {et~al.}(2022)\citenamefont {Grabowska}, \citenamefont {Kane}, \citenamefont {Nachman},\ and\ \citenamefont {Bauer}}]{Grabowska:2022uos}%
  \BibitemOpen
  \bibfield  {author} {\bibinfo {author} {\bibfnamefont {D.~M.}\ \bibnamefont {Grabowska}}, \bibinfo {author} {\bibfnamefont {C.}~\bibnamefont {Kane}}, \bibinfo {author} {\bibfnamefont {B.}~\bibnamefont {Nachman}}, \ and\ \bibinfo {author} {\bibfnamefont {C.~W.}\ \bibnamefont {Bauer}},\ }\href@noop {} {\  (\bibinfo {year} {2022})},\ \Eprint {http://arxiv.org/abs/2208.03333} {arXiv:2208.03333 [quant-ph]} \BibitemShut {NoStop}%
\bibitem [{\citenamefont {Zhang}\ \emph {et~al.}(2023)\citenamefont {Zhang}, \citenamefont {Liu}, \citenamefont {Cheng}, \citenamefont {He}, \citenamefont {Wang}, \citenamefont {Wang}, \citenamefont {Zhu}, \citenamefont {Su}, \citenamefont {Zhou}, \citenamefont {Zheng}, \citenamefont {Sun}, \citenamefont {Yang}, \citenamefont {Hauke}, \citenamefont {Zheng}, \citenamefont {Halimeh}, \citenamefont {Yuan},\ and\ \citenamefont {Pan}}]{zhang2023observation}%
  \BibitemOpen
  \bibfield  {author} {\bibinfo {author} {\bibfnamefont {W.-Y.}\ \bibnamefont {Zhang}}, \bibinfo {author} {\bibfnamefont {Y.}~\bibnamefont {Liu}}, \bibinfo {author} {\bibfnamefont {Y.}~\bibnamefont {Cheng}}, \bibinfo {author} {\bibfnamefont {M.-G.}\ \bibnamefont {He}}, \bibinfo {author} {\bibfnamefont {H.-Y.}\ \bibnamefont {Wang}}, \bibinfo {author} {\bibfnamefont {T.-Y.}\ \bibnamefont {Wang}}, \bibinfo {author} {\bibfnamefont {Z.-H.}\ \bibnamefont {Zhu}}, \bibinfo {author} {\bibfnamefont {G.-X.}\ \bibnamefont {Su}}, \bibinfo {author} {\bibfnamefont {Z.-Y.}\ \bibnamefont {Zhou}}, \bibinfo {author} {\bibfnamefont {Y.-G.}\ \bibnamefont {Zheng}}, \bibinfo {author} {\bibfnamefont {H.}~\bibnamefont {Sun}}, \bibinfo {author} {\bibfnamefont {B.}~\bibnamefont {Yang}}, \bibinfo {author} {\bibfnamefont {P.}~\bibnamefont {Hauke}}, \bibinfo {author} {\bibfnamefont {W.}~\bibnamefont {Zheng}}, \bibinfo {author} {\bibfnamefont {J.~C.}\ \bibnamefont {Halimeh}}, \bibinfo {author} {\bibfnamefont {Z.-S.}\ \bibnamefont {Yuan}},
  \ and\ \bibinfo {author} {\bibfnamefont {J.-W.}\ \bibnamefont {Pan}},\ }\href@noop {} {\enquote {\bibinfo {title} {Observation of microscopic confinement dynamics by a tunable topological $\theta$-angle},}\ } (\bibinfo {year} {2023}),\ \Eprint {http://arxiv.org/abs/2306.11794} {arXiv:2306.11794 [cond-mat.quant-gas]} \BibitemShut {NoStop}%
\bibitem [{\citenamefont {Farrell}\ \emph {et~al.}(2024{\natexlab{a}})\citenamefont {Farrell}, \citenamefont {Illa}, \citenamefont {Ciavarella},\ and\ \citenamefont {Savage}}]{Farrell:2023fgd}%
  \BibitemOpen
  \bibfield  {author} {\bibinfo {author} {\bibfnamefont {R.~C.}\ \bibnamefont {Farrell}}, \bibinfo {author} {\bibfnamefont {M.}~\bibnamefont {Illa}}, \bibinfo {author} {\bibfnamefont {A.~N.}\ \bibnamefont {Ciavarella}}, \ and\ \bibinfo {author} {\bibfnamefont {M.~J.}\ \bibnamefont {Savage}},\ }\href {\doibase 10.1103/PRXQuantum.5.020315} {\bibfield  {journal} {\bibinfo  {journal} {PRX Quantum}\ }\textbf {\bibinfo {volume} {5}},\ \bibinfo {pages} {020315} (\bibinfo {year} {2024}{\natexlab{a}})},\ \Eprint {http://arxiv.org/abs/2308.04481} {arXiv:2308.04481 [quant-ph]} \BibitemShut {NoStop}%
\bibitem [{\citenamefont {Nagano}\ \emph {et~al.}(2023)\citenamefont {Nagano}, \citenamefont {Bapat},\ and\ \citenamefont {Bauer}}]{Nagano:2023uaq}%
  \BibitemOpen
  \bibfield  {author} {\bibinfo {author} {\bibfnamefont {L.}~\bibnamefont {Nagano}}, \bibinfo {author} {\bibfnamefont {A.}~\bibnamefont {Bapat}}, \ and\ \bibinfo {author} {\bibfnamefont {C.~W.}\ \bibnamefont {Bauer}},\ }\href {\doibase 10.1103/PhysRevD.108.034501} {\bibfield  {journal} {\bibinfo  {journal} {Phys. Rev. D}\ }\textbf {\bibinfo {volume} {108}},\ \bibinfo {pages} {034501} (\bibinfo {year} {2023})},\ \Eprint {http://arxiv.org/abs/2302.10933} {arXiv:2302.10933 [hep-ph]} \BibitemShut {NoStop}%
\bibitem [{\citenamefont {Zohar}\ \emph {et~al.}(2013{\natexlab{b}})\citenamefont {Zohar}, \citenamefont {Cirac},\ and\ \citenamefont {Reznik}}]{Zohar:2012xf}%
  \BibitemOpen
  \bibfield  {author} {\bibinfo {author} {\bibfnamefont {E.}~\bibnamefont {Zohar}}, \bibinfo {author} {\bibfnamefont {J.~I.}\ \bibnamefont {Cirac}}, \ and\ \bibinfo {author} {\bibfnamefont {B.}~\bibnamefont {Reznik}},\ }\href {\doibase 10.1103/PhysRevLett.110.125304} {\bibfield  {journal} {\bibinfo  {journal} {Phys. Rev. Lett.}\ }\textbf {\bibinfo {volume} {110}},\ \bibinfo {pages} {125304} (\bibinfo {year} {2013}{\natexlab{b}})},\ \Eprint {http://arxiv.org/abs/1211.2241} {arXiv:1211.2241 [quant-ph]} \BibitemShut {NoStop}%
\bibitem [{\citenamefont {Stannigel}\ \emph {et~al.}(2014)\citenamefont {Stannigel}, \citenamefont {Hauke}, \citenamefont {Marcos}, \citenamefont {Hafezi}, \citenamefont {Diehl}, \citenamefont {Dalmonte},\ and\ \citenamefont {Zoller}}]{Stannigel:2013zka}%
  \BibitemOpen
  \bibfield  {author} {\bibinfo {author} {\bibfnamefont {K.}~\bibnamefont {Stannigel}}, \bibinfo {author} {\bibfnamefont {P.}~\bibnamefont {Hauke}}, \bibinfo {author} {\bibfnamefont {D.}~\bibnamefont {Marcos}}, \bibinfo {author} {\bibfnamefont {M.}~\bibnamefont {Hafezi}}, \bibinfo {author} {\bibfnamefont {S.}~\bibnamefont {Diehl}}, \bibinfo {author} {\bibfnamefont {M.}~\bibnamefont {Dalmonte}}, \ and\ \bibinfo {author} {\bibfnamefont {P.}~\bibnamefont {Zoller}},\ }\href {\doibase 10.1103/PhysRevLett.112.120406} {\bibfield  {journal} {\bibinfo  {journal} {Phys. Rev. Lett.}\ }\textbf {\bibinfo {volume} {112}},\ \bibinfo {pages} {120406} (\bibinfo {year} {2014})},\ \Eprint {http://arxiv.org/abs/1308.0528} {arXiv:1308.0528 [quant-ph]} \BibitemShut {NoStop}%
\bibitem [{\citenamefont {Mezzacapo}\ \emph {et~al.}(2015)\citenamefont {Mezzacapo}, \citenamefont {Rico}, \citenamefont {Sabin}, \citenamefont {Egusquiza}, \citenamefont {Lamata},\ and\ \citenamefont {Solano}}]{Mezzacapo:2015bra}%
  \BibitemOpen
  \bibfield  {author} {\bibinfo {author} {\bibfnamefont {A.}~\bibnamefont {Mezzacapo}}, \bibinfo {author} {\bibfnamefont {E.}~\bibnamefont {Rico}}, \bibinfo {author} {\bibfnamefont {C.}~\bibnamefont {Sabin}}, \bibinfo {author} {\bibfnamefont {I.~L.}\ \bibnamefont {Egusquiza}}, \bibinfo {author} {\bibfnamefont {L.}~\bibnamefont {Lamata}}, \ and\ \bibinfo {author} {\bibfnamefont {E.}~\bibnamefont {Solano}},\ }\href {\doibase 10.1103/PhysRevLett.115.240502} {\bibfield  {journal} {\bibinfo  {journal} {Phys. Rev. Lett.}\ }\textbf {\bibinfo {volume} {115}},\ \bibinfo {pages} {240502} (\bibinfo {year} {2015})},\ \Eprint {http://arxiv.org/abs/1505.04720} {arXiv:1505.04720 [quant-ph]} \BibitemShut {NoStop}%
\bibitem [{\citenamefont {Raychowdhury}\ and\ \citenamefont {Stryker}(2020{\natexlab{a}})}]{Raychowdhury:2018osk}%
  \BibitemOpen
  \bibfield  {author} {\bibinfo {author} {\bibfnamefont {I.}~\bibnamefont {Raychowdhury}}\ and\ \bibinfo {author} {\bibfnamefont {J.~R.}\ \bibnamefont {Stryker}},\ }\href {\doibase 10.1103/PhysRevResearch.2.033039} {\bibfield  {journal} {\bibinfo  {journal} {Phys. Rev. Res.}\ }\textbf {\bibinfo {volume} {2}},\ \bibinfo {pages} {033039} (\bibinfo {year} {2020}{\natexlab{a}})},\ \Eprint {http://arxiv.org/abs/1812.07554} {arXiv:1812.07554 [hep-lat]} \BibitemShut {NoStop}%
\bibitem [{\citenamefont {Raychowdhury}\ and\ \citenamefont {Stryker}(2020{\natexlab{b}})}]{Raychowdhury:2019iki}%
  \BibitemOpen
  \bibfield  {author} {\bibinfo {author} {\bibfnamefont {I.}~\bibnamefont {Raychowdhury}}\ and\ \bibinfo {author} {\bibfnamefont {J.~R.}\ \bibnamefont {Stryker}},\ }\href {\doibase 10.1103/PhysRevD.101.114502} {\bibfield  {journal} {\bibinfo  {journal} {Phys. Rev. D}\ }\textbf {\bibinfo {volume} {101}},\ \bibinfo {pages} {114502} (\bibinfo {year} {2020}{\natexlab{b}})},\ \Eprint {http://arxiv.org/abs/1912.06133} {arXiv:1912.06133 [hep-lat]} \BibitemShut {NoStop}%
\bibitem [{\citenamefont {Klco}\ \emph {et~al.}(2020)\citenamefont {Klco}, \citenamefont {Stryker},\ and\ \citenamefont {Savage}}]{Klco:2019evd}%
  \BibitemOpen
  \bibfield  {author} {\bibinfo {author} {\bibfnamefont {N.}~\bibnamefont {Klco}}, \bibinfo {author} {\bibfnamefont {J.~R.}\ \bibnamefont {Stryker}}, \ and\ \bibinfo {author} {\bibfnamefont {M.~J.}\ \bibnamefont {Savage}},\ }\href {\doibase 10.1103/PhysRevD.101.074512} {\bibfield  {journal} {\bibinfo  {journal} {Phys. Rev. D}\ }\textbf {\bibinfo {volume} {101}},\ \bibinfo {pages} {074512} (\bibinfo {year} {2020})},\ \Eprint {http://arxiv.org/abs/1908.06935} {arXiv:1908.06935 [quant-ph]} \BibitemShut {NoStop}%
\bibitem [{\citenamefont {Dasgupta}\ and\ \citenamefont {Raychowdhury}(2022)}]{Dasgupta:2020itb}%
  \BibitemOpen
  \bibfield  {author} {\bibinfo {author} {\bibfnamefont {R.}~\bibnamefont {Dasgupta}}\ and\ \bibinfo {author} {\bibfnamefont {I.}~\bibnamefont {Raychowdhury}},\ }\href {\doibase 10.1103/PhysRevA.105.023322} {\bibfield  {journal} {\bibinfo  {journal} {Phys. Rev. A}\ }\textbf {\bibinfo {volume} {105}},\ \bibinfo {pages} {023322} (\bibinfo {year} {2022})},\ \Eprint {http://arxiv.org/abs/2009.13969} {arXiv:2009.13969 [hep-lat]} \BibitemShut {NoStop}%
\bibitem [{\citenamefont {Davoudi}\ \emph {et~al.}(2021)\citenamefont {Davoudi}, \citenamefont {Raychowdhury},\ and\ \citenamefont {Shaw}}]{Davoudi:2020yln}%
  \BibitemOpen
  \bibfield  {author} {\bibinfo {author} {\bibfnamefont {Z.}~\bibnamefont {Davoudi}}, \bibinfo {author} {\bibfnamefont {I.}~\bibnamefont {Raychowdhury}}, \ and\ \bibinfo {author} {\bibfnamefont {A.}~\bibnamefont {Shaw}},\ }\href {\doibase 10.1103/PhysRevD.104.074505} {\bibfield  {journal} {\bibinfo  {journal} {Phys. Rev. D}\ }\textbf {\bibinfo {volume} {104}},\ \bibinfo {pages} {074505} (\bibinfo {year} {2021})},\ \Eprint {http://arxiv.org/abs/2009.11802} {arXiv:2009.11802 [hep-lat]} \BibitemShut {NoStop}%
\bibitem [{\citenamefont {Atas}\ \emph {et~al.}(2021)\citenamefont {Atas}, \citenamefont {Zhang}, \citenamefont {Lewis}, \citenamefont {Jahanpour}, \citenamefont {Haase},\ and\ \citenamefont {Muschik}}]{Atas:2021ext}%
  \BibitemOpen
  \bibfield  {author} {\bibinfo {author} {\bibfnamefont {Y.~Y.}\ \bibnamefont {Atas}}, \bibinfo {author} {\bibfnamefont {J.}~\bibnamefont {Zhang}}, \bibinfo {author} {\bibfnamefont {R.}~\bibnamefont {Lewis}}, \bibinfo {author} {\bibfnamefont {A.}~\bibnamefont {Jahanpour}}, \bibinfo {author} {\bibfnamefont {J.~F.}\ \bibnamefont {Haase}}, \ and\ \bibinfo {author} {\bibfnamefont {C.~A.}\ \bibnamefont {Muschik}},\ }\href {\doibase 10.1038/s41467-021-26825-4} {\bibfield  {journal} {\bibinfo  {journal} {Nature Commun.}\ }\textbf {\bibinfo {volume} {12}},\ \bibinfo {pages} {6499} (\bibinfo {year} {2021})},\ \Eprint {http://arxiv.org/abs/2102.08920} {arXiv:2102.08920 [quant-ph]} \BibitemShut {NoStop}%
\bibitem [{\citenamefont {A~Rahman}\ \emph {et~al.}(2021)\citenamefont {A~Rahman}, \citenamefont {Lewis}, \citenamefont {Mendicelli},\ and\ \citenamefont {Powell}}]{ARahman:2021ktn}%
  \BibitemOpen
  \bibfield  {author} {\bibinfo {author} {\bibfnamefont {S.}~\bibnamefont {A~Rahman}}, \bibinfo {author} {\bibfnamefont {R.}~\bibnamefont {Lewis}}, \bibinfo {author} {\bibfnamefont {E.}~\bibnamefont {Mendicelli}}, \ and\ \bibinfo {author} {\bibfnamefont {S.}~\bibnamefont {Powell}},\ }\href {\doibase 10.1103/PhysRevD.104.034501} {\bibfield  {journal} {\bibinfo  {journal} {Phys. Rev. D}\ }\textbf {\bibinfo {volume} {104}},\ \bibinfo {pages} {034501} (\bibinfo {year} {2021})},\ \Eprint {http://arxiv.org/abs/2103.08661} {arXiv:2103.08661 [hep-lat]} \BibitemShut {NoStop}%
\bibitem [{\citenamefont {Osborne}\ \emph {et~al.}(2022)\citenamefont {Osborne}, \citenamefont {McCulloch}, \citenamefont {Yang}, \citenamefont {Hauke},\ and\ \citenamefont {Halimeh}}]{Osborne:2022jxq}%
  \BibitemOpen
  \bibfield  {author} {\bibinfo {author} {\bibfnamefont {J.}~\bibnamefont {Osborne}}, \bibinfo {author} {\bibfnamefont {I.~P.}\ \bibnamefont {McCulloch}}, \bibinfo {author} {\bibfnamefont {B.}~\bibnamefont {Yang}}, \bibinfo {author} {\bibfnamefont {P.}~\bibnamefont {Hauke}}, \ and\ \bibinfo {author} {\bibfnamefont {J.~C.}\ \bibnamefont {Halimeh}},\ }\href@noop {} {\  (\bibinfo {year} {2022})},\ \Eprint {http://arxiv.org/abs/2211.01380} {arXiv:2211.01380 [cond-mat.quant-gas]} \BibitemShut {NoStop}%
\bibitem [{\citenamefont {Halimeh}\ \emph {et~al.}(2022)\citenamefont {Halimeh}, \citenamefont {Lang},\ and\ \citenamefont {Hauke}}]{halimeh2022gauge}%
  \BibitemOpen
  \bibfield  {author} {\bibinfo {author} {\bibfnamefont {J.~C.}\ \bibnamefont {Halimeh}}, \bibinfo {author} {\bibfnamefont {H.}~\bibnamefont {Lang}}, \ and\ \bibinfo {author} {\bibfnamefont {P.}~\bibnamefont {Hauke}},\ }\href {\doibase 10.1088/1367-2630/ac5564} {\bibfield  {journal} {\bibinfo  {journal} {New Journal of Physics}\ }\textbf {\bibinfo {volume} {24}},\ \bibinfo {pages} {033015} (\bibinfo {year} {2022})}\BibitemShut {NoStop}%
\bibitem [{\citenamefont {A~Rahman}\ \emph {et~al.}(2022)\citenamefont {A~Rahman}, \citenamefont {Lewis}, \citenamefont {Mendicelli},\ and\ \citenamefont {Powell}}]{ARahman:2022tkr}%
  \BibitemOpen
  \bibfield  {author} {\bibinfo {author} {\bibfnamefont {S.}~\bibnamefont {A~Rahman}}, \bibinfo {author} {\bibfnamefont {R.}~\bibnamefont {Lewis}}, \bibinfo {author} {\bibfnamefont {E.}~\bibnamefont {Mendicelli}}, \ and\ \bibinfo {author} {\bibfnamefont {S.}~\bibnamefont {Powell}},\ }\href {\doibase 10.1103/PhysRevD.106.074502} {\bibfield  {journal} {\bibinfo  {journal} {Phys. Rev. D}\ }\textbf {\bibinfo {volume} {106}},\ \bibinfo {pages} {074502} (\bibinfo {year} {2022})},\ \Eprint {http://arxiv.org/abs/2205.09247} {arXiv:2205.09247 [hep-lat]} \BibitemShut {NoStop}%
\bibitem [{\citenamefont {Zache}\ \emph {et~al.}(2023)\citenamefont {Zache}, \citenamefont {Gonz\'alez-Cuadra},\ and\ \citenamefont {Zoller}}]{zache2023quantum}%
  \BibitemOpen
  \bibfield  {author} {\bibinfo {author} {\bibfnamefont {T.~V.}\ \bibnamefont {Zache}}, \bibinfo {author} {\bibfnamefont {D.}~\bibnamefont {Gonz\'alez-Cuadra}}, \ and\ \bibinfo {author} {\bibfnamefont {P.}~\bibnamefont {Zoller}},\ }\href {\doibase 10.1103/PhysRevLett.131.171902} {\bibfield  {journal} {\bibinfo  {journal} {Phys. Rev. Lett.}\ }\textbf {\bibinfo {volume} {131}},\ \bibinfo {pages} {171902} (\bibinfo {year} {2023})}\BibitemShut {NoStop}%
\bibitem [{\citenamefont {Alexandru}\ \emph {et~al.}(2024)\citenamefont {Alexandru}, \citenamefont {Bedaque}, \citenamefont {Carosso}, \citenamefont {Cervia}, \citenamefont {Murairi},\ and\ \citenamefont {Sheng}}]{Alexandru:2023qzd}%
  \BibitemOpen
  \bibfield  {author} {\bibinfo {author} {\bibfnamefont {A.}~\bibnamefont {Alexandru}}, \bibinfo {author} {\bibfnamefont {P.~F.}\ \bibnamefont {Bedaque}}, \bibinfo {author} {\bibfnamefont {A.}~\bibnamefont {Carosso}}, \bibinfo {author} {\bibfnamefont {M.~J.}\ \bibnamefont {Cervia}}, \bibinfo {author} {\bibfnamefont {E.~M.}\ \bibnamefont {Murairi}}, \ and\ \bibinfo {author} {\bibfnamefont {A.}~\bibnamefont {Sheng}},\ }\href {\doibase 10.1103/PhysRevD.109.094502} {\bibfield  {journal} {\bibinfo  {journal} {Phys. Rev. D}\ }\textbf {\bibinfo {volume} {109}},\ \bibinfo {pages} {094502} (\bibinfo {year} {2024})},\ \Eprint {http://arxiv.org/abs/2308.05253} {arXiv:2308.05253 [hep-lat]} \BibitemShut {NoStop}%
\bibitem [{\citenamefont {D'Andrea}\ \emph {et~al.}(2024)\citenamefont {D'Andrea}, \citenamefont {Bauer}, \citenamefont {Grabowska},\ and\ \citenamefont {Freytsis}}]{DAndrea:2023qnr}%
  \BibitemOpen
  \bibfield  {author} {\bibinfo {author} {\bibfnamefont {I.}~\bibnamefont {D'Andrea}}, \bibinfo {author} {\bibfnamefont {C.~W.}\ \bibnamefont {Bauer}}, \bibinfo {author} {\bibfnamefont {D.~M.}\ \bibnamefont {Grabowska}}, \ and\ \bibinfo {author} {\bibfnamefont {M.}~\bibnamefont {Freytsis}},\ }\href {\doibase 10.1103/PhysRevD.109.074501} {\bibfield  {journal} {\bibinfo  {journal} {Phys. Rev. D}\ }\textbf {\bibinfo {volume} {109}},\ \bibinfo {pages} {074501} (\bibinfo {year} {2024})},\ \Eprint {http://arxiv.org/abs/2307.11829} {arXiv:2307.11829 [hep-ph]} \BibitemShut {NoStop}%
\bibitem [{\citenamefont {Turro}\ \emph {et~al.}(2024)\citenamefont {Turro}, \citenamefont {Ciavarella},\ and\ \citenamefont {Yao}}]{Turro:2024pxu}%
  \BibitemOpen
  \bibfield  {author} {\bibinfo {author} {\bibfnamefont {F.}~\bibnamefont {Turro}}, \bibinfo {author} {\bibfnamefont {A.}~\bibnamefont {Ciavarella}}, \ and\ \bibinfo {author} {\bibfnamefont {X.}~\bibnamefont {Yao}},\ }\href {\doibase 10.1103/PhysRevD.109.114511} {\bibfield  {journal} {\bibinfo  {journal} {Phys. Rev. D}\ }\textbf {\bibinfo {volume} {109}},\ \bibinfo {pages} {114511} (\bibinfo {year} {2024})},\ \Eprint {http://arxiv.org/abs/2402.04221} {arXiv:2402.04221 [hep-lat]} \BibitemShut {NoStop}%
\bibitem [{\citenamefont {Anishetty}\ \emph {et~al.}(2010)\citenamefont {Anishetty}, \citenamefont {Mathur},\ and\ \citenamefont {Raychowdhury}}]{Anishetty:2009nh}%
  \BibitemOpen
  \bibfield  {author} {\bibinfo {author} {\bibfnamefont {R.}~\bibnamefont {Anishetty}}, \bibinfo {author} {\bibfnamefont {M.}~\bibnamefont {Mathur}}, \ and\ \bibinfo {author} {\bibfnamefont {I.}~\bibnamefont {Raychowdhury}},\ }\href {\doibase 10.1088/1751-8113/43/3/035403} {\bibfield  {journal} {\bibinfo  {journal} {J. Phys. A}\ }\textbf {\bibinfo {volume} {43}},\ \bibinfo {pages} {035403} (\bibinfo {year} {2010})},\ \Eprint {http://arxiv.org/abs/0909.2394} {arXiv:0909.2394 [hep-lat]} \BibitemShut {NoStop}%
\bibitem [{\citenamefont {Alexandru}\ \emph {et~al.}(2019)\citenamefont {Alexandru}, \citenamefont {Bedaque}, \citenamefont {Harmalkar}, \citenamefont {Lamm}, \citenamefont {Lawrence},\ and\ \citenamefont {Warrington}}]{Alexandru:2019nsa}%
  \BibitemOpen
  \bibfield  {author} {\bibinfo {author} {\bibfnamefont {A.}~\bibnamefont {Alexandru}}, \bibinfo {author} {\bibfnamefont {P.~F.}\ \bibnamefont {Bedaque}}, \bibinfo {author} {\bibfnamefont {S.}~\bibnamefont {Harmalkar}}, \bibinfo {author} {\bibfnamefont {H.}~\bibnamefont {Lamm}}, \bibinfo {author} {\bibfnamefont {S.}~\bibnamefont {Lawrence}}, \ and\ \bibinfo {author} {\bibfnamefont {N.~C.}\ \bibnamefont {Warrington}} (\bibinfo {collaboration} {NuQS}),\ }\href {\doibase 10.1103/PhysRevD.100.114501} {\bibfield  {journal} {\bibinfo  {journal} {Phys. Rev. D}\ }\textbf {\bibinfo {volume} {100}},\ \bibinfo {pages} {114501} (\bibinfo {year} {2019})},\ \Eprint {http://arxiv.org/abs/1906.11213} {arXiv:1906.11213 [hep-lat]} \BibitemShut {NoStop}%
\bibitem [{\citenamefont {Ciavarella}\ \emph {et~al.}(2021)\citenamefont {Ciavarella}, \citenamefont {Klco},\ and\ \citenamefont {Savage}}]{Ciavarella:2021nmj}%
  \BibitemOpen
  \bibfield  {author} {\bibinfo {author} {\bibfnamefont {A.}~\bibnamefont {Ciavarella}}, \bibinfo {author} {\bibfnamefont {N.}~\bibnamefont {Klco}}, \ and\ \bibinfo {author} {\bibfnamefont {M.~J.}\ \bibnamefont {Savage}},\ }\href {\doibase 10.1103/PhysRevD.103.094501} {\bibfield  {journal} {\bibinfo  {journal} {Phys. Rev. D}\ }\textbf {\bibinfo {volume} {103}},\ \bibinfo {pages} {094501} (\bibinfo {year} {2021})},\ \Eprint {http://arxiv.org/abs/2101.10227} {arXiv:2101.10227 [quant-ph]} \BibitemShut {NoStop}%
\bibitem [{\citenamefont {Farrell}\ \emph {et~al.}(2023{\natexlab{a}})\citenamefont {Farrell}, \citenamefont {Chernyshev}, \citenamefont {Powell}, \citenamefont {Zemlevskiy}, \citenamefont {Illa},\ and\ \citenamefont {Savage}}]{Farrell:2022wyt}%
  \BibitemOpen
  \bibfield  {author} {\bibinfo {author} {\bibfnamefont {R.~C.}\ \bibnamefont {Farrell}}, \bibinfo {author} {\bibfnamefont {I.~A.}\ \bibnamefont {Chernyshev}}, \bibinfo {author} {\bibfnamefont {S.~J.~M.}\ \bibnamefont {Powell}}, \bibinfo {author} {\bibfnamefont {N.~A.}\ \bibnamefont {Zemlevskiy}}, \bibinfo {author} {\bibfnamefont {M.}~\bibnamefont {Illa}}, \ and\ \bibinfo {author} {\bibfnamefont {M.~J.}\ \bibnamefont {Savage}},\ }\href {\doibase 10.1103/PhysRevD.107.054512} {\bibfield  {journal} {\bibinfo  {journal} {Phys. Rev. D}\ }\textbf {\bibinfo {volume} {107}},\ \bibinfo {pages} {054512} (\bibinfo {year} {2023}{\natexlab{a}})},\ \Eprint {http://arxiv.org/abs/2207.01731} {arXiv:2207.01731 [quant-ph]} \BibitemShut {NoStop}%
\bibitem [{\citenamefont {Farrell}\ \emph {et~al.}(2023{\natexlab{b}})\citenamefont {Farrell}, \citenamefont {Chernyshev}, \citenamefont {Powell}, \citenamefont {Zemlevskiy}, \citenamefont {Illa},\ and\ \citenamefont {Savage}}]{Farrell:2022vyh}%
  \BibitemOpen
  \bibfield  {author} {\bibinfo {author} {\bibfnamefont {R.~C.}\ \bibnamefont {Farrell}}, \bibinfo {author} {\bibfnamefont {I.~A.}\ \bibnamefont {Chernyshev}}, \bibinfo {author} {\bibfnamefont {S.~J.~M.}\ \bibnamefont {Powell}}, \bibinfo {author} {\bibfnamefont {N.~A.}\ \bibnamefont {Zemlevskiy}}, \bibinfo {author} {\bibfnamefont {M.}~\bibnamefont {Illa}}, \ and\ \bibinfo {author} {\bibfnamefont {M.~J.}\ \bibnamefont {Savage}},\ }\href {\doibase 10.1103/PhysRevD.107.054513} {\bibfield  {journal} {\bibinfo  {journal} {Phys. Rev. D}\ }\textbf {\bibinfo {volume} {107}},\ \bibinfo {pages} {054513} (\bibinfo {year} {2023}{\natexlab{b}})},\ \Eprint {http://arxiv.org/abs/2209.10781} {arXiv:2209.10781 [quant-ph]} \BibitemShut {NoStop}%
\bibitem [{\citenamefont {Atas}\ \emph {et~al.}(2023)\citenamefont {Atas}, \citenamefont {Haase}, \citenamefont {Zhang}, \citenamefont {Wei}, \citenamefont {Pfaendler}, \citenamefont {Lewis},\ and\ \citenamefont {Muschik}}]{Atas:2022dqm}%
  \BibitemOpen
  \bibfield  {author} {\bibinfo {author} {\bibfnamefont {Y.~Y.}\ \bibnamefont {Atas}}, \bibinfo {author} {\bibfnamefont {J.~F.}\ \bibnamefont {Haase}}, \bibinfo {author} {\bibfnamefont {J.}~\bibnamefont {Zhang}}, \bibinfo {author} {\bibfnamefont {V.}~\bibnamefont {Wei}}, \bibinfo {author} {\bibfnamefont {S.~M.~L.}\ \bibnamefont {Pfaendler}}, \bibinfo {author} {\bibfnamefont {R.}~\bibnamefont {Lewis}}, \ and\ \bibinfo {author} {\bibfnamefont {C.~A.}\ \bibnamefont {Muschik}},\ }\href {\doibase 10.1103/PhysRevResearch.5.033184} {\bibfield  {journal} {\bibinfo  {journal} {Phys. Rev. Res.}\ }\textbf {\bibinfo {volume} {5}},\ \bibinfo {pages} {033184} (\bibinfo {year} {2023})},\ \Eprint {http://arxiv.org/abs/2207.03473} {arXiv:2207.03473 [quant-ph]} \BibitemShut {NoStop}%
\bibitem [{\citenamefont {Ciavarella}\ and\ \citenamefont {Chernyshev}(2022)}]{Ciavarella:2021lel}%
  \BibitemOpen
  \bibfield  {author} {\bibinfo {author} {\bibfnamefont {A.~N.}\ \bibnamefont {Ciavarella}}\ and\ \bibinfo {author} {\bibfnamefont {I.~A.}\ \bibnamefont {Chernyshev}},\ }\href {\doibase 10.1103/PhysRevD.105.074504} {\bibfield  {journal} {\bibinfo  {journal} {Phys. Rev. D}\ }\textbf {\bibinfo {volume} {105}},\ \bibinfo {pages} {074504} (\bibinfo {year} {2022})},\ \Eprint {http://arxiv.org/abs/2112.09083} {arXiv:2112.09083 [quant-ph]} \BibitemShut {NoStop}%
\bibitem [{\citenamefont {Ciavarella}(2023)}]{Ciavarella:2023mfc}%
  \BibitemOpen
  \bibfield  {author} {\bibinfo {author} {\bibfnamefont {A.~N.}\ \bibnamefont {Ciavarella}},\ }\href {\doibase 10.1103/PhysRevD.108.094513} {\bibfield  {journal} {\bibinfo  {journal} {Phys. Rev. D}\ }\textbf {\bibinfo {volume} {108}},\ \bibinfo {pages} {094513} (\bibinfo {year} {2023})},\ \Eprint {http://arxiv.org/abs/2307.05593} {arXiv:2307.05593 [hep-lat]} \BibitemShut {NoStop}%
\bibitem [{\citenamefont {Hayata}\ and\ \citenamefont {Hidaka}(2023)}]{hayata2023qdeformedformulationhamiltonian}%
  \BibitemOpen
  \bibfield  {author} {\bibinfo {author} {\bibfnamefont {T.}~\bibnamefont {Hayata}}\ and\ \bibinfo {author} {\bibfnamefont {Y.}~\bibnamefont {Hidaka}},\ }\href {https://arxiv.org/abs/2306.12324} {\enquote {\bibinfo {title} {$q$ deformed formulation of hamiltonian su(3) yang-mills theory},}\ } (\bibinfo {year} {2023}),\ \Eprint {http://arxiv.org/abs/2306.12324} {arXiv:2306.12324 [hep-lat]} \BibitemShut {NoStop}%
\bibitem [{\citenamefont {Farrell}\ \emph {et~al.}(2024{\natexlab{b}})\citenamefont {Farrell}, \citenamefont {Illa}, \citenamefont {Ciavarella},\ and\ \citenamefont {Savage}}]{Farrell:2024fit}%
  \BibitemOpen
  \bibfield  {author} {\bibinfo {author} {\bibfnamefont {R.~C.}\ \bibnamefont {Farrell}}, \bibinfo {author} {\bibfnamefont {M.}~\bibnamefont {Illa}}, \bibinfo {author} {\bibfnamefont {A.~N.}\ \bibnamefont {Ciavarella}}, \ and\ \bibinfo {author} {\bibfnamefont {M.~J.}\ \bibnamefont {Savage}},\ }\href {\doibase 10.1103/PhysRevD.109.114510} {\bibfield  {journal} {\bibinfo  {journal} {Phys. Rev. D}\ }\textbf {\bibinfo {volume} {109}},\ \bibinfo {pages} {114510} (\bibinfo {year} {2024}{\natexlab{b}})},\ \Eprint {http://arxiv.org/abs/2401.08044} {arXiv:2401.08044 [quant-ph]} \BibitemShut {NoStop}%
\bibitem [{\citenamefont {Ciavarella}\ and\ \citenamefont {Bauer}(2024)}]{Ciavarella:2024fzw}%
  \BibitemOpen
  \bibfield  {author} {\bibinfo {author} {\bibfnamefont {A.~N.}\ \bibnamefont {Ciavarella}}\ and\ \bibinfo {author} {\bibfnamefont {C.~W.}\ \bibnamefont {Bauer}},\ }\href {\doibase 10.1103/PhysRevLett.133.111901} {\bibfield  {journal} {\bibinfo  {journal} {Phys. Rev. Lett.}\ }\textbf {\bibinfo {volume} {133}},\ \bibinfo {pages} {111901} (\bibinfo {year} {2024})},\ \Eprint {http://arxiv.org/abs/2402.10265} {arXiv:2402.10265 [hep-ph]} \BibitemShut {NoStop}%
\bibitem [{\citenamefont {Pichler}\ \emph {et~al.}(2016)\citenamefont {Pichler}, \citenamefont {Dalmonte}, \citenamefont {Rico}, \citenamefont {Zoller},\ and\ \citenamefont {Montangero}}]{Pichler:2015yqa}%
  \BibitemOpen
  \bibfield  {author} {\bibinfo {author} {\bibfnamefont {T.}~\bibnamefont {Pichler}}, \bibinfo {author} {\bibfnamefont {M.}~\bibnamefont {Dalmonte}}, \bibinfo {author} {\bibfnamefont {E.}~\bibnamefont {Rico}}, \bibinfo {author} {\bibfnamefont {P.}~\bibnamefont {Zoller}}, \ and\ \bibinfo {author} {\bibfnamefont {S.}~\bibnamefont {Montangero}},\ }\href {\doibase 10.1103/PhysRevX.6.011023} {\bibfield  {journal} {\bibinfo  {journal} {Phys. Rev. X}\ }\textbf {\bibinfo {volume} {6}},\ \bibinfo {pages} {011023} (\bibinfo {year} {2016})},\ \Eprint {http://arxiv.org/abs/1505.04440} {arXiv:1505.04440 [cond-mat.quant-gas]} \BibitemShut {NoStop}%
\bibitem [{\citenamefont {Ba\~nuls}\ \emph {et~al.}(2017)\citenamefont {Ba\~nuls}, \citenamefont {Cichy}, \citenamefont {Cirac}, \citenamefont {Jansen},\ and\ \citenamefont {K\"uhn}}]{Banuls:2017ena}%
  \BibitemOpen
  \bibfield  {author} {\bibinfo {author} {\bibfnamefont {M.~C.}\ \bibnamefont {Ba\~nuls}}, \bibinfo {author} {\bibfnamefont {K.}~\bibnamefont {Cichy}}, \bibinfo {author} {\bibfnamefont {J.~I.}\ \bibnamefont {Cirac}}, \bibinfo {author} {\bibfnamefont {K.}~\bibnamefont {Jansen}}, \ and\ \bibinfo {author} {\bibfnamefont {S.}~\bibnamefont {K\"uhn}},\ }\href {\doibase 10.1103/PhysRevX.7.041046} {\bibfield  {journal} {\bibinfo  {journal} {Phys. Rev. X}\ }\textbf {\bibinfo {volume} {7}},\ \bibinfo {pages} {041046} (\bibinfo {year} {2017})},\ \Eprint {http://arxiv.org/abs/1707.06434} {arXiv:1707.06434 [hep-lat]} \BibitemShut {NoStop}%
\bibitem [{\citenamefont {Ba\~nuls}\ \emph {et~al.}(2018)\citenamefont {Ba\~nuls}, \citenamefont {Cichy}, \citenamefont {Cirac}, \citenamefont {Jansen},\ and\ \citenamefont {K\"uhn}}]{Banuls:2018jag}%
  \BibitemOpen
  \bibfield  {author} {\bibinfo {author} {\bibfnamefont {M.~C.}\ \bibnamefont {Ba\~nuls}}, \bibinfo {author} {\bibfnamefont {K.}~\bibnamefont {Cichy}}, \bibinfo {author} {\bibfnamefont {J.~I.}\ \bibnamefont {Cirac}}, \bibinfo {author} {\bibfnamefont {K.}~\bibnamefont {Jansen}}, \ and\ \bibinfo {author} {\bibfnamefont {S.}~\bibnamefont {K\"uhn}},\ }\href {\doibase 10.22323/1.334.0022} {\bibfield  {journal} {\bibinfo  {journal} {PoS}\ }\textbf {\bibinfo {volume} {LATTICE2018}},\ \bibinfo {pages} {022} (\bibinfo {year} {2018})},\ \Eprint {http://arxiv.org/abs/1810.12838} {arXiv:1810.12838 [hep-lat]} \BibitemShut {NoStop}%
\bibitem [{\citenamefont {Ba\~nuls}(2023)}]{Banuls:2022vxp}%
  \BibitemOpen
  \bibfield  {author} {\bibinfo {author} {\bibfnamefont {M.~C.}\ \bibnamefont {Ba\~nuls}},\ }\href {\doibase 10.1146/annurev-conmatphys-040721-022705} {\bibfield  {journal} {\bibinfo  {journal} {Ann. Rev. Condensed Matter Phys.}\ }\textbf {\bibinfo {volume} {14}},\ \bibinfo {pages} {173} (\bibinfo {year} {2023})},\ \Eprint {http://arxiv.org/abs/2205.10345} {arXiv:2205.10345 [quant-ph]} \BibitemShut {NoStop}%
\bibitem [{\citenamefont {Kan}\ and\ \citenamefont {Nam}(2021)}]{Kan:2021xfc}%
  \BibitemOpen
  \bibfield  {author} {\bibinfo {author} {\bibfnamefont {A.}~\bibnamefont {Kan}}\ and\ \bibinfo {author} {\bibfnamefont {Y.}~\bibnamefont {Nam}},\ }\href@noop {} {\  (\bibinfo {year} {2021})},\ \Eprint {http://arxiv.org/abs/2107.12769} {arXiv:2107.12769 [quant-ph]} \BibitemShut {NoStop}%
\bibitem [{\citenamefont {Alexandru}\ \emph {et~al.}(2023)\citenamefont {Alexandru}, \citenamefont {Bedaque}, \citenamefont {Carosso}, \citenamefont {Cervia},\ and\ \citenamefont {Sheng}}]{Alexandru:2022son}%
  \BibitemOpen
  \bibfield  {author} {\bibinfo {author} {\bibfnamefont {A.}~\bibnamefont {Alexandru}}, \bibinfo {author} {\bibfnamefont {P.~F.}\ \bibnamefont {Bedaque}}, \bibinfo {author} {\bibfnamefont {A.}~\bibnamefont {Carosso}}, \bibinfo {author} {\bibfnamefont {M.~J.}\ \bibnamefont {Cervia}}, \ and\ \bibinfo {author} {\bibfnamefont {A.}~\bibnamefont {Sheng}},\ }\href {\doibase 10.1103/PhysRevD.107.034503} {\bibfield  {journal} {\bibinfo  {journal} {Phys. Rev. D}\ }\textbf {\bibinfo {volume} {107}},\ \bibinfo {pages} {034503} (\bibinfo {year} {2023})},\ \Eprint {http://arxiv.org/abs/2209.00098} {arXiv:2209.00098 [hep-lat]} \BibitemShut {NoStop}%
\bibitem [{\citenamefont {Tong}\ \emph {et~al.}(2022)\citenamefont {Tong}, \citenamefont {Albert}, \citenamefont {McClean}, \citenamefont {Preskill},\ and\ \citenamefont {Su}}]{tong2022provably}%
  \BibitemOpen
  \bibfield  {author} {\bibinfo {author} {\bibfnamefont {Y.}~\bibnamefont {Tong}}, \bibinfo {author} {\bibfnamefont {V.~V.}\ \bibnamefont {Albert}}, \bibinfo {author} {\bibfnamefont {J.~R.}\ \bibnamefont {McClean}}, \bibinfo {author} {\bibfnamefont {J.}~\bibnamefont {Preskill}}, \ and\ \bibinfo {author} {\bibfnamefont {Y.}~\bibnamefont {Su}},\ }\href {\doibase 10.22331/q-2022-09-22-816} {\bibfield  {journal} {\bibinfo  {journal} {Quantum}\ }\textbf {\bibinfo {volume} {6}},\ \bibinfo {pages} {816} (\bibinfo {year} {2022})}\BibitemShut {NoStop}%
\bibitem [{\citenamefont {Davoudi}\ \emph {et~al.}(2023)\citenamefont {Davoudi}, \citenamefont {Shaw},\ and\ \citenamefont {Stryker}}]{Davoudi:2022xmb}%
  \BibitemOpen
  \bibfield  {author} {\bibinfo {author} {\bibfnamefont {Z.}~\bibnamefont {Davoudi}}, \bibinfo {author} {\bibfnamefont {A.~F.}\ \bibnamefont {Shaw}}, \ and\ \bibinfo {author} {\bibfnamefont {J.~R.}\ \bibnamefont {Stryker}},\ }\href {\doibase 10.22331/q-2023-12-20-1213} {\bibfield  {journal} {\bibinfo  {journal} {Quantum}\ }\textbf {\bibinfo {volume} {7}},\ \bibinfo {pages} {1213} (\bibinfo {year} {2023})},\ \Eprint {http://arxiv.org/abs/2212.14030} {arXiv:2212.14030 [hep-lat]} \BibitemShut {NoStop}%
\bibitem [{\citenamefont {Kane}\ \emph {et~al.}(2024{\natexlab{a}})\citenamefont {Kane}, \citenamefont {Gomes},\ and\ \citenamefont {Kreshchuk}}]{Kane:2023jdo}%
  \BibitemOpen
  \bibfield  {author} {\bibinfo {author} {\bibfnamefont {C.~F.}\ \bibnamefont {Kane}}, \bibinfo {author} {\bibfnamefont {N.}~\bibnamefont {Gomes}}, \ and\ \bibinfo {author} {\bibfnamefont {M.}~\bibnamefont {Kreshchuk}},\ }\href {\doibase 10.1103/PhysRevA.110.012455} {\bibfield  {journal} {\bibinfo  {journal} {Phys. Rev. A}\ }\textbf {\bibinfo {volume} {110}},\ \bibinfo {pages} {012455} (\bibinfo {year} {2024}{\natexlab{a}})},\ \Eprint {http://arxiv.org/abs/2310.13757} {arXiv:2310.13757 [quant-ph]} \BibitemShut {NoStop}%
\bibitem [{\citenamefont {Hariprakash}\ \emph {et~al.}(2023)\citenamefont {Hariprakash}, \citenamefont {Modi}, \citenamefont {Kreshchuk}, \citenamefont {Kane},\ and\ \citenamefont {Bauer}}]{Hariprakash:2023tla}%
  \BibitemOpen
  \bibfield  {author} {\bibinfo {author} {\bibfnamefont {S.}~\bibnamefont {Hariprakash}}, \bibinfo {author} {\bibfnamefont {N.~S.}\ \bibnamefont {Modi}}, \bibinfo {author} {\bibfnamefont {M.}~\bibnamefont {Kreshchuk}}, \bibinfo {author} {\bibfnamefont {C.~F.}\ \bibnamefont {Kane}}, \ and\ \bibinfo {author} {\bibfnamefont {C.~W.}\ \bibnamefont {Bauer}},\ }\href@noop {} {\  (\bibinfo {year} {2023})},\ \Eprint {http://arxiv.org/abs/2312.11637} {arXiv:2312.11637 [quant-ph]} \BibitemShut {NoStop}%
\bibitem [{\citenamefont {Rhodes}\ \emph {et~al.}(2024)\citenamefont {Rhodes}, \citenamefont {Kreshchuk},\ and\ \citenamefont {Pathak}}]{Rhodes:2024zbr}%
  \BibitemOpen
  \bibfield  {author} {\bibinfo {author} {\bibfnamefont {M.}~\bibnamefont {Rhodes}}, \bibinfo {author} {\bibfnamefont {M.}~\bibnamefont {Kreshchuk}}, \ and\ \bibinfo {author} {\bibfnamefont {S.}~\bibnamefont {Pathak}},\ }\href@noop {} {\  (\bibinfo {year} {2024})},\ \Eprint {http://arxiv.org/abs/2405.10416} {arXiv:2405.10416 [quant-ph]} \BibitemShut {NoStop}%
\bibitem [{\citenamefont {Du}\ and\ \citenamefont {Vary}(2024)}]{Du:2024ixj}%
  \BibitemOpen
  \bibfield  {author} {\bibinfo {author} {\bibfnamefont {W.}~\bibnamefont {Du}}\ and\ \bibinfo {author} {\bibfnamefont {J.~P.}\ \bibnamefont {Vary}},\ }\href@noop {} {\  (\bibinfo {year} {2024})},\ \Eprint {http://arxiv.org/abs/2407.13672} {arXiv:2407.13672 [quant-ph]} \BibitemShut {NoStop}%
\bibitem [{\citenamefont {Li}\ \emph {et~al.}(2024)\citenamefont {Li}, \citenamefont {Grabowska},\ and\ \citenamefont {Savage}}]{Li:2024lrl}%
  \BibitemOpen
  \bibfield  {author} {\bibinfo {author} {\bibfnamefont {Z.}~\bibnamefont {Li}}, \bibinfo {author} {\bibfnamefont {D.~M.}\ \bibnamefont {Grabowska}}, \ and\ \bibinfo {author} {\bibfnamefont {M.~J.}\ \bibnamefont {Savage}},\ }\href@noop {} {\  (\bibinfo {year} {2024})},\ \Eprint {http://arxiv.org/abs/2407.13835} {arXiv:2407.13835 [quant-ph]} \BibitemShut {NoStop}%
\bibitem [{\citenamefont {Kane}\ \emph {et~al.}(2024{\natexlab{b}})\citenamefont {Kane}, \citenamefont {Hariprakash}, \citenamefont {Modi}, \citenamefont {Kreshchuk},\ and\ \citenamefont {Bauer}}]{Kane:2024odt}%
  \BibitemOpen
  \bibfield  {author} {\bibinfo {author} {\bibfnamefont {C.~F.}\ \bibnamefont {Kane}}, \bibinfo {author} {\bibfnamefont {S.}~\bibnamefont {Hariprakash}}, \bibinfo {author} {\bibfnamefont {N.~S.}\ \bibnamefont {Modi}}, \bibinfo {author} {\bibfnamefont {M.}~\bibnamefont {Kreshchuk}}, \ and\ \bibinfo {author} {\bibfnamefont {C.~W.}\ \bibnamefont {Bauer}},\ }\href@noop {} {\  (\bibinfo {year} {2024}{\natexlab{b}})},\ \Eprint {http://arxiv.org/abs/2408.16824} {arXiv:2408.16824 [quant-ph]} \BibitemShut {NoStop}%
\bibitem [{\citenamefont {Bauer}\ \emph {et~al.}(2021)\citenamefont {Bauer}, \citenamefont {de~Jong}, \citenamefont {Nachman},\ and\ \citenamefont {Provasoli}}]{Bauer:2019qxa}%
  \BibitemOpen
  \bibfield  {author} {\bibinfo {author} {\bibfnamefont {C.~W.}\ \bibnamefont {Bauer}}, \bibinfo {author} {\bibfnamefont {W.~A.}\ \bibnamefont {de~Jong}}, \bibinfo {author} {\bibfnamefont {B.}~\bibnamefont {Nachman}}, \ and\ \bibinfo {author} {\bibfnamefont {D.}~\bibnamefont {Provasoli}},\ }\href {\doibase 10.1103/PhysRevLett.126.062001} {\bibfield  {journal} {\bibinfo  {journal} {Phys. Rev. Lett.}\ }\textbf {\bibinfo {volume} {126}},\ \bibinfo {pages} {062001} (\bibinfo {year} {2021})},\ \Eprint {http://arxiv.org/abs/1904.03196} {arXiv:1904.03196 [hep-ph]} \BibitemShut {NoStop}%
\bibitem [{\citenamefont {Bauer}\ \emph {et~al.}(2024)\citenamefont {Bauer}, \citenamefont {Chigusa},\ and\ \citenamefont {Yamazaki}}]{Bauer:2023ujy}%
  \BibitemOpen
  \bibfield  {author} {\bibinfo {author} {\bibfnamefont {C.~W.}\ \bibnamefont {Bauer}}, \bibinfo {author} {\bibfnamefont {S.}~\bibnamefont {Chigusa}}, \ and\ \bibinfo {author} {\bibfnamefont {M.}~\bibnamefont {Yamazaki}},\ }\href {\doibase 10.1103/PhysRevA.109.032432} {\bibfield  {journal} {\bibinfo  {journal} {Phys. Rev. A}\ }\textbf {\bibinfo {volume} {109}},\ \bibinfo {pages} {032432} (\bibinfo {year} {2024})},\ \Eprint {http://arxiv.org/abs/2310.19881} {arXiv:2310.19881 [hep-ph]} \BibitemShut {NoStop}%
\bibitem [{\citenamefont {Chigusa}\ and\ \citenamefont {Yamazaki}(2022)}]{Chigusa:2022act}%
  \BibitemOpen
  \bibfield  {author} {\bibinfo {author} {\bibfnamefont {S.}~\bibnamefont {Chigusa}}\ and\ \bibinfo {author} {\bibfnamefont {M.}~\bibnamefont {Yamazaki}},\ }\href {\doibase 10.1016/j.physletb.2022.137466} {\bibfield  {journal} {\bibinfo  {journal} {Phys. Lett. B}\ }\textbf {\bibinfo {volume} {834}},\ \bibinfo {pages} {137466} (\bibinfo {year} {2022})},\ \Eprint {http://arxiv.org/abs/2204.12500} {arXiv:2204.12500 [hep-ph]} \BibitemShut {NoStop}%
\bibitem [{\citenamefont {Tagliacozzo}\ \emph {et~al.}(2013)\citenamefont {Tagliacozzo}, \citenamefont {Celi}, \citenamefont {Orland},\ and\ \citenamefont {Lewenstein}}]{Tagliacozzo:2012df}%
  \BibitemOpen
  \bibfield  {author} {\bibinfo {author} {\bibfnamefont {L.}~\bibnamefont {Tagliacozzo}}, \bibinfo {author} {\bibfnamefont {A.}~\bibnamefont {Celi}}, \bibinfo {author} {\bibfnamefont {P.}~\bibnamefont {Orland}}, \ and\ \bibinfo {author} {\bibfnamefont {M.}~\bibnamefont {Lewenstein}},\ }\href {\doibase 10.1038/ncomms3615} {\bibfield  {journal} {\bibinfo  {journal} {Nature Commun.}\ }\textbf {\bibinfo {volume} {4}},\ \bibinfo {pages} {2615} (\bibinfo {year} {2013})},\ \Eprint {http://arxiv.org/abs/1211.2704} {arXiv:1211.2704 [cond-mat.quant-gas]} \BibitemShut {NoStop}%
\bibitem [{\citenamefont {Bazavov}\ \emph {et~al.}(2015)\citenamefont {Bazavov}, \citenamefont {Meurice}, \citenamefont {Tsai}, \citenamefont {Unmuth-Yockey},\ and\ \citenamefont {Zhang}}]{Bazavov:2015kka}%
  \BibitemOpen
  \bibfield  {author} {\bibinfo {author} {\bibfnamefont {A.}~\bibnamefont {Bazavov}}, \bibinfo {author} {\bibfnamefont {Y.}~\bibnamefont {Meurice}}, \bibinfo {author} {\bibfnamefont {S.-W.}\ \bibnamefont {Tsai}}, \bibinfo {author} {\bibfnamefont {J.}~\bibnamefont {Unmuth-Yockey}}, \ and\ \bibinfo {author} {\bibfnamefont {J.}~\bibnamefont {Zhang}},\ }\href {\doibase 10.1103/PhysRevD.92.076003} {\bibfield  {journal} {\bibinfo  {journal} {Phys. Rev. D}\ }\textbf {\bibinfo {volume} {92}},\ \bibinfo {pages} {076003} (\bibinfo {year} {2015})},\ \Eprint {http://arxiv.org/abs/1503.08354} {arXiv:1503.08354 [hep-lat]} \BibitemShut {NoStop}%
\bibitem [{\citenamefont {Jordan}\ \emph {et~al.}(2018)\citenamefont {Jordan}, \citenamefont {Krovi}, \citenamefont {Lee},\ and\ \citenamefont {Preskill}}]{Jordan:2017lea}%
  \BibitemOpen
  \bibfield  {author} {\bibinfo {author} {\bibfnamefont {S.~P.}\ \bibnamefont {Jordan}}, \bibinfo {author} {\bibfnamefont {H.}~\bibnamefont {Krovi}}, \bibinfo {author} {\bibfnamefont {K.~S.~M.}\ \bibnamefont {Lee}}, \ and\ \bibinfo {author} {\bibfnamefont {J.}~\bibnamefont {Preskill}},\ }\href {\doibase 10.22331/q-2018-01-08-44} {\bibfield  {journal} {\bibinfo  {journal} {Quantum}\ }\textbf {\bibinfo {volume} {2}},\ \bibinfo {pages} {44} (\bibinfo {year} {2018})},\ \Eprint {http://arxiv.org/abs/1703.00454} {arXiv:1703.00454 [quant-ph]} \BibitemShut {NoStop}%
\bibitem [{\citenamefont {Gonz\'alez-Cuadra}\ \emph {et~al.}(2017)\citenamefont {Gonz\'alez-Cuadra}, \citenamefont {Zohar},\ and\ \citenamefont {Cirac}}]{Gonzalez-Cuadra:2017lvz}%
  \BibitemOpen
  \bibfield  {author} {\bibinfo {author} {\bibfnamefont {D.}~\bibnamefont {Gonz\'alez-Cuadra}}, \bibinfo {author} {\bibfnamefont {E.}~\bibnamefont {Zohar}}, \ and\ \bibinfo {author} {\bibfnamefont {J.~I.}\ \bibnamefont {Cirac}},\ }\href {\doibase 10.1088/1367-2630/aa6f37} {\bibfield  {journal} {\bibinfo  {journal} {New J. Phys.}\ }\textbf {\bibinfo {volume} {19}},\ \bibinfo {pages} {063038} (\bibinfo {year} {2017})},\ \Eprint {http://arxiv.org/abs/1702.05492} {arXiv:1702.05492 [quant-ph]} \BibitemShut {NoStop}%
\bibitem [{\citenamefont {G\"org}\ \emph {et~al.}(2019)\citenamefont {G\"org}, \citenamefont {Sandholzer}, \citenamefont {Minguzzi}, \citenamefont {Desbuquois}, \citenamefont {Messer},\ and\ \citenamefont {Esslinger}}]{Gorg:2018xyc}%
  \BibitemOpen
  \bibfield  {author} {\bibinfo {author} {\bibfnamefont {F.}~\bibnamefont {G\"org}}, \bibinfo {author} {\bibfnamefont {K.}~\bibnamefont {Sandholzer}}, \bibinfo {author} {\bibfnamefont {J.}~\bibnamefont {Minguzzi}}, \bibinfo {author} {\bibfnamefont {R.}~\bibnamefont {Desbuquois}}, \bibinfo {author} {\bibfnamefont {M.}~\bibnamefont {Messer}}, \ and\ \bibinfo {author} {\bibfnamefont {T.}~\bibnamefont {Esslinger}},\ }\href {\doibase 10.1038/s41567-019-0615-4} {\bibfield  {journal} {\bibinfo  {journal} {Nature Phys.}\ }\textbf {\bibinfo {volume} {15}},\ \bibinfo {pages} {1161} (\bibinfo {year} {2019})},\ \Eprint {http://arxiv.org/abs/1812.05895} {arXiv:1812.05895 [cond-mat.quant-gas]} \BibitemShut {NoStop}%
\bibitem [{\citenamefont {Lamm}\ \emph {et~al.}(2019)\citenamefont {Lamm}, \citenamefont {Lawrence},\ and\ \citenamefont {Yamauchi}}]{Lamm:2019bik}%
  \BibitemOpen
  \bibfield  {author} {\bibinfo {author} {\bibfnamefont {H.}~\bibnamefont {Lamm}}, \bibinfo {author} {\bibfnamefont {S.}~\bibnamefont {Lawrence}}, \ and\ \bibinfo {author} {\bibfnamefont {Y.}~\bibnamefont {Yamauchi}} (\bibinfo {collaboration} {NuQS}),\ }\href {\doibase 10.1103/PhysRevD.100.034518} {\bibfield  {journal} {\bibinfo  {journal} {Phys. Rev. D}\ }\textbf {\bibinfo {volume} {100}},\ \bibinfo {pages} {034518} (\bibinfo {year} {2019})},\ \Eprint {http://arxiv.org/abs/1903.08807} {arXiv:1903.08807 [hep-lat]} \BibitemShut {NoStop}%
\bibitem [{\citenamefont {Zohar}\ and\ \citenamefont {Cirac}(2019)}]{Zohar:2019ygc}%
  \BibitemOpen
  \bibfield  {author} {\bibinfo {author} {\bibfnamefont {E.}~\bibnamefont {Zohar}}\ and\ \bibinfo {author} {\bibfnamefont {J.~I.}\ \bibnamefont {Cirac}},\ }\href {\doibase 10.1103/PhysRevD.99.114511} {\bibfield  {journal} {\bibinfo  {journal} {Phys. Rev. D}\ }\textbf {\bibinfo {volume} {99}},\ \bibinfo {pages} {114511} (\bibinfo {year} {2019})},\ \Eprint {http://arxiv.org/abs/1905.00652} {arXiv:1905.00652 [quant-ph]} \BibitemShut {NoStop}%
\bibitem [{\citenamefont {Buser}\ \emph {et~al.}(2021)\citenamefont {Buser}, \citenamefont {Gharibyan}, \citenamefont {Hanada}, \citenamefont {Honda},\ and\ \citenamefont {Liu}}]{Buser:2020cvn}%
  \BibitemOpen
  \bibfield  {author} {\bibinfo {author} {\bibfnamefont {A.~J.}\ \bibnamefont {Buser}}, \bibinfo {author} {\bibfnamefont {H.}~\bibnamefont {Gharibyan}}, \bibinfo {author} {\bibfnamefont {M.}~\bibnamefont {Hanada}}, \bibinfo {author} {\bibfnamefont {M.}~\bibnamefont {Honda}}, \ and\ \bibinfo {author} {\bibfnamefont {J.}~\bibnamefont {Liu}},\ }\href {\doibase 10.1007/JHEP09(2021)034} {\bibfield  {journal} {\bibinfo  {journal} {JHEP}\ }\textbf {\bibinfo {volume} {09}},\ \bibinfo {pages} {034} (\bibinfo {year} {2021})},\ \Eprint {http://arxiv.org/abs/2011.06576} {arXiv:2011.06576 [hep-th]} \BibitemShut {NoStop}%
\bibitem [{\citenamefont {Barata}\ \emph {et~al.}(2021)\citenamefont {Barata}, \citenamefont {Mueller}, \citenamefont {Tarasov},\ and\ \citenamefont {Venugopalan}}]{Barata:2020jtq}%
  \BibitemOpen
  \bibfield  {author} {\bibinfo {author} {\bibfnamefont {J.~a.}\ \bibnamefont {Barata}}, \bibinfo {author} {\bibfnamefont {N.}~\bibnamefont {Mueller}}, \bibinfo {author} {\bibfnamefont {A.}~\bibnamefont {Tarasov}}, \ and\ \bibinfo {author} {\bibfnamefont {R.}~\bibnamefont {Venugopalan}},\ }\href {\doibase 10.1103/PhysRevA.103.042410} {\bibfield  {journal} {\bibinfo  {journal} {Phys. Rev. A}\ }\textbf {\bibinfo {volume} {103}},\ \bibinfo {pages} {042410} (\bibinfo {year} {2021})},\ \Eprint {http://arxiv.org/abs/2012.00020} {arXiv:2012.00020 [hep-th]} \BibitemShut {NoStop}%
\bibitem [{\citenamefont {Stryker}(2021)}]{Stryker:2021asy}%
  \BibitemOpen
  \bibfield  {author} {\bibinfo {author} {\bibfnamefont {J.~R.}\ \bibnamefont {Stryker}},\ }\href@noop {} {\  (\bibinfo {year} {2021})},\ \Eprint {http://arxiv.org/abs/2105.11548} {arXiv:2105.11548 [hep-lat]} \BibitemShut {NoStop}%
\bibitem [{\citenamefont {Kreshchuk}\ \emph {et~al.}(2021)\citenamefont {Kreshchuk}, \citenamefont {Jia}, \citenamefont {Kirby}, \citenamefont {Goldstein}, \citenamefont {Vary},\ and\ \citenamefont {Love}}]{kreshchuk2021light}%
  \BibitemOpen
  \bibfield  {author} {\bibinfo {author} {\bibfnamefont {M.}~\bibnamefont {Kreshchuk}}, \bibinfo {author} {\bibfnamefont {S.}~\bibnamefont {Jia}}, \bibinfo {author} {\bibfnamefont {W.~M.}\ \bibnamefont {Kirby}}, \bibinfo {author} {\bibfnamefont {G.}~\bibnamefont {Goldstein}}, \bibinfo {author} {\bibfnamefont {J.~P.}\ \bibnamefont {Vary}}, \ and\ \bibinfo {author} {\bibfnamefont {P.~J.}\ \bibnamefont {Love}},\ }\href {\doibase 10.3390/e23050597} {\bibfield  {journal} {\bibinfo  {journal} {Entropy}\ }\textbf {\bibinfo {volume} {23}} (\bibinfo {year} {2021}),\ 10.3390/e23050597}\BibitemShut {NoStop}%
\bibitem [{\citenamefont {Kreshchuk}\ \emph {et~al.}(2022)\citenamefont {Kreshchuk}, \citenamefont {Kirby}, \citenamefont {Goldstein}, \citenamefont {Beauchemin},\ and\ \citenamefont {Love}}]{kreshchuk2022quantum}%
  \BibitemOpen
  \bibfield  {author} {\bibinfo {author} {\bibfnamefont {M.}~\bibnamefont {Kreshchuk}}, \bibinfo {author} {\bibfnamefont {W.~M.}\ \bibnamefont {Kirby}}, \bibinfo {author} {\bibfnamefont {G.}~\bibnamefont {Goldstein}}, \bibinfo {author} {\bibfnamefont {H.}~\bibnamefont {Beauchemin}}, \ and\ \bibinfo {author} {\bibfnamefont {P.~J.}\ \bibnamefont {Love}},\ }\href {\doibase 10.1103/PhysRevA.105.032418} {\bibfield  {journal} {\bibinfo  {journal} {Phys. Rev. A}\ }\textbf {\bibinfo {volume} {105}},\ \bibinfo {pages} {032418} (\bibinfo {year} {2022})}\BibitemShut {NoStop}%
\bibitem [{\citenamefont {Kreshchuk}\ \emph {et~al.}(2023)\citenamefont {Kreshchuk}, \citenamefont {Vary},\ and\ \citenamefont {Love}}]{kreshchuk2023simulatingscatteringcompositeparticles}%
  \BibitemOpen
  \bibfield  {author} {\bibinfo {author} {\bibfnamefont {M.}~\bibnamefont {Kreshchuk}}, \bibinfo {author} {\bibfnamefont {J.~P.}\ \bibnamefont {Vary}}, \ and\ \bibinfo {author} {\bibfnamefont {P.~J.}\ \bibnamefont {Love}},\ }\href {https://arxiv.org/abs/2310.13742} {\enquote {\bibinfo {title} {Simulating scattering of composite particles},}\ } (\bibinfo {year} {2023}),\ \Eprint {http://arxiv.org/abs/2310.13742} {arXiv:2310.13742 [quant-ph]} \BibitemShut {NoStop}%
\bibitem [{\citenamefont {Kogut}\ and\ \citenamefont {Susskind}(1975)}]{PhysRevD.11.395}%
  \BibitemOpen
  \bibfield  {author} {\bibinfo {author} {\bibfnamefont {J.}~\bibnamefont {Kogut}}\ and\ \bibinfo {author} {\bibfnamefont {L.}~\bibnamefont {Susskind}},\ }\href {\doibase 10.1103/PhysRevD.11.395} {\bibfield  {journal} {\bibinfo  {journal} {Phys. Rev. D}\ }\textbf {\bibinfo {volume} {11}},\ \bibinfo {pages} {395} (\bibinfo {year} {1975})}\BibitemShut {NoStop}%
\bibitem [{\citenamefont {Carena}\ \emph {et~al.}(2022)\citenamefont {Carena}, \citenamefont {Lamm}, \citenamefont {Li},\ and\ \citenamefont {Liu}}]{Carena:2022kpg}%
  \BibitemOpen
  \bibfield  {author} {\bibinfo {author} {\bibfnamefont {M.}~\bibnamefont {Carena}}, \bibinfo {author} {\bibfnamefont {H.}~\bibnamefont {Lamm}}, \bibinfo {author} {\bibfnamefont {Y.-Y.}\ \bibnamefont {Li}}, \ and\ \bibinfo {author} {\bibfnamefont {W.}~\bibnamefont {Liu}},\ }\href {\doibase 10.1103/PhysRevLett.129.051601} {\bibfield  {journal} {\bibinfo  {journal} {Phys. Rev. Lett.}\ }\textbf {\bibinfo {volume} {129}},\ \bibinfo {pages} {051601} (\bibinfo {year} {2022})},\ \Eprint {http://arxiv.org/abs/2203.02823} {arXiv:2203.02823 [hep-lat]} \BibitemShut {NoStop}%
\bibitem [{\citenamefont {Gustafson}\ and\ \citenamefont {Van~de Water}(2024)}]{Gustafson:2023aai}%
  \BibitemOpen
  \bibfield  {author} {\bibinfo {author} {\bibfnamefont {E.}~\bibnamefont {Gustafson}}\ and\ \bibinfo {author} {\bibfnamefont {R.}~\bibnamefont {Van~de Water}},\ }\href {\doibase 10.22323/1.453.0215} {\bibfield  {journal} {\bibinfo  {journal} {PoS}\ }\textbf {\bibinfo {volume} {LATTICE2023}},\ \bibinfo {pages} {215} (\bibinfo {year} {2024})},\ \Eprint {http://arxiv.org/abs/2402.04317} {arXiv:2402.04317 [hep-lat]} \BibitemShut {NoStop}%
\bibitem [{\citenamefont {Zache}\ \emph {et~al.}(2018)\citenamefont {Zache}, \citenamefont {Hebenstreit}, \citenamefont {Jendrzejewski}, \citenamefont {Oberthaler}, \citenamefont {Berges},\ and\ \citenamefont {Hauke}}]{Zache:2018jbt}%
  \BibitemOpen
  \bibfield  {author} {\bibinfo {author} {\bibfnamefont {T.~V.}\ \bibnamefont {Zache}}, \bibinfo {author} {\bibfnamefont {F.}~\bibnamefont {Hebenstreit}}, \bibinfo {author} {\bibfnamefont {F.}~\bibnamefont {Jendrzejewski}}, \bibinfo {author} {\bibfnamefont {M.~K.}\ \bibnamefont {Oberthaler}}, \bibinfo {author} {\bibfnamefont {J.}~\bibnamefont {Berges}}, \ and\ \bibinfo {author} {\bibfnamefont {P.}~\bibnamefont {Hauke}},\ }\href {\doibase 10.1088/2058-9565/aac33b} {\bibfield  {journal} {\bibinfo  {journal} {Quantum Sci. Technol.}\ }\textbf {\bibinfo {volume} {3}},\ \bibinfo {pages} {034010} (\bibinfo {year} {2018})},\ \Eprint {http://arxiv.org/abs/1802.06704} {arXiv:1802.06704 [cond-mat.quant-gas]} \BibitemShut {NoStop}%
\bibitem [{\citenamefont {Creutz}\ \emph {et~al.}(2002)\citenamefont {Creutz}, \citenamefont {Horvath},\ and\ \citenamefont {Neuberger}}]{Creutz:2001wp}%
  \BibitemOpen
  \bibfield  {author} {\bibinfo {author} {\bibfnamefont {M.}~\bibnamefont {Creutz}}, \bibinfo {author} {\bibfnamefont {I.}~\bibnamefont {Horvath}}, \ and\ \bibinfo {author} {\bibfnamefont {H.}~\bibnamefont {Neuberger}},\ }\href {\doibase 10.1016/S0920-5632(01)01836-9} {\bibfield  {journal} {\bibinfo  {journal} {Nucl. Phys. B Proc. Suppl.}\ }\textbf {\bibinfo {volume} {106}},\ \bibinfo {pages} {760} (\bibinfo {year} {2002})},\ \Eprint {http://arxiv.org/abs/hep-lat/0110009} {arXiv:hep-lat/0110009} \BibitemShut {NoStop}%
\bibitem [{\citenamefont {Hayata}\ \emph {et~al.}(2023)\citenamefont {Hayata}, \citenamefont {Nakayama},\ and\ \citenamefont {Yamamoto}}]{Hayata:2023zuk}%
  \BibitemOpen
  \bibfield  {author} {\bibinfo {author} {\bibfnamefont {T.}~\bibnamefont {Hayata}}, \bibinfo {author} {\bibfnamefont {K.}~\bibnamefont {Nakayama}}, \ and\ \bibinfo {author} {\bibfnamefont {A.}~\bibnamefont {Yamamoto}},\ }\href {\doibase 10.1103/PhysRevD.108.034511} {\bibfield  {journal} {\bibinfo  {journal} {Phys. Rev. D}\ }\textbf {\bibinfo {volume} {108}},\ \bibinfo {pages} {034511} (\bibinfo {year} {2023})},\ \Eprint {http://arxiv.org/abs/2305.18934} {arXiv:2305.18934 [hep-lat]} \BibitemShut {NoStop}%
\bibitem [{\citenamefont {Kavaki}\ and\ \citenamefont {Lewis}(2024)}]{Kavaki:2024ijd}%
  \BibitemOpen
  \bibfield  {author} {\bibinfo {author} {\bibfnamefont {A.~H.~Z.}\ \bibnamefont {Kavaki}}\ and\ \bibinfo {author} {\bibfnamefont {R.}~\bibnamefont {Lewis}},\ }\href {\doibase 10.1038/s42005-024-01697-4} {\bibfield  {journal} {\bibinfo  {journal} {Commun. Phys.}\ }\textbf {\bibinfo {volume} {7}},\ \bibinfo {pages} {208} (\bibinfo {year} {2024})},\ \Eprint {http://arxiv.org/abs/2401.14570} {arXiv:2401.14570 [hep-lat]} \BibitemShut {NoStop}%
\bibitem [{\citenamefont {Hidalgo}\ and\ \citenamefont {Draper}(2024)}]{PhysRevD.109.076004}%
  \BibitemOpen
  \bibfield  {author} {\bibinfo {author} {\bibfnamefont {L.}~\bibnamefont {Hidalgo}}\ and\ \bibinfo {author} {\bibfnamefont {P.}~\bibnamefont {Draper}},\ }\href {\doibase 10.1103/PhysRevD.109.076004} {\bibfield  {journal} {\bibinfo  {journal} {Phys. Rev. D}\ }\textbf {\bibinfo {volume} {109}},\ \bibinfo {pages} {076004} (\bibinfo {year} {2024})}\BibitemShut {NoStop}%
\bibitem [{\citenamefont {Barata}\ \emph {et~al.}(2022)\citenamefont {Barata}, \citenamefont {Du}, \citenamefont {Li}, \citenamefont {Qian},\ and\ \citenamefont {Salgado}}]{PhysRevD.106.074013}%
  \BibitemOpen
  \bibfield  {author} {\bibinfo {author} {\bibfnamefont {J.~a.}\ \bibnamefont {Barata}}, \bibinfo {author} {\bibfnamefont {X.}~\bibnamefont {Du}}, \bibinfo {author} {\bibfnamefont {M.}~\bibnamefont {Li}}, \bibinfo {author} {\bibfnamefont {W.}~\bibnamefont {Qian}}, \ and\ \bibinfo {author} {\bibfnamefont {C.~A.}\ \bibnamefont {Salgado}},\ }\href {\doibase 10.1103/PhysRevD.106.074013} {\bibfield  {journal} {\bibinfo  {journal} {Phys. Rev. D}\ }\textbf {\bibinfo {volume} {106}},\ \bibinfo {pages} {074013} (\bibinfo {year} {2022})}\BibitemShut {NoStop}%
\bibitem [{\citenamefont {Mathur}(2005)}]{Mathur:2004kr}%
  \BibitemOpen
  \bibfield  {author} {\bibinfo {author} {\bibfnamefont {M.}~\bibnamefont {Mathur}},\ }\href {\doibase 10.1088/0305-4470/38/46/008} {\bibfield  {journal} {\bibinfo  {journal} {J. Phys. A}\ }\textbf {\bibinfo {volume} {38}},\ \bibinfo {pages} {10015} (\bibinfo {year} {2005})},\ \Eprint {http://arxiv.org/abs/hep-lat/0403029} {arXiv:hep-lat/0403029} \BibitemShut {NoStop}%
\bibitem [{\citenamefont {Mathur}(2006)}]{mathur2006loop}%
  \BibitemOpen
  \bibfield  {author} {\bibinfo {author} {\bibfnamefont {M.}~\bibnamefont {Mathur}},\ }\href@noop {} {\bibfield  {journal} {\bibinfo  {journal} {Physics Letters B}\ }\textbf {\bibinfo {volume} {640}},\ \bibinfo {pages} {292} (\bibinfo {year} {2006})}\BibitemShut {NoStop}%
\bibitem [{\citenamefont {Mathur}(2007)}]{mathur2007loop}%
  \BibitemOpen
  \bibfield  {author} {\bibinfo {author} {\bibfnamefont {M.}~\bibnamefont {Mathur}},\ }\href@noop {} {\bibfield  {journal} {\bibinfo  {journal} {Nuclear Physics B}\ }\textbf {\bibinfo {volume} {779}},\ \bibinfo {pages} {32} (\bibinfo {year} {2007})}\BibitemShut {NoStop}%
\bibitem [{\citenamefont {Kadam}\ \emph {et~al.}(2023)\citenamefont {Kadam}, \citenamefont {Raychowdhury},\ and\ \citenamefont {Stryker}}]{Kadam:2022ipf}%
  \BibitemOpen
  \bibfield  {author} {\bibinfo {author} {\bibfnamefont {S.~V.}\ \bibnamefont {Kadam}}, \bibinfo {author} {\bibfnamefont {I.}~\bibnamefont {Raychowdhury}}, \ and\ \bibinfo {author} {\bibfnamefont {J.~R.}\ \bibnamefont {Stryker}},\ }\href {\doibase 10.1103/PhysRevD.107.094513} {\bibfield  {journal} {\bibinfo  {journal} {Phys. Rev. D}\ }\textbf {\bibinfo {volume} {107}},\ \bibinfo {pages} {094513} (\bibinfo {year} {2023})},\ \Eprint {http://arxiv.org/abs/2212.04490} {arXiv:2212.04490 [hep-lat]} \BibitemShut {NoStop}%
\bibitem [{\citenamefont {Kadam}\ \emph {et~al.}(2024)\citenamefont {Kadam}, \citenamefont {Naskar}, \citenamefont {Raychowdhury},\ and\ \citenamefont {Stryker}}]{Kadam:2024zkj}%
  \BibitemOpen
  \bibfield  {author} {\bibinfo {author} {\bibfnamefont {S.~V.}\ \bibnamefont {Kadam}}, \bibinfo {author} {\bibfnamefont {A.}~\bibnamefont {Naskar}}, \bibinfo {author} {\bibfnamefont {I.}~\bibnamefont {Raychowdhury}}, \ and\ \bibinfo {author} {\bibfnamefont {J.~R.}\ \bibnamefont {Stryker}},\ }\href@noop {} {\  (\bibinfo {year} {2024})},\ \Eprint {http://arxiv.org/abs/2407.19181} {arXiv:2407.19181 [hep-lat]} \BibitemShut {NoStop}%
\bibitem [{\citenamefont {Alam}\ \emph {et~al.}(2022)\citenamefont {Alam}, \citenamefont {Hadfield}, \citenamefont {Lamm},\ and\ \citenamefont {Li}}]{Alam:2021uuq}%
  \BibitemOpen
  \bibfield  {author} {\bibinfo {author} {\bibfnamefont {M.~S.}\ \bibnamefont {Alam}}, \bibinfo {author} {\bibfnamefont {S.}~\bibnamefont {Hadfield}}, \bibinfo {author} {\bibfnamefont {H.}~\bibnamefont {Lamm}}, \ and\ \bibinfo {author} {\bibfnamefont {A.~C.~Y.}\ \bibnamefont {Li}} (\bibinfo {collaboration} {SQMS}),\ }\href {\doibase 10.1103/PhysRevD.105.114501} {\bibfield  {journal} {\bibinfo  {journal} {Phys. Rev. D}\ }\textbf {\bibinfo {volume} {105}},\ \bibinfo {pages} {114501} (\bibinfo {year} {2022})},\ \Eprint {http://arxiv.org/abs/2108.13305} {arXiv:2108.13305 [quant-ph]} \BibitemShut {NoStop}%
\bibitem [{\citenamefont {Alexandru}\ \emph {et~al.}(2022)\citenamefont {Alexandru}, \citenamefont {Bedaque}, \citenamefont {Brett},\ and\ \citenamefont {Lamm}}]{Alexandru:2021jpm}%
  \BibitemOpen
  \bibfield  {author} {\bibinfo {author} {\bibfnamefont {A.}~\bibnamefont {Alexandru}}, \bibinfo {author} {\bibfnamefont {P.~F.}\ \bibnamefont {Bedaque}}, \bibinfo {author} {\bibfnamefont {R.}~\bibnamefont {Brett}}, \ and\ \bibinfo {author} {\bibfnamefont {H.}~\bibnamefont {Lamm}},\ }\href {\doibase 10.1103/PhysRevD.105.114508} {\bibfield  {journal} {\bibinfo  {journal} {Phys. Rev. D}\ }\textbf {\bibinfo {volume} {105}},\ \bibinfo {pages} {114508} (\bibinfo {year} {2022})},\ \Eprint {http://arxiv.org/abs/2112.08482} {arXiv:2112.08482 [hep-lat]} \BibitemShut {NoStop}%
\bibitem [{\citenamefont {Gustafson}\ \emph {et~al.}(2022)\citenamefont {Gustafson}, \citenamefont {Lamm}, \citenamefont {Lovelace},\ and\ \citenamefont {Musk}}]{Gustafson:2022xdt}%
  \BibitemOpen
  \bibfield  {author} {\bibinfo {author} {\bibfnamefont {E.~J.}\ \bibnamefont {Gustafson}}, \bibinfo {author} {\bibfnamefont {H.}~\bibnamefont {Lamm}}, \bibinfo {author} {\bibfnamefont {F.}~\bibnamefont {Lovelace}}, \ and\ \bibinfo {author} {\bibfnamefont {D.}~\bibnamefont {Musk}},\ }\href {\doibase 10.1103/PhysRevD.106.114501} {\bibfield  {journal} {\bibinfo  {journal} {Phys. Rev. D}\ }\textbf {\bibinfo {volume} {106}},\ \bibinfo {pages} {114501} (\bibinfo {year} {2022})},\ \Eprint {http://arxiv.org/abs/2208.12309} {arXiv:2208.12309 [quant-ph]} \BibitemShut {NoStop}%
\bibitem [{\citenamefont {Gustafson}\ \emph {et~al.}(2024{\natexlab{a}})\citenamefont {Gustafson}, \citenamefont {Lamm},\ and\ \citenamefont {Lovelace}}]{Gustafson:2023kvd}%
  \BibitemOpen
  \bibfield  {author} {\bibinfo {author} {\bibfnamefont {E.~J.}\ \bibnamefont {Gustafson}}, \bibinfo {author} {\bibfnamefont {H.}~\bibnamefont {Lamm}}, \ and\ \bibinfo {author} {\bibfnamefont {F.}~\bibnamefont {Lovelace}},\ }\href {\doibase 10.1103/PhysRevD.109.054503} {\bibfield  {journal} {\bibinfo  {journal} {Phys. Rev. D}\ }\textbf {\bibinfo {volume} {109}},\ \bibinfo {pages} {054503} (\bibinfo {year} {2024}{\natexlab{a}})},\ \Eprint {http://arxiv.org/abs/2312.10285} {arXiv:2312.10285 [hep-lat]} \BibitemShut {NoStop}%
\bibitem [{\citenamefont {Gustafson}\ \emph {et~al.}(2024{\natexlab{b}})\citenamefont {Gustafson}, \citenamefont {Ji}, \citenamefont {Lamm}, \citenamefont {Murairi}, \citenamefont {Perez},\ and\ \citenamefont {Zhu}}]{Gustafson:2024kym}%
  \BibitemOpen
  \bibfield  {author} {\bibinfo {author} {\bibfnamefont {E.~J.}\ \bibnamefont {Gustafson}}, \bibinfo {author} {\bibfnamefont {Y.}~\bibnamefont {Ji}}, \bibinfo {author} {\bibfnamefont {H.}~\bibnamefont {Lamm}}, \bibinfo {author} {\bibfnamefont {E.~M.}\ \bibnamefont {Murairi}}, \bibinfo {author} {\bibfnamefont {S.~O.}\ \bibnamefont {Perez}}, \ and\ \bibinfo {author} {\bibfnamefont {S.}~\bibnamefont {Zhu}},\ }\href {\doibase 10.1103/PhysRevD.110.034515} {\bibfield  {journal} {\bibinfo  {journal} {Phys. Rev. D}\ }\textbf {\bibinfo {volume} {110}},\ \bibinfo {pages} {034515} (\bibinfo {year} {2024}{\natexlab{b}})},\ \Eprint {http://arxiv.org/abs/2405.05973} {arXiv:2405.05973 [hep-lat]} \BibitemShut {NoStop}%
\bibitem [{\citenamefont {Assi}\ and\ \citenamefont {Lamm}(2024)}]{Assi:2024pdn}%
  \BibitemOpen
  \bibfield  {author} {\bibinfo {author} {\bibfnamefont {B.}~\bibnamefont {Assi}}\ and\ \bibinfo {author} {\bibfnamefont {H.}~\bibnamefont {Lamm}},\ }\href@noop {} {\  (\bibinfo {year} {2024})},\ \Eprint {http://arxiv.org/abs/2405.12204} {arXiv:2405.12204 [hep-lat]} \BibitemShut {NoStop}%
\bibitem [{\citenamefont {Lamm}\ \emph {et~al.}(2024)\citenamefont {Lamm}, \citenamefont {Li}, \citenamefont {Shu}, \citenamefont {Wang},\ and\ \citenamefont {Xu}}]{Lamm:2024jnl}%
  \BibitemOpen
  \bibfield  {author} {\bibinfo {author} {\bibfnamefont {H.}~\bibnamefont {Lamm}}, \bibinfo {author} {\bibfnamefont {Y.-Y.}\ \bibnamefont {Li}}, \bibinfo {author} {\bibfnamefont {J.}~\bibnamefont {Shu}}, \bibinfo {author} {\bibfnamefont {Y.-L.}\ \bibnamefont {Wang}}, \ and\ \bibinfo {author} {\bibfnamefont {B.}~\bibnamefont {Xu}},\ }\href {\doibase 10.1103/PhysRevD.110.054505} {\bibfield  {journal} {\bibinfo  {journal} {Phys. Rev. D}\ }\textbf {\bibinfo {volume} {110}},\ \bibinfo {pages} {054505} (\bibinfo {year} {2024})},\ \Eprint {http://arxiv.org/abs/2405.12890} {arXiv:2405.12890 [hep-lat]} \BibitemShut {NoStop}%
\bibitem [{\citenamefont {Muarari}\ \emph {et~al.}(2024)\citenamefont {Muarari}, \citenamefont {Sohaib~Alam}, \citenamefont {Lamm}, \citenamefont {Hadfield},\ and\ \citenamefont {Gustafson}}]{Muarari:2024dqx}%
  \BibitemOpen
  \bibfield  {author} {\bibinfo {author} {\bibfnamefont {E.}~\bibnamefont {Muarari}}, \bibinfo {author} {\bibfnamefont {M.}~\bibnamefont {Sohaib~Alam}}, \bibinfo {author} {\bibfnamefont {H.}~\bibnamefont {Lamm}}, \bibinfo {author} {\bibfnamefont {S.}~\bibnamefont {Hadfield}}, \ and\ \bibinfo {author} {\bibfnamefont {E.}~\bibnamefont {Gustafson}},\ }\href@noop {} {\  (\bibinfo {year} {2024})},\ \Eprint {http://arxiv.org/abs/2408.00075} {arXiv:2408.00075 [quant-ph]} \BibitemShut {NoStop}%
\bibitem [{\citenamefont {Bender}\ and\ \citenamefont {Zohar}(2020)}]{Bender:2020ztu}%
  \BibitemOpen
  \bibfield  {author} {\bibinfo {author} {\bibfnamefont {J.}~\bibnamefont {Bender}}\ and\ \bibinfo {author} {\bibfnamefont {E.}~\bibnamefont {Zohar}},\ }\href {\doibase 10.1103/PhysRevD.102.114517} {\bibfield  {journal} {\bibinfo  {journal} {Phys. Rev. D}\ }\textbf {\bibinfo {volume} {102}},\ \bibinfo {pages} {114517} (\bibinfo {year} {2020})},\ \Eprint {http://arxiv.org/abs/2008.01349} {arXiv:2008.01349 [quant-ph]} \BibitemShut {NoStop}%
\bibitem [{\citenamefont {Romiti}\ and\ \citenamefont {Urbach}(2024)}]{Romiti:2023hbd}%
  \BibitemOpen
  \bibfield  {author} {\bibinfo {author} {\bibfnamefont {S.}~\bibnamefont {Romiti}}\ and\ \bibinfo {author} {\bibfnamefont {C.}~\bibnamefont {Urbach}},\ }\href {\doibase 10.1140/epjc/s10052-024-13037-5} {\bibfield  {journal} {\bibinfo  {journal} {Eur. Phys. J. C}\ }\textbf {\bibinfo {volume} {84}},\ \bibinfo {pages} {708} (\bibinfo {year} {2024})},\ \Eprint {http://arxiv.org/abs/2311.11928} {arXiv:2311.11928 [hep-lat]} \BibitemShut {NoStop}%
\bibitem [{\citenamefont {Fontana}\ \emph {et~al.}(2024)\citenamefont {Fontana}, \citenamefont {Riaza},\ and\ \citenamefont {Celi}}]{Fontana:2024rux}%
  \BibitemOpen
  \bibfield  {author} {\bibinfo {author} {\bibfnamefont {P.}~\bibnamefont {Fontana}}, \bibinfo {author} {\bibfnamefont {M.~M.}\ \bibnamefont {Riaza}}, \ and\ \bibinfo {author} {\bibfnamefont {A.}~\bibnamefont {Celi}},\ }\href@noop {} {\  (\bibinfo {year} {2024})},\ \Eprint {http://arxiv.org/abs/2409.04441} {arXiv:2409.04441 [quant-ph]} \BibitemShut {NoStop}%
\bibitem [{\citenamefont {Garofalo}\ \emph {et~al.}(2024)\citenamefont {Garofalo}, \citenamefont {Hartung}, \citenamefont {Jakobs}, \citenamefont {Jansen}, \citenamefont {Ostmeyer}, \citenamefont {Rolfes}, \citenamefont {Romiti},\ and\ \citenamefont {Urbach}}]{Garofalo:2023zkd}%
  \BibitemOpen
  \bibfield  {author} {\bibinfo {author} {\bibfnamefont {M.}~\bibnamefont {Garofalo}}, \bibinfo {author} {\bibfnamefont {T.}~\bibnamefont {Hartung}}, \bibinfo {author} {\bibfnamefont {T.}~\bibnamefont {Jakobs}}, \bibinfo {author} {\bibfnamefont {K.}~\bibnamefont {Jansen}}, \bibinfo {author} {\bibfnamefont {J.}~\bibnamefont {Ostmeyer}}, \bibinfo {author} {\bibfnamefont {D.}~\bibnamefont {Rolfes}}, \bibinfo {author} {\bibfnamefont {S.}~\bibnamefont {Romiti}}, \ and\ \bibinfo {author} {\bibfnamefont {C.}~\bibnamefont {Urbach}},\ }\href {\doibase 10.22323/1.453.0231} {\bibfield  {journal} {\bibinfo  {journal} {PoS}\ }\textbf {\bibinfo {volume} {LATTICE2023}},\ \bibinfo {pages} {231} (\bibinfo {year} {2024})},\ \Eprint {http://arxiv.org/abs/2311.15926} {arXiv:2311.15926 [hep-lat]} \BibitemShut {NoStop}%
\bibitem [{\citenamefont {Dong}\ \emph {et~al.}(2022)\citenamefont {Dong}, \citenamefont {Lin},\ and\ \citenamefont {Tong}}]{Dong:2022mmq}%
  \BibitemOpen
  \bibfield  {author} {\bibinfo {author} {\bibfnamefont {Y.}~\bibnamefont {Dong}}, \bibinfo {author} {\bibfnamefont {L.}~\bibnamefont {Lin}}, \ and\ \bibinfo {author} {\bibfnamefont {Y.}~\bibnamefont {Tong}},\ }\href {\doibase 10.1103/PRXQuantum.3.040305} {\bibfield  {journal} {\bibinfo  {journal} {PRX Quantum}\ }\textbf {\bibinfo {volume} {3}},\ \bibinfo {pages} {040305} (\bibinfo {year} {2022})},\ \Eprint {http://arxiv.org/abs/2204.05955} {arXiv:2204.05955 [quant-ph]} \BibitemShut {NoStop}%
\bibitem [{\citenamefont {Carena}\ \emph {et~al.}(2024)\citenamefont {Carena}, \citenamefont {Lamm}, \citenamefont {Li},\ and\ \citenamefont {Liu}}]{Carena:2024dzu}%
  \BibitemOpen
  \bibfield  {author} {\bibinfo {author} {\bibfnamefont {M.}~\bibnamefont {Carena}}, \bibinfo {author} {\bibfnamefont {H.}~\bibnamefont {Lamm}}, \bibinfo {author} {\bibfnamefont {Y.-Y.}\ \bibnamefont {Li}}, \ and\ \bibinfo {author} {\bibfnamefont {W.}~\bibnamefont {Liu}},\ }\href@noop {} {\  (\bibinfo {year} {2024})},\ \Eprint {http://arxiv.org/abs/2402.16780} {arXiv:2402.16780 [hep-lat]} \BibitemShut {NoStop}%
\bibitem [{\citenamefont {Creutz}(1977)}]{PhysRevD.15.1128}%
  \BibitemOpen
  \bibfield  {author} {\bibinfo {author} {\bibfnamefont {M.}~\bibnamefont {Creutz}},\ }\href {\doibase 10.1103/PhysRevD.15.1128} {\bibfield  {journal} {\bibinfo  {journal} {Phys. Rev. D}\ }\textbf {\bibinfo {volume} {15}},\ \bibinfo {pages} {1128} (\bibinfo {year} {1977})}\BibitemShut {NoStop}%
\bibitem [{\citenamefont {Littlejohn}\ and\ \citenamefont {Reinsch}(1997)}]{Littlejohn:1997qb}%
  \BibitemOpen
  \bibfield  {author} {\bibinfo {author} {\bibfnamefont {R.~G.}\ \bibnamefont {Littlejohn}}\ and\ \bibinfo {author} {\bibfnamefont {M.}~\bibnamefont {Reinsch}},\ }\href {\doibase 10.1103/RevModPhys.69.213} {\bibfield  {journal} {\bibinfo  {journal} {Rev. Mod. Phys.}\ }\textbf {\bibinfo {volume} {69}},\ \bibinfo {pages} {213} (\bibinfo {year} {1997})}\BibitemShut {NoStop}%
\bibitem [{\citenamefont {Childs}\ \emph {et~al.}(2021)\citenamefont {Childs}, \citenamefont {Su}, \citenamefont {Tran}, \citenamefont {Wiebe},\ and\ \citenamefont {Zhu}}]{Childs:2019hts}%
  \BibitemOpen
  \bibfield  {author} {\bibinfo {author} {\bibfnamefont {A.~M.}\ \bibnamefont {Childs}}, \bibinfo {author} {\bibfnamefont {Y.}~\bibnamefont {Su}}, \bibinfo {author} {\bibfnamefont {M.~C.}\ \bibnamefont {Tran}}, \bibinfo {author} {\bibfnamefont {N.}~\bibnamefont {Wiebe}}, \ and\ \bibinfo {author} {\bibfnamefont {S.}~\bibnamefont {Zhu}},\ }\href {\doibase 10.1103/PhysRevX.11.011020} {\bibfield  {journal} {\bibinfo  {journal} {Phys. Rev. X}\ }\textbf {\bibinfo {volume} {11}},\ \bibinfo {pages} {011020} (\bibinfo {year} {2021})},\ \Eprint {http://arxiv.org/abs/1912.08854} {arXiv:1912.08854 [quant-ph]} \BibitemShut {NoStop}%
\bibitem [{\citenamefont {Low}\ and\ \citenamefont {Chuang}(2017)}]{Low:2016sck}%
  \BibitemOpen
  \bibfield  {author} {\bibinfo {author} {\bibfnamefont {G.~H.}\ \bibnamefont {Low}}\ and\ \bibinfo {author} {\bibfnamefont {I.~L.}\ \bibnamefont {Chuang}},\ }\href {\doibase 10.1103/PhysRevLett.118.010501} {\bibfield  {journal} {\bibinfo  {journal} {Phys. Rev. Lett.}\ }\textbf {\bibinfo {volume} {118}},\ \bibinfo {pages} {010501} (\bibinfo {year} {2017})},\ \Eprint {http://arxiv.org/abs/1606.02685} {arXiv:1606.02685 [quant-ph]} \BibitemShut {NoStop}%
\bibitem [{\citenamefont {Low}\ and\ \citenamefont {Chuang}(2019)}]{Low:2016znh}%
  \BibitemOpen
  \bibfield  {author} {\bibinfo {author} {\bibfnamefont {G.~H.}\ \bibnamefont {Low}}\ and\ \bibinfo {author} {\bibfnamefont {I.~L.}\ \bibnamefont {Chuang}},\ }\href {\doibase 10.22331/q-2019-07-12-163} {\bibfield  {journal} {\bibinfo  {journal} {Quantum}\ }\textbf {\bibinfo {volume} {3}},\ \bibinfo {pages} {163} (\bibinfo {year} {2019})},\ \Eprint {http://arxiv.org/abs/1610.06546} {arXiv:1610.06546 [quant-ph]} \BibitemShut {NoStop}%
\bibitem [{\citenamefont {Motlagh}\ and\ \citenamefont {Wiebe}(2024)}]{Motlagh:2023oqc}%
  \BibitemOpen
  \bibfield  {author} {\bibinfo {author} {\bibfnamefont {D.}~\bibnamefont {Motlagh}}\ and\ \bibinfo {author} {\bibfnamefont {N.}~\bibnamefont {Wiebe}},\ }\href {\doibase 10.1103/PRXQuantum.5.020368} {\bibfield  {journal} {\bibinfo  {journal} {PRX Quantum}\ }\textbf {\bibinfo {volume} {5}},\ \bibinfo {pages} {020368} (\bibinfo {year} {2024})},\ \Eprint {http://arxiv.org/abs/2308.01501} {arXiv:2308.01501 [quant-ph]} \BibitemShut {NoStop}%
\bibitem [{\citenamefont {Kikuchi}\ \emph {et~al.}(2023)\citenamefont {Kikuchi}, \citenamefont {Mc~Keever}, \citenamefont {Coopmans}, \citenamefont {Lubasch},\ and\ \citenamefont {Benedetti}}]{Kikuchi:2023qbb}%
  \BibitemOpen
  \bibfield  {author} {\bibinfo {author} {\bibfnamefont {Y.}~\bibnamefont {Kikuchi}}, \bibinfo {author} {\bibfnamefont {C.}~\bibnamefont {Mc~Keever}}, \bibinfo {author} {\bibfnamefont {L.}~\bibnamefont {Coopmans}}, \bibinfo {author} {\bibfnamefont {M.}~\bibnamefont {Lubasch}}, \ and\ \bibinfo {author} {\bibfnamefont {M.}~\bibnamefont {Benedetti}},\ }\href {\doibase 10.1038/s41534-023-00762-0} {\bibfield  {journal} {\bibinfo  {journal} {npj Quantum Inf.}\ }\textbf {\bibinfo {volume} {9}},\ \bibinfo {pages} {93} (\bibinfo {year} {2023})},\ \Eprint {http://arxiv.org/abs/2303.05533} {arXiv:2303.05533 [quant-ph]} \BibitemShut {NoStop}%
\end{thebibliography}%
\appendix

\section{Conventions and Normalizations}
\label{app:normalizations}
\subsection{Wigner D Functions}

The Wigner D matrix, whose components are called the Wigner D functions, used in this paper are defined as the matrix elements of the rotation operator $\hat D(\alpha, \beta, \gamma)$ in the $JM$-representation, albeit with a slightly unusual normalization
\begin{align}
    \bra{\LTot\MTot}\hat D(\alpha, \beta, \gamma) \ket{\LTot\NTot} = \sqrt{\frac{8\pi^2}{2L+1}} \WD
    \,.
\end{align}
Here $\alpha$, $\beta$ and $\gamma$ are the Euler angles specifying the rotation operator.
Using the fact that all representations of \sutwo are self dual, there is an isomorphism between a representation and its dual, which can be written as 
\begin{align}
    \bra{j, m} \to (-1)^{j-m} \ket{j, -m}
    \,.
\end{align}
This implies that one can also define the ${\cal D}$ through 
\begin{align}
    \sqrt{\frac{8\pi^2}{2L+1}} \WDSymbol^L_{MN}(\alpha, \beta, \gamma) = (-1)^{\LTot+\MTot}\bra{\alpha, \beta, \gamma} \ket{\LTot (-\MTot) \NTot}
    \,.
\end{align}
Note that the atypical factors of $\sqrt{8\pi^2/(2\LTot+1})$ in all of these relations are due to the fact that the functions $\WDSymbol^L_{MN}(\alpha, \beta, \gamma)$ is related to the `typical' Wigner D functions, $D$, via
\begin{align}
    \WDSymbol^L_{MN}(\alpha, \beta, \gamma) = \sqrt{\frac{8\pi^2}{2L+1}} D^L_{MN}(\alpha, \beta, \gamma)
    \,.
\end{align}
The unusual normalization in our definition is chosen such that the $\WDSymbol$ functions satisfy the orthogonality relation
\begin{align}
\int \left[ d\alpha d(\cos \beta) d\gamma \right]\WD \WDpDag= \delta_{\LTot, \LTot'} \delta_{\MTot, \MTot'} \;\delta_{\NTot, \NTot'} \, .
\end{align}
An important feature is that they form an orthonormal and complete basis of states, implying that
\begin{align}
\sum_{\LTot,\MTot, \NTot} \WD \WDDagP = \delta(\alpha - \alpha')\delta(\cos \beta - \cos \beta') \delta(\gamma - \gamma ')
\,.
\end{align}
The action of the differential total momentum operators on the Wigner $D$ functions give
\begin{align}
\LTotOpSq\WD &= \LTot\left(\LTot+1\right)\WD \nonumber \\
\LTotOpComp{z} \WD &= - \MTot \WD \nonumber \\
\LpTotOpComp{z} \WD &= -\NTot \WD \, .
\end{align}

\subsection{Spherical Harmonics and Associated Legendre Polynomials}
The associate Legendre polynomials $\ALPSymbol_{\ell}^m(\cos \theta)$ satisfies the differential equation
\begin{align}
\label{eq:Plmdef}
\left(-\pdv[2]{\theta} - \cot\thetaDef \pdv{\theta} + m^2 \csc^2\theta\right) \ALPSymbol_{\ell}^m(\cos \theta) = \ell(\ell+1)\ALPSymbol_{\ell}^m(\cos \theta) \, .
\end{align}
It is important to note that associate Legendre polynomials are not all mutually orthogonal, but for a given $m$, they are orthogonal to one another. 
We choose to normalize our functions such that they form an orthonormal (not just orthogonal) and complete set of functions, 
\begin{align}
    \int d(\cos \theta) \ALPSymbol_{\ell}^m(\cos \theta) \ALPSymbol_{\ell'}^m(\cos \theta) &= \delta_{\ell, \ell'} \qquad \text{for}\,\, \ell \geq |m| \nonumber 
    \\
    \sum_{\ell} \ALPSymbol_{\ell}^m(\cos \theta) \ALPSymbol_{\ell}^m(\cos \theta') &= \delta(\cos \theta - \cos \theta') \,.
\end{align}
This means that they are related to the conventional associated Legendre polynomials $P^m_{\ell} (\cos \theta)$ via
\begin{align}
    \ALPSymbol_{\ell}^m (\cos \theta) = \sqrt{\frac{2\ell+1}{2}\frac{(\ell-m)!}{(\ell+m)!}}P^m_{\ell} (\cos \theta)
    \,.
\end{align}
Associated Legendre polynomials can be written as derivatives of regular Legendre Polynomials
\begin{align}
    P_{\ell}^m(\cos \theta) = (-1)^m (\sin\theta)^m \frac{\rm d^m}{\rm d (\cos\theta)^m} P_\ell(\cos\theta)
    \,.
\end{align}
Since Legendre Polynomials are polynomials of rank $\ell$, one immediately finds
\begin{align}
\label{eq:ASPZero}
    \ALPSymbol_{\ell}^m(\cos\theta) = 0 \qquad {\rm for} \quad \abs{m} > \ell
\end{align}

Spherical Harmonics are eigenfunctions of the angular momentum operators. In particular
\begin{align}
    -\left(\pdv[2]{}{\theta}+\cot \theta \pdv{}{\theta}+\csc^2 \theta \pdv[2]{}{\phi}\right) Y_{\ell m}(\theta, \phi) &= \ell(\ell+1) \, Y_{\ell m}(\theta, \phi)\nonumber
    \\
    -i \pdv{}{\phi}Y_{\ell m}(\theta, \phi) &= m \, Y_{\ell m}(\theta, \phi)
    \,.
\end{align}
They also form an orthonormal and complete set of functions
\begin{align}
    \int \! \dd(\cos \theta) \dd \phi Y_{\ell m}(\theta, \phi)Y^{*}_{\ell' m'}(\theta, \phi) &= \delta_{\ell\ell'} \delta_{mm'} \nonumber \\
    \sum_{\ell m} Y_{\ell m}(\theta, \phi)Y_{\ell m}(\theta', \phi') &= \delta(\cos \theta - \cos \theta') \delta(\phi - \phi') \,.
\end{align}

The two functions are related to one-another via
\begin{align}
    Y_{\ell m}(\theta, \phi) = (-1)^m \sqrt{\frac{1}{2\pi}}\ALPSymbol_\ell^m(\theta) e^{i m \phi}
    \,.
\end{align}
\subsection{ Mixed-Basis Eigenstates}
\label{sec:MBEig}
A key property of any set of states that is chosen to span a Hilbert space is that the set is complete and orthonormal. While Spherical Harmonics, Legendre polynomials and Wigner D functions are all orthonormal and complete, associate Legendre polynomials are not orthonormal and in fact are over-complete. Therefore, one might worry that the sequestered mixed basis is not a proper basis for spanning the Hilbert space. However it is simple enough to prove that different states are orthonormal: 
\begin{align}
\braket{\{\omega'_\kappa\};\Omega'}{\{\omega_\kappa\};\Omega}=&\,\left(\prod_{\kappa=1}^{\NL}\delta(\omega_\kappa-\omega'_\kappa)\right)\int d\alpha d\gamma d[\cos \beta] d[\cos \thetaDef]d[\cos \thetaSeq{\mu}] \nonumber
\\
&\times \left[\ALPSymbol_{n'}^{\sigma'}(\cos \thetaDef)\CD^{\LTot'}_{\MTot' \NTot'}(\alpha,\beta,\gamma) \left(\prod_{\mu=3}^{\NL} Y_{\ell'_\mu, m'_\mu}(\thetaSeq{\mu}, \phiSeq{\mu})\right) \right] \nonumber
\\
&\times \left[\ALPSymbol_{n}^{\sigma}(\cos \thetaDef)\CD^{\LTot}_{\MTot \NTot}(\alpha,\beta,\gamma) \left(\prod_{\mu=3}^{\NL} Y_{\ell_\mu, m_\mu}(\thetaSeq{\mu}, \phiSeq{\mu})\right) \right]  \nonumber
\\
=&\, \left(\prod_{\kappa=1}^{\NL}\delta(\omega_\kappa-\omega'_\kappa)\right) \delta_{\LTot,\LTot'}\delta_{\MTot, \MTot'}\delta_{\NTot, \NTot'} \left(\prod_{\mu=3}^{\NL} \delta_{\LSeq{\mu} \ell'_\mu }\delta_{\MSeq{\mu} m'_\mu }\right)\int d[\cos \thetaDef] \ALPSymbol_{n'}^{\sigma'}(\cos \thetaDef)\ALPSymbol_{n}^{\sigma}(\cos \thetaDef) \nonumber
\\
=&\, \left(\prod_{\kappa=1}^{\NL}\delta(\omega_\kappa-\omega'_\kappa)\right)\delta_{\LTot,\LTot'}\delta_{\MTot, \MTot'}\delta_{\NTot, \NTot'} \left(\prod_{\mu=3}^{\NL} \delta_{\LSeq{\mu} \ell'_\mu }\delta_{\MSeq{\mu} m'_\mu }\right)\delta_{nn'}
\end{align}
where in the last line we have used $\sigma = \sigma'$ due to $\NTot = \NTot'$ and $\MSeq{\mu} = \MSeq{\mu}'$. Completeness can be proved in a similar manner, using the completenes relations of spherical harmonics, associate Legendre polynomials and Wigner-D functions.

\section{Electric Bilinear Results}
\label{sec:EEResults}
In this appendix, we give the expressions for all the bilinears that can occur in the Hamiltonian. 
To obtain the expressions, one starts from 
\begin{align}
\label{eq:ELRdefApp}
\hat\CE^a_{\kappa R} = \frac{1}{2}\left(\hat\Sigma^a_\kappa + \LOpComp{a}_\kappa\right)
\,,\qquad
\hat\CE^a_{\kappa L} = \frac{1}{2}\left(\hat\Sigma^a_\kappa - \LOpComp{a}_\kappa\right)
\,.
\end{align}
and the expressions for $\LOp_\kappa$ and $\boldsymbol{\hat \Sigma_\kappa}$ given in \cref{eq:Ldefsec3,eq:Sigmadef}, and are restated here for convenience.
\begin{align}
\boldsymbol{\hat \Sigma_\kappa} &= \RotE \cdot \left(2 i \nSeqBody{\kappa} \pdv{}{\omega_\kappa}+ \cot \left(\frac{\omega_\kappa}{2}\right)\left(\nSeqBody{\kappa}\cross \LSeqOp_\kappa\right)\right) \nonumber \\
\LOp_\kappa &= \RotE \cdot \LSeqOp_\kappa
\end{align}
The expressions for the rotation matrix $\RotE$ and the directions $\nSeqBody{\kappa}$ are given in \cref{eq:nkappa_seq,eq:RotE}, while those of the angular momentum operator in the body frame $\LSeqOp_{\kappa}$ are given in \cref{eq:LSeqOpdef,eq:LSeq12def}.

Using these expressions, the bilinears that only depend on one physical link can be written as
\begin{align}
\boldsymbol{\CE}_{\mu \zeta}\cdot\boldsymbol{\CE}_{\mu \zeta'} =& - \pdv[2]{}{\omega_\mu} - \cot\left(\frac{\omega_\mu}{2}\right)\pdv{}{\omega_\mu}+ \Gamma^{\mu}_{\zeta \zeta'}\LSeqOpSq_\mu \nonumber 
\\
\boldsymbol{\CE}_{2 \zeta}\cdot\boldsymbol{\CE}_{2 \zeta'} =& - \pdv[2]{}{\omega_2} - \cot\left(\frac{\omega_2}{2}\right)\pdv{}{\omega_2}+ \Gamma^{2}_{\zeta \zeta'}\NSeqSq \nonumber  
\\
\boldsymbol{\CE}_{1 \zeta}\cdot \boldsymbol{\CE}_{1 \zeta'} =& - \pdv[2]{}{\omega_1} - \cot\left(\frac{\omega_1}{2}\right)\pdv{}{\omega_1}+ \Gamma^{1}_{\zeta \zeta'} \left(\LSigmaOpSq+\NSeqSq -2 \LSigmaOp \cdot \LSeqOp_{2}\right),
\end{align}
where $\zeta_\kappa \in \{L, R\}$,  the coefficient $\Gamma$ is given by
\begin{align}
\Gamma^{\kappa}_{\zeta \zeta'}&=\frac{1}{4}\csc^2\left( \frac{\omega_\kappa}{2}\right)  \begin{cases}
1 \quad & \zeta = L \,\,\text{and}\,\, \zeta' = L \\ \\
1 \quad & \zeta = R \,\,\text{and}\,\, \zeta' = R\\ \\
\cos \omega_\kappa \quad & \zeta = R \,\,\text{and}\,\, \zeta' = L \\ \\
\cos \omega_\kappa \quad & \zeta = L \,\,\text{and}\,\, \zeta' = R
\end{cases}
\,,
\end{align}
and $\NSeqSq$ is again given by
\begin{align}
\NSeqSq&=-\pdv[2]{\thetaDef} - \cot\thetaDef \pdv{\thetaDef} + \csc^2\thetaDef \LSigmaOpComp{z}\LSigmaOpComp{z}.
\end{align}
Moving on to the more complicated bilinears, the easiest combinations are those that do not involve Rod 1 and Rod 2. 
These can be written as
\begin{align}
\boldsymbol{\CE}_{\mu \zeta_\mu}\cdot \boldsymbol{\CE}_{\nu \zeta_\nu} =&  -\nSeqBody{\mu}\cdot \nSeqBody{\nu}\pdv{}{\omega_\mu}{\omega_\nu}+ \frac{i}{2}\left[\cot\left(\frac{\omega_\nu}{2}\right)\nSeqBody{\mu}\cdot \left(\nSeqBody{\nu}\times \LSeqOp_{\nu}\right)-\Delta_{\zeta_\nu}\nSeqBody{\mu}\cdot \LSeqOp_{\nu}\right]\pdv{}{\omega_\mu}\nonumber \\
&+\frac{i}{2}\left[\cot\left(\frac{\omega_\mu}{2}\right)\nSeqBody{\nu}\cdot \left(\nSeqBody{\mu}\times \LSeqOp_{\mu}\right)-\Delta_{\zeta_\mu}\nSeqBody{\nu}\cdot \LSeqOp_{\mu}\right]\pdv{}{\omega_\nu}+ \frac{1}{4}\cot\left(\frac{\omega_\nu}{2}\right)\cot\left(\frac{\omega_\mu}{2}\right)\left(\nSeqBody{\mu}\times \LSeqOp_{\mu}\right)\cdot \left(\nSeqBody{\nu}\times \LSeqOp_{\nu}\right)\nonumber \\
&-\frac{1}{4} \Delta_{\zeta_\nu}\cot\left(\frac{\omega_\mu}{2}\right)\LSeqOp_{\nu}\cdot \left(\nSeqBody{\mu}\times \LSeqOp_{\mu}\right)- \frac{1}{4}\Delta_{\zeta_\mu}\cot\left(\frac{\omega_\nu}{2}\right)\LSeqOp_{\mu}\cdot \left(\nSeqBody{\nu}\times \LSeqOp_{\nu}\right)+ \frac{1}{4}\Delta_{\zeta_\mu}\Delta_{\zeta_\nu}\LSeqOp_{\mu}\cdot \LSeqOp_{\nu}
\,,
\end{align}
where $\mu \neq \nu$ and the coefficients $\Delta_{\zeta_\kappa}$ are defined to be
\begin{align}
\Delta_{\zeta_\kappa} &= \begin{cases}
+1 & \quad \zeta_\kappa = L \\
-1 & \quad \zeta_\kappa = R 
\end{cases}
\,.
\end{align}
The most complicated bilinears are those that involve rods 1 and 2. 
For the bilinear that involves rod 2 and rod $\mu$ we find
\begin{align}
\boldsymbol{\CE}_{2\zeta_2}\cdot \boldsymbol{\CE}_{\mu \zeta_\mu}&= - \nSeqBody{2}\cdot \nSeqBody{\mu}\pdv{}{\omega_2}{\omega_\mu}+\frac{i}{2}\left[\cotw{2}\,\nSeqBody{\mu} \cdot \crossvL{2}{2}- \Delta_{\zeta_2}\nSeqBody{\mu}\cdot \LSeqOp_{2}\right]\pdv{}{\omega_\mu}\nonumber \\
&+\frac{i}{2}\left[\cotw{\mu}\,\nSeqBody{2}\cdot \crossvL{\mu}{\mu}-\Delta_{\zeta_\mu}\, \nSeqBody{2}\cdot \LSeqOp_{\mu}\right]\pdv{}{\omega_2}\nonumber \\
&+\frac{1}{4}\cotw{2}\cotw{\mu}\crossvL{\mu}{\mu} \cdot \crossvL{2}{2} -\frac{1}{4}\Delta_{\zeta_2}\cotw{\mu}\crossvL{\mu}{\mu}\cdot \LSeqOp_{2} \nonumber\\
&-\frac{1}{4}\Delta_{\zeta_\mu}\cotw{2}\LSeqOp_{\mu}\cdot \crossvL{2}{2} +\frac{1}{4}\Delta_{\zeta_2}\Delta_{\zeta_\mu}\,\LSeqOp_{\mu}\cdot \LSeqOp_{2}
\end{align}
where repeated $\mu$ indices are NOT summed over. Notice that $\boldsymbol{\CE}_{2\zeta_2}\cdot \boldsymbol{\CE}_{\nu \zeta_\nu}$ and $\boldsymbol{\CE}_{\mu\zeta_\mu}\cdot \boldsymbol{\CE}_{\nu \zeta_\nu}$ have a very similar form, with $\mu \rightarrow 2$, along with a few sign changes.
The bilinears that involve Rod 1 and Rod $\mu$ are given by
\begin{align}
\boldsymbol{\CE}_{1\zeta_1}\cdot \boldsymbol{\CE}_{\mu \zeta_\mu}&= - \nSeqBody{1}\cdot \nSeqBody{\mu}\pdv{}{\omega_1}{\omega_\mu}+\frac{i}{2}\left[\cotw{1}\,\nSeqBody{\mu} \cdot \crossvL{1}{1}- \Delta_{\zeta_1}\nSeqBody{\mu}\cdot \LSeqOp_{1}\right]\pdv{}{\omega_\mu}\nonumber \\
&+\frac{i}{2}\left[\cotw{\mu}\,\nSeqBody{1}\cdot \crossvL{\mu}{\mu}-\Delta_{\zeta_\mu}\, \nSeqBody{1}\cdot \LSeqOp_{\mu}\right]\pdv{}{\omega_1}\nonumber \\
&+\frac{1}{4}\cotw{1}\cotw{\mu}\crossvL{\mu}{\mu}\cdot \crossvL{1}{1} -\frac{1}{4}\Delta_{\zeta_1}\cotw{\mu}\crossvL{\mu}{\mu}\cdot \LSeqOp_{1} \nonumber\\
&-\frac{1}{4}\Delta_{\zeta_\mu}\cotw{1}\LSeqOp_{\mu}\times \crossvL{1}{1} +\frac{1}{4}\Delta_{\zeta_1}\Delta_{\zeta_\mu}\,\LSeqOp_{\mu}\cdot \LSeqOp_{1}
\end{align}
where $\boldsymbol{\CE}_{1\zeta_1}\cdot \boldsymbol{\CE}_{\mu \zeta_\mu}$ and $\boldsymbol{\CE}_{2\zeta_2}\cdot \boldsymbol{\CE}_{\mu \zeta_\mu}$ are identical, when written in this notation, with the substitution $(1 \leftrightarrow 2)$. However, since $\nSeqBody{1}, \nSeqBody{2}, \LSeqOp_1$ and $\LSeqOp_2$ have dramatically different effects on the eigenstates in the sequestered mixed-basis (see \cref{app:RecursionRel}),
we choose to present these bilinears separately. 

Lastly, for the bilinear that involves only Rod 1 and Rod 2 we obtain
\begin{align}
\boldsymbol{\CE}_{1 {\zeta_1}}\boldsymbol{\CE}_{2 {\zeta_2}} &=- \nSeqBody{1}\cdot \nSeqBody{2}\pdv{}{\omega_1}\pdv{}{\omega_2}+ \frac{i}{2}\left[\cotw{1}\nSeqBody{2}\cdot \crossvL{1}{1}-\Delta_{\zeta_1}\nSeqBody{2}\cdot \LSeqOp_{1}\right]\pdv{}{\omega_2}\nonumber \\
&+ \frac{i}{2}\left[\cotw{2}\nSeqBody{1}\cdot \crossvL{2}{2}-\Delta_{\zeta_2}\nSeqBody{1}\cdot \LSeqOp_{2}\right]\pdv{}{\omega_1}\nonumber \\
&+ \frac{1}{4}\cotw{1}\cotw{2}\left[\crossvL{2}{2} \cdot \crossvL{1}{1}+ i \csc^2 \thetaDef \, \nSeqBody{2}\times \crossvL{1}{1}\right]\nonumber \\
&-\frac{1}{4}\Delta_{\zeta_1}\cotw{2}\left[\crossvL{2}{2}\cdot \LSeqOp_1+ i \csc^2\thetaDef\, \nSeqBody{2}\cdot \LSeqOp_1\right]\nonumber \\
&- \frac{1}{4}\Delta_{\zeta_2}\cotw{1}\left[\LSeqOp_{2}\cdot \crossvL{1}{1}- i \cot\thetaDef\,\csc\thetaDef\, \nSeqBody{2}\cdot \LSeqOp_1\right]\nonumber \\
&+\frac{1}{4}\Delta_{\zeta_1}\Delta_{\zeta_2}\left[\LSeqOp_2 \cdot \LSeqOp_1+i \cot \thetaDef\, \csc \thetaDef\, \nSeqBody{2}\cdot \crossvL{1}{1}\right].
\end{align}

\section{Recursion Relations}
\label{app:RecursionRel}
In this appendix, we tabulate how every operator which appears in either an electric bilinaer or a loop variable acts on the mixed-basis eigenstates, as defined in \cref{sec:mixed}. Since the spherical harmonics, Wigner D functions and associate Legendre polynomials are all related, their recursion relations share many common coefficients. We will therefore introduce shorthand notation for these coefficients. First are the coefficients of the canonical raising and lowering operators, as well as the projected angular momentum,
\begin{align}
\OpCoeff{0}{+}{\ell}{m}&= \sqrt{\ell(\ell+1)-m(m+1)} \nonumber \\
\OpCoeff{0}{-}{\ell}{m}&= \sqrt{\ell(\ell+1)-m(m-1)}\nonumber \\
\OpCoeff{0}{0}{\ell}{m}&= m \, ,
\end{align}
where the canonical raising and lowering operators raise and lower only the projected angular momentum quantum number; we include $\OpCoeff{0}{0}{\ell}{m}$ as it will make certain expressions much simpler to write. Next are the coefficients that raise and lower only the total angular momentum quantum number,
\begin{align}
\OpCoeff{+}{0}{\ell}{m}&= \sqrt{\frac{\left(\ell-m+1\right)\left(\ell+m+1\right)}{\left(2\ell+1\right)\left(2\ell+3\right)}} \\
\OpCoeff{-}{0}{\ell}{m}&= \sqrt{\frac{\left(\ell-m\right)\left(\ell+m\right)}{\left(2\ell+1\right)\left(2\ell-1\right)}} \, .
\end{align}
Last are the operators that raise and lower both the total and the projected angular momentum quantum numbers:
\begin{align}
\OpCoeff{+}{+}{\ell}{m}&= \sqrt{\frac{\left(\ell+m+1\right)\left(\ell+m+2\right)}{\left(2\ell+1\right)\left(2\ell+3\right)}} \\
\OpCoeff{+}{-}{\ell}{m}&= \sqrt{\frac{\left(\ell-m+1\right)\left(\ell-m+2\right)}{\left(2\ell+1\right)\left(2\ell+3\right)}} \nonumber \\
\OpCoeff{-}{+}{\ell}{m}&= \sqrt{\frac{\left(\ell-m\right)\left(\ell-m-1\right)}{\left(2\ell+1\right)\left(2\ell-1\right)}} \\
\OpCoeff{-}{-}{\ell}{m}&= \sqrt{\frac{\left(\ell+m-1\right)\left(\ell+m\right)}{\left(2\ell+1\right)\left(2\ell-1\right)}} \, .
\end{align}
We also introduce the Casimir operator,
\begin{align}
\cas{\ell} &= \ell(\ell+1).
\end{align}

For ease of reading, we will handle the electric bilinears first, followed by the loop variables. We will separate out the operators based on which bilinear they appear in, as well as classify how they affect the initial state. Lastly, in order to cut down the size of the expression, we define the ket on which these operators will always act, namely
\begin{align}
\ket{\Omega} &=\left(\ket{\NSeq}\otimes\left(\prod_{\nu=3}^{\NL} \ket{\LSeq{\nu}, \MSeq{\nu}}\right)\otimes\ket{ \LTot, \MTot, \NTot}\right),
\end{align}
and our notation is to be understood as
\begin{align}
\left(\prod_{\nu= 3}^{\NL} \ket{\LSeq{\nu}, \MSeq{\nu}}\right) &= \ket{\LSeq{3}, \MSeq{3}}\otimes  \ket{\LSeq{4}, \MSeq{4}}\otimes \cdots \otimes  \ket{\LSeq{\NL-1}, \MSeq{\NL-1}}\otimes  \ket{\LSeq{\NL}, \MSeq{\NL}}
\end{align}
As discussed in \cref{sec:mixed} in the main text, electric bilinears in the mixed sequestered basis are composed of operators that can only change quantum numbers associated with at most three rods.
To simplify the notation and improve ease of reading, we use a simplified notation to indicate what quantum numbers in the state $\ket{\Omega}$ are not changed by a given operator;
some relevant examples are:  
\begin{align}
    \ket{\Omega \notin \{\mu, \nu\}} &\equiv \ket{n} \otimes \left(\prod_{\rho \neq \mu, \nu} \ket{\LSeq{\rho}, \MSeq{\rho}} \right) \otimes \ket{L, M, N},
    \\
    \ket{\Omega \notin \{ \LTot, \mu \}} &\equiv \ket{n} \otimes \left( \prod_{\rho \neq \mu} \ket{\LSeq{\rho}, \MSeq{\rho}} \right),
    \\
    \ket{\Omega \notin \{n, \mu \}} &\equiv \left( \prod_{\rho \neq \mu} \ket{\LSeq{\rho}, \MSeq{\rho}} \right)\otimes \ket{\LTot, \MTot, \NTot}
    \\
    \ket{\Omega \notin \{n, L \}} &\equiv \left( \prod_{\rho} \ket{\LSeq{\rho}, \MSeq{\rho}} \right)
\end{align}
To balance simplifying the notation while maintaining the intuition for the physical meaning of the quantum numbers, we the quantum numbers associated with a given rod $\mu$ will always appear in the same ket, \ie, $\ket{\LSeq{\mu}, \MSeq{\mu}}$, and similarly for the total angular momentum quantum numbers, \ie, $\ket{\LTot, \MTot, \NTot}$.
Additionally, for ease of readability, we will often write products of states with quantum numbers in different order.
However, it is understood that all states are combined in a consistent manner according to the convention in \cref{eq:seq_mixed_basis_state}
For example,
\begin{equation}
    \ket{L, M, N} \ket{\LSeq{\mu}, \MSeq{\mu}} \ket{\Omega \notin \{L, \mu\}} = \ket{\NSeq} \otimes \left(\prod_\rho \ket{\LSeq{\rho}, \MSeq{\rho}} \right)\otimes \ket{L, M, N}
\end{equation}

Nevertheless, these expressions will not be incredibly compact, though intuition can be gained to more easily understand them; we will try to add this in when possible.

\subsection{Operators appearing in Electric Bilinears}
In order to make this appendix as useful to the reader as possible, we have grouped together operators in a way that not only matches the groupings found in \cref{sec:EEResults}, but also makes it easier to understand how each bilinear acts. 

As we will be separating operators by which quantum numbers they change, we will present our results as a list. 
\begin{itemize}
\item Operators that appear in $\boldsymbol{\CE}_{\kappa \zeta}\cdot \boldsymbol{\CE}_{\kappa \zeta'} $
\begin{itemize}
\item{These operators do not change any quantum numbers}
\begin{align}
\LSeqOpSq_\mu \ket{\Omega} &= \cas{\LSeq{\mu}}\ket{\Omega} \qquad &\LSeqOpComp{z}_\mu \ket{\Omega}&= \OpCoeff{0}{0}{\LSeq{\mu}}{\MSeq{\mu}} \ket{\Omega} =\MSeq{\mu} \ket{\Omega} \nonumber \\
\NSeqSq \ket{\Omega} &= \cas{\NSeq} \ket{\Omega}\qquad &\LSigmaOpComp{z} \ket{\Omega} &=\left(-\OpCoeff{0}{0}{\LTot}{\NTot}-\sum_\mu \OpCoeff{0}{0}{\LSeq{\mu}}{\MSeq{\mu}}\right) \ket{\Omega} = -\sigma \ket{\Omega}   \nonumber \\
\LTotOpSq \ket{\Omega} &= \cas{\LTot} \ket{\Omega}\qquad &\LpTotOpComp{z} \ket{\Omega} &=-\OpCoeff{0}{0}{\LTot}{\NTot}\ket{\Omega}= -\NTot \ket{\Omega}\, .
\end{align}
\item This operator only changes the projected quantum numbers, $\MSeq{\mu}$ and $\NTot$, but do not change $\sigma$: 
\begin{align}
\LSigmaOpSq \ket{\Omega} =& \left(\cas{\LTot}+\sum_\mu\cas{\LSeq{\mu}}\right)\ket{\Omega} +\sum_{\substack{\mu, \nu \\\nu > \mu}} \,\sum_{\delta_m =\pm }\bigg(\OpCoeff{0}{0}{\LSeq{\mu}}{\MSeq{\mu}}\OpCoeff{0}{0}{\LSeq{\nu}}{\MSeq{\nu}}\ket{\LSeq{\mu}, \MSeq{\mu}}\ket{\LSeq{\nu}, \MSeq{\nu}} \nonumber
\\
&+ \OpCoeff{0}{\delta_m}{\LSeq{\mu}}{\MSeq{\mu}}\OpCoeff{0}{-\delta_m}{\LSeq{\nu}}{\MSeq{\nu}}\ket{\LSeq{\mu}, \MSeq{\mu}+\delta_m}\ket{\LSeq{\nu}, \MSeq{\nu}-\delta_m}\bigg)\otimes \ket{\Omega \notin \{\mu, \nu\}} \nonumber
\\
&+\sum_{\mu} \sum_{\delta_m =\pm }\bigg(\OpCoeff{0}{0}{\LTot}{\NTot}\OpCoeff{0}{0}{\LSeq{\mu}}{\MSeq{\mu}}\ket{\LTot, \MTot, \NTot}\ket{\LSeq{\mu}, \MSeq{\mu}} \nonumber
\\
&+\OpCoeff{0}{\delta_m}{\LTot}{\NTot}\OpCoeff{0}{-\delta_m}{\LSeq{\mu}}{\MSeq{\mu}}\ket{\LTot, \MTot, \NTot+\delta_m}\ket{\LSeq{\mu}, \MSeq{\mu}-\delta_m}\bigg)\otimes \ket{\Omega \notin \{ \LTot, \mu \}} \nonumber
\end{align}
where we use the shorthand that, for example,
\begin{align}
\ket{\ell,\, m + \delta_m} \equiv \begin{cases}
\ket{\ell,\, m + 1} \qquad &\text{if}\,\,\, \delta_m = + \\
\ket{\ell,\, m - 1} \qquad &\text{if}
\,\,\, \delta_m = -
\end{cases} \, .
\end{align}
Notice that in order to preserve $\sigma$, this operator raises, for example, $\MSeq{3}$ while lowering $\MSeq{4}$. The fact that no other quantum number is changed can be intuited by noticing that both $\LSeqOp_{\mu}$ and $\LTotOp$ can only change $\MSeq{\mu}$ and $\NTot$, respectively, and nothing else.  

\item This operator can change $\MSeq{\mu}$ and  $\NTot$, either in a way that preserves $\sigma$ or allows $\sigma$ to change:
\begin{align}
\LSigmaOp\cdot \LSeqOp_{2}\ket{\Omega} =& \frac{1}{2}\sum_{\mu}\sum_{\delta_m =\pm}\left(\OpCoeff{0}{0}{\NSeq}{\sigma}\OpCoeff{0}{0}{\LSeq{\mu}}{\MSeq{\mu}}\ket{\LSeq{\mu}, \MSeq{\mu}}+\OpCoeff{0}{\delta_m}{\NSeq}{\sigma}\OpCoeff{0}{\delta_m}{\LSeq{\mu}}{\MSeq{\mu}}\ket{\LSeq{\mu}, \MSeq{\mu}+\delta_m}\right)\otimes \ket{\Omega \notin \{\mu\}}\nonumber 
\\
&+\frac{1}{2}\sum_{\delta_m =\pm}\left(\OpCoeff{0}{0}{\NSeq}{\sigma}\OpCoeff{0}{0}{\LTot}{\NTot}\ket{\LTot, \MTot, \NTot}+\OpCoeff{0}{\delta_m}{\NSeq}{\sigma}\OpCoeff{0}{\delta_m}{\LTot}{\NTot}\ket{\LTot, \MTot, \NTot+\delta_m}\right) \otimes \ket{\Omega \notin \{ L \}}
\end{align}
Even though $\LSeqOp_{2}$ is quite complicated when written in terms of differential operators, it behaves in an analagous way to $\LSeqOp_{\mu}$. Said in the different way, because this operator involves only $\LSeqOp_\sigma$ and $\LSeqOp_2$, it can only affect projected angular momentum quantum numbers, not total angular momentum quantum numbers. The fact that $\OpCoeff{0}{\delta_m}{\NSeq}{\sigma}$ depends on $\sigma$, and its ramifications on the resources required for simulating this theory, is addressed in \cref{sec:ResourceScaling}.
\end{itemize}

\item Operators that appear in $\boldsymbol{\CE}_{\mu \zeta_\mu}\cdot \boldsymbol{\CE}_{\nu \zeta_\nu}$. Note that these operators always preserve $\sigma$.
Furthermore, operators of this type acting on different rods commute, \ie, $[\nSeqBody{\mu}, \nSeqBody{\nu}] = [\nSeqBody{\mu}, \LSeqOp_{\nu}] = [\LSeqOp_{\mu}, \LSeqOp_{\nu}] = 0$.

\begin{itemize}
\item These operators always change both $\LSeq{\mu}$ and $\LSeq{\nu}$ and can also change both $\MSeq{\mu}$ and $\MSeq{\nu}$
\begin{align}
\nSeqBody{\mu}\cdot \nSeqBody{\nu}\ket{\Omega} =&\frac{1}{2} \sum_{\substack{\delta_{\LSeq{\mu}}=\pm\\\delta_{\LSeq{\nu}}=\pm}}\sum_{\delta_{m}=\pm} \bigg[\OpCoeff{\delta_{\LSeq{\mu}}}{0}{\LSeq{\mu}}{\MSeq{\mu}}\OpCoeff{\delta_{\LSeq{\nu}}}{0}{\LSeq{\nu}}{\MSeq{\nu}}\ket{\LSeq{\mu}+\delta_{\LSeq{\mu}}, \MSeq{\mu}}\ket{\LSeq{\nu}+\delta_{\LSeq{\nu}}, \MSeq{\nu}} 
\\
&-\delta_{\LSeq{\mu}}\delta_{\LSeq{\nu}}\OpCoeff{\delta_{\LSeq{\mu}}}{\delta_m}{\LSeq{\mu}}{\MSeq{\mu}}\OpCoeff{\delta_{\LSeq{\nu}}}{-\delta_m}{\LSeq{\nu}}{\MSeq{\nu}}\ket{\LSeq{\mu}+\delta_{\LSeq{\mu}}, \MSeq{\mu}+\delta_m}\ket{\LSeq{\nu}+\delta_{\LSeq{\nu}}, \MSeq{\nu}-\delta_m}\bigg] \otimes \ket{\Omega \notin \{ \mu, \nu \}} \nonumber 
\end{align}

\begin{align}
\label{eq:v3dotv4crossL4}
\nSeqBody{\mu}\cdot \left(\nSeqBody{\nu}\times \LSeqOp_{\nu}\right)\ket{\Omega} =& -\frac{i}{2}\sum_{\substack{\delta_{\LSeq{\mu}}=\pm\\\delta_{\LSeq{\nu}}=\pm}}\sum_{\delta_m=\pm} \Bigg[\delta_{\LSeq{\nu}}\OpCoeff{\delta_{\LSeq{\mu}}}{0}{\LSeq{\mu}}{\MSeq{\mu}}\OpCoeff{\delta_{\LSeq{\nu}}}{-\delta_m}{\LSeq{\nu}}{\MSeq{\nu}+\delta_m}\OpCoeff{0}{\delta_m}{\LSeq{\nu}}{\MSeq{\nu}}\ket{\LSeq{\mu}+\delta_{\LSeq{\mu}}, \MSeq{\mu}}\ket{\LSeq{\nu}+\delta_{\LSeq{\nu}}, \MSeq{\nu}}\nonumber 
\\
&-\delta_{\LSeq{\mu}}\OpCoeff{\delta_{\LSeq{\mu}}}{-\delta_m}{\LSeq{\mu}}{\MSeq{\mu}}\left(\OpCoeff{\delta_\LSeq{\nu}}{0}{\LSeq{\nu}}{\MSeq{\nu}+\delta_{m}}\OpCoeff{0}{\delta_m}{\LSeq{\nu}}{\MSeq{\nu}}+\delta_{\LSeq{\nu}}\delta_m\OpCoeff{\delta_\LSeq{\nu}}{\delta_m}{\LSeq{\nu}}{\MSeq{\nu}}\OpCoeff{0}{0}{\LSeq{\nu}}{\MSeq{\nu}}\right)
\\
&\times \ket{\LSeq{\mu}+\delta_{\LSeq{\mu}}, \MSeq{\mu}-\delta_m}\ket{\LSeq{\nu}+\delta_{\LSeq{\nu}}, \MSeq{\nu}+\delta_m}\Bigg] \otimes \ket{\Omega \notin \{ \mu, \nu\}} \nonumber 
\end{align}

\begin{align}
\label{Eq.N3dN4}
\left(\nSeqBody{\mu}\times \LSeqOp_{\mu}\right)\cdot \left(\nSeqBody{\nu}\times \LSeqOp_{\nu}\right)\ket{\Omega} =& -\frac{1}{2}\sum_{\substack{\delta_\LSeq{\mu}=\pm\\\delta_\LSeq{\nu}=\pm}}\sum_{\delta_m = \pm}\bigg[\frac{\delta_{\LSeq{\mu}}\delta_{\LSeq{\nu}}}{4}\left(\OpCoeff{\delta_{\LSeq{\mu}}}{\delta_m}{\LSeq{\mu}}{\MSeq{\mu}-\delta_m}\OpCoeff{0}{-\delta_m}{\LSeq{\mu}}{\MSeq{\mu}}+\OpCoeff{\delta_{\LSeq{\mu}}}{-\delta_m}{\LSeq{\mu}}{\MSeq{\mu}+\delta_m}\OpCoeff{0}{\delta_m}{\LSeq{\mu}}{\MSeq{\mu}}\right)\nonumber
\\
&\times \left(\OpCoeff{\delta_{\LSeq{\nu}}}{\delta_m}{\LSeq{\nu}}{\MSeq{\nu}-\delta_m}\OpCoeff{0}{-\delta_m}{\LSeq{\nu}}{\MSeq{\nu}}+\OpCoeff{\delta_{\LSeq{\nu}}}{-\delta_m}{\LSeq{\nu}}{\MSeq{\nu}+\delta_m}\OpCoeff{0}{\delta_m}{\LSeq{\nu}}{\MSeq{\nu}}\right)\ket{\LSeq{\mu}+\delta_{\LSeq{\mu}}, \MSeq{\mu}}\ket{\LSeq{\nu}+\delta_{\LSeq{\nu}}, \MSeq{\nu}} \nonumber
\\
&-\left(\OpCoeff{\delta_\LSeq{\mu}}{0}{\LSeq{\mu}}{\MSeq{\mu}+\delta_m}\OpCoeff{0}{\delta_m}{\LSeq{\mu}}{\MSeq{\mu}}+\delta_m \delta_{\LSeq{\mu}}\OpCoeff{\delta_\LSeq{\mu}}{\delta_m}{\LSeq{\mu}}{\MSeq{\mu}}\OpCoeff{0}{0}{\LSeq{\mu}}{\MSeq{\mu}}\right)
\\
&\times \left(\OpCoeff{\delta_\LSeq{\nu}}{0}{\LSeq{\nu}}{\MSeq{\nu}-\delta_m}\OpCoeff{0}{-\delta_m}{\LSeq{\nu}}{\MSeq{\nu}}-\delta_m \delta_{\LSeq{\nu}}\OpCoeff{\delta_\LSeq{\nu}}{-\delta_m}{\LSeq{\nu}}{\MSeq{\nu}}\OpCoeff{0}{0}{\LSeq{\nu}}{\MSeq{\nu}}\right)\nonumber 
\\
&\times \ket{\LSeq{\mu}+\delta_{\LSeq{\mu}}, \MSeq{\mu}+\delta_m}\ket{\LSeq{\nu}+\delta_{\LSeq{\nu}}, \MSeq{\nu}-\delta_m}\bigg]\otimes \ket{\Omega \notin \{ \mu, \nu \}} \nonumber
\end{align}
Looking at the first operator, notice that the appearance of $\nSeqBody{\mu}$ allows for $\LSeq{\mu}$ to be incremented.

The second operator is more complicated, and it is worth discussing in more detail the pattern in the quantum numbers that are modified. 
In the left-hand side, the operator $\nSeqBody{\nu}\times \LSeqOp_{\nu}$ appears. While $\LSeqOp_{\nu}$ can only change $\MSeq{\nu}$ by $\pm 1$, $\nSeqBody{\nu}$ can change both $\LSeq{\nu}$ and $\MSeq{\nu}$, each by $\pm 1$, with uncorrolated signs. Furthermore, if $\MSeq{\nu}$ is changed, there are two paths to reach the same final state. The first path has $\LSeqOp_{\nu}$ changing $\MSeq{\nu}$ and then $\nSeqBody{\nu}$ only changing only $\LSeq{\nu}$; the second has $\LSeqOp_{\nu}$ leaving $\MSeq{\nu}$ unchanged and then $\nSeqBody{\nu}$ changing both $\LSeq{\nu}$ and $\MSeq{\nu}$:
\begin{align}
\begin{array}{c@{}c@{}c}
\ket{\LSeq{\nu}, \MSeq{\nu}} & \overset{\LSeqOp_\nu}{\xRightarrow{\hspace{1cm}}} & \ket{\LSeq{\nu}, \MSeq{\nu}+\delta_m} \\
\mathrel{\rotatebox[origin=c]{90}{$\overset{\LSeqOp_\nu}{\xLeftarrow{\rule{0.8cm}{0pt}}}$}} & & \mathrel{\rotatebox[origin=c]{-90}{$\overset{\nSeqBody{\nu}}{\xRightarrow{\rule{0.8cm}{0pt}}}$}} \\
\ket{\delta_{\LSeq{\nu}}, \MSeq{\nu}} \,\,\,\,& \overset{\nSeqBody{\nu}}{\xRightarrow{\hspace{1cm}}} & \,\,\,\,\ket{\LSeq{\nu}+\delta_{\LSeq{\nu}}, \MSeq{\nu}+\delta_m} 
\end{array}
\end{align}
One might assume that a similar situation arises when the final and initial state has the same $\MSeq{\nu}$. In this case, the first path has $\LSeqOp_{\nu}$ changing $\MSeq{\nu}$ and then $\nSeqBody{\nu}$ changing $\LSeq{\nu}$ and changing $\MSeq{\nu}$ with the opposite sign as $\LSeqOp_{\nu}$ did; the second has $\LSeqOp_{\nu}$ leaving $\MSeq{\nu}$ unchanged and then have $\nSeqBody{\nu}$ changing only $\LSeq{\nu}$. However, this second path only arises from the operator
\begin{align}
\eta_\nu^z\, \LSeqOpComp{z} = i \cos \thetaSeq{\nu}\pdv{}{\phiSeq{\nu}} 
\end{align}
which does not appear in $\nSeqBody{\nu}\times \LSeqOp_{\nu}$.

The third operator has a similar structure to the second operator.

\item This operator changes only $\LSeq{\mu}$ and can also change both $\MSeq{\mu}$ and $\MSeq{\nu}$
\begin{align}
\nSeqBody{\mu}\cdot \LSeqOp_{\nu}\ket{\Omega} =\frac{1}{2}\sum_{\delta_{\LSeq{}}=\pm}\sum_{\delta_m =\pm} & \bigg[\delta_{\LSeq{}}\delta_m \OpCoeff{\delta_{\LSeq{}}}{-\delta_m}{\LSeq{\mu}}{\MSeq{\mu}}\OpCoeff{0}{\delta_m}{\LSeq{\nu}}{\MSeq{\nu}}\ket{\LSeq{\mu}+\delta_{\LSeq{}}, \MSeq{\mu}-\delta_m}\ket{\LSeq{\nu}, \MSeq{\nu}+\delta_m} \nonumber
\\
&+ \OpCoeff{\delta_{\LSeq{}}}{0}{\LSeq{\mu}}{\MSeq{\mu}}\OpCoeff{0}{0}{\LSeq{\nu}}{\MSeq{\nu}}\ket{\LSeq{\mu}+\delta_{\LSeq{}}, \MSeq{\mu}}\ket{\LSeq{\nu}, \MSeq{\nu}}\bigg] \otimes \ket{\Omega \notin \{ \mu, \nu \} } 
\end{align}
Once again, we can intuit that only $\LSeq{\mu}$ can be changed by this operator and $\LSeq{\nu}$ remains unchanged, by noticing that only $\nSeqBody{\mu}$ and $\LSeqOp_\nu$ appear.

\item This operator only changes $\LSeq{\nu}$ and can also change both $\MSeq{\mu}$ and $\MSeq{\nu}$ 
\begin{align}
\LSeqOp_{\mu}\cdot \left(\nSeqBody{\nu}\times \LSeqOp_{\nu}\right)\ket{\Omega} =& -\frac{i}{2}\sum_{\delta_{\LSeq{\nu}}=\pm}\sum_{\delta_m=\pm}\bigg[\delta_{\LSeq{\nu}}\,\OpCoeff{0}{0}{\LSeq{\mu}}{\MSeq{\mu}}\OpCoeff{\delta_{\LSeq{\nu}}}{-\delta_m}{\LSeq{\nu}}{\MSeq{\nu}+\delta_m} \OpCoeff{0}{\delta_m}{\LSeq{\nu}}{\MSeq{\nu}}\ket{\LSeq{\mu}, \MSeq{\mu}}\ket{\LSeq{\nu}+\delta_{\LSeq{\nu}}, \MSeq{\nu}} \nonumber 
\\
&-\OpCoeff{0}{-\delta_m}{\LSeq{\mu}}{\MSeq{\mu}}\left(\delta_m \, \OpCoeff{\delta_\LSeq{\nu}}{0}{\LSeq{\nu}}{\MSeq{\nu}+\delta_m}\OpCoeff{0}{\delta_m}{\LSeq{\nu}}{\MSeq{\nu}}+\delta_\LSeq{\nu} \, \OpCoeff{\delta_\LSeq{\nu}}{\delta_m}{\LSeq{\nu}}{\MSeq{\nu}}\OpCoeff{0}{0}{\LSeq{\nu}}{\MSeq{\nu}}\right) \nonumber
\\
&\times \ket{\LSeq{\mu}, \MSeq{\mu}-\delta_m}\ket{\LSeq{\nu}+\delta_{\LSeq{\nu}}, \MSeq{\nu}+\delta_m}\bigg] \otimes \ket{\Omega \notin \{\mu, \nu\}} 
\end{align}

\item This operator only changes both $\MSeq{\mu}$ and $\MSeq{\nu}$
\begin{align}
\LSeqOp_{\mu}\cdot \LSeqOp_{\nu}\ket{\Omega} =& \frac{1}{2}\sum_{\delta_m=\pm}\bigg[ \OpCoeff{0}{\delta_m}{\LSeq{\mu}}{\MSeq{\mu}}\OpCoeff{0}{-\delta_m}{\LSeq{\nu}}{\MSeq{\nu}}\ket{\LSeq{\mu}, \MSeq{\mu}+\delta_m}\ket{\LSeq{\nu}, \MSeq{\nu}-\delta_m} \nonumber
\\
&+ \OpCoeff{0}{0}{\LSeq{\mu}}{\MSeq{\mu}}\OpCoeff{0}{0}{\LSeq{\nu}}{\MSeq{\nu}}\ket{\LSeq{\mu}, \MSeq{\mu}}\ket{\LSeq{\nu}, \MSeq{\nu}}\bigg] \otimes \ket{\Omega \notin \{\mu, \nu \}} 
\label{eq:L3dotL4}
\end{align}
\end{itemize}

\item Operators that appear in $\boldsymbol{\CE}_{2 \zeta_2}\cdot \boldsymbol{\CE}_{\mu \zeta_\mu}$

In this case, many of the same arguments made above about what quantum numbers are or are not changed based on whether $\nSeqBody{\mu}$ and $\LSeqOp_\mu$ appears in the operator can be applied. The same is true for $\nSeqBody{2}$ and $\LSeqOp_2$, though it is important to remember that $\sigma$ is not a true quantum number.
\begin{itemize}
\item These operators change both $\NSeq$ and $\LSeq{\mu}$, and can also change $\MSeq{\mu}$ and therefore $\sigma$ 
\begin{align}
\nSeqBody{2}\cdot \nSeqBody{\mu}\ket{\Omega}=&\frac{1}{2}\sum_{\substack{\delta_n =\pm\\\delta_\ell=\pm}}\ket{\NSeq+\delta_n} \otimes \sum_{\delta m = \pm}\Big(\OpCoeff{\delta_n}{0}{\NSeq}{\sigma}\OpCoeff{\delta_\ell}{0}{\LSeq{\mu}}{\MSeq{\mu}}\ket{\LSeq{\mu}+\delta_\ell,\MSeq{\mu}} \nonumber
\\
&+\delta_n \delta_\ell\, \OpCoeff{\delta_n}{\delta_m}{\NSeq}{\sigma}\OpCoeff{\delta_\ell}{\delta_m}{\LSeq{\mu}}{\MSeq{\mu}}\ket{\LSeq{\mu}+\delta_\ell, \MSeq{\mu}+\delta_m}\Big)\otimes \ket{\Omega \notin \{n, \mu\}}
\end{align}

\begin{align}
\nSeqBody{2}\cdot \left(\nSeqBody{\mu}\times \LSeqOp_{\mu}\right)\ket{\Omega} =& -\frac{i}{2}\sum_{\substack{\delta_{\NSeq}=\pm\\\delta_{\LSeq{}}=\pm}}\sum_{\delta_m=\pm}\bigg[\delta_n \OpCoeff{\delta_n}{\delta_m}{\NSeq}{\sigma}\left(\OpCoeff{\delta_{\LSeq{}}}{0}{\LSeq{\mu}}{\MSeq{\mu}+\delta_m}\OpCoeff{0}{\delta_m}{\LSeq{\mu}}{\MSeq{\mu}}+\delta_m \delta_\LSeq{}\,\,\OpCoeff{\delta_\LSeq{}}{\delta_m}{\LSeq{\mu}}{\MSeq{\mu}} \OpCoeff{0}{0}{\LSeq{\mu}}{\MSeq{\mu}}\right)\ket{\LSeq{\mu}+\delta_\LSeq{},\MSeq{\mu}+\delta_m}\nonumber
\\
&+\delta_\LSeq{}\, \OpCoeff{\delta_n}{0}{\NSeq}{\sigma}\OpCoeff{\delta_\LSeq{}}{\delta_m}{\LSeq{\mu}}{\MSeq{\mu}-\delta_m}\OpCoeff{0}{-\delta_m}{\LSeq{\mu}}{\MSeq{\mu}}\ket{\LSeq{\mu}+\delta_\LSeq{},\MSeq{\mu}}\bigg]\otimes \ket{\NSeq+\delta_n} \otimes \ket{\Omega \notin \{\NSeq, \mu \}},
\end{align}

\begin{align}
\nSeqBody{\mu}\cdot \crossvL{2}{2}\ket{\Omega} =&\, -\frac{i}{2}\sum_{\substack{\delta_n = \pm\\\delta_\ell = \pm}}\sum_{\delta_m=\pm}\left[\frac{\delta_n}{2}\left(\OpCoeff{\delta_n}{-\delta_m}{\NSeq}{\sigma+\delta_m}\OpCoeff{0}{\delta_m}{\NSeq}{\sigma}+\OpCoeff{\delta_n}{\delta_m}{\NSeq}{\sigma-\delta_m}\OpCoeff{0}{-\delta_m}{\NSeq}{\sigma}\right)\OpCoeff{\delta_\ell}{0}{\LSeq{\mu}}{\MSeq{\mu}}\ket{\LSeq{\mu}+\delta_\ell, \MSeq{\mu}}\right. \nonumber 
\\
&+\left.\delta_\ell \left(\OpCoeff{\delta_n}{0}{\NSeq}{\sigma+\delta_m}\OpCoeff{0}{\delta_m}{\NSeq}{\sigma}+\delta_n \delta_m\,\,\OpCoeff{\delta_n}{\delta_m}{\NSeq}{\sigma}\OpCoeff{0}{0}{\NSeq}{\sigma}\right)\OpCoeff{\delta_\ell}{\delta_m}{\LSeq{\mu}}{\MSeq{mu}}\ket{\LSeq{\mu}+\delta_\ell, \MSeq{\mu}+\delta_m}\right]\nonumber 
\\
&\otimes \ket{n+\delta_n} \otimes \ket{\Omega \notin \{n, \mu \} },
\label{Eq.N2dN3}
\end{align}

\begin{align}
\crossvL{\mu}{\mu} \cdot \crossvL{2}{2}\ket{\Omega} =&\, -\frac{1}{2}\sum_{\substack{\delta_n = \pm\\\delta_\ell=\pm}}\sum_{\delta_m = \pm} \Bigg[ \frac{\delta_n \delta_\ell}{4}\left(\OpCoeff{\delta_n}{\delta_m}{\NSeq}{\sigma - \delta_m}\OpCoeff{0}{-\delta_m}{\NSeq}{\sigma}+\OpCoeff{\delta_n}{-\delta_m}{\NSeq}{\sigma + \delta_m}\OpCoeff{0}{\delta_m}{\NSeq}{\sigma}\right)\nonumber 
\\
&\times\left(\OpCoeff{\delta_\ell}{\delta_m}{\LSeq{\mu}}{\MSeq{\mu} - \delta_m}\OpCoeff{0}{-\delta_m}{\LSeq{\mu}}{\MSeq{\mu}}+\OpCoeff{\delta_\ell}{-\delta_m}{\LSeq{\mu}}{\MSeq{\mu} + \delta_m}\OpCoeff{0}{\delta_m}{\LSeq{\mu}}{\MSeq{\mu}}\right)\ket{\LSeq{\mu}+\delta_\ell, \MSeq{\mu}}\nonumber
\\
&+\left(\OpCoeff{\delta_n}{0}{\NSeq}{\sigma + \delta_m}\OpCoeff{0}{\delta_m}{\NSeq}{\sigma}+\delta_m \delta_n\,\,\OpCoeff{\delta_n}{\delta_m}{\NSeq}{\sigma}\OpCoeff{0}{0}{\NSeq}{\sigma}\right) \nonumber
\\
&\times \left(\OpCoeff{\delta_\ell}{0}{\LSeq{\mu}}{\MSeq{\mu} + \delta_m}\OpCoeff{0}{\delta_m}{\LSeq{\mu}}{\MSeq{\mu}}+\delta_m \delta_\ell\,\,\OpCoeff{\delta_\ell}{\delta_m}{\LSeq{\mu}}{\MSeq{\mu}}\OpCoeff{0}{0}{\LSeq{\mu}}{\MSeq{\mu}}\right) \nonumber 
\\
&\times \ket{\LSeq{\mu}+\delta_\ell, \MSeq{\mu}+\delta_m}\Bigg] \otimes \ket{n+\delta_n}\otimes \ket{\Omega \notin \{n, \mu\}},
\end{align}

\begin{align}
\crossvL{\mu}{\mu}\cdot \LSeqOp_{2}=&\,\frac{i}{2}\sum_{\delta_\ell = \pm}\sum_{\delta_m=\pm}\bigg[\delta_\ell \, \OpCoeff{0}{0}{\NSeq}{\sigma}\OpCoeff{\delta_\ell}{-\delta_m}{\LSeq{\mu}}{\MSeq{\mu}+\delta_m}\OpCoeff{0}{\delta_m}{\LSeq{\mu}}{\MSeq{\mu}}\ket{\LSeq{\mu}+\delta_\ell, \MSeq{\mu}}\nonumber \\
&- \delta_\ell \OpCoeff{0}{\delta_m}{\NSeq}{\sigma}\left(\delta_m \delta_\ell\,\, \OpCoeff{\delta_\ell}{0}{\LSeq{\mu}}{\MSeq{\mu}+\delta_m}\OpCoeff{0}{\delta_m}{\LSeq{\mu}}{\MSeq{\mu}}+\OpCoeff{\delta_\ell}{\delta_m}{\LSeq{\mu}}{\MSeq{\mu}}\OpCoeff{0}{0}{\LSeq{\mu}}{\MSeq{\mu}}\right)\ket{\LSeq{\mu}+\delta_\ell, \MSeq{\mu}+\delta_m}\bigg]\nonumber \\
&\otimes \ket{n+\delta_n}\otimes \ket{\Omega \notin \{ n, \mu \}}
\label{eq:Cross33L2}
\end{align}

\item These operators change $\NSeq$ and can also change $\MSeq{\mu}$ and therefore $\sigma$
\begin{align}
\hspace{-0.3in} \nSeqBody{2}\cdot \LSeqOp_{\mu}\ket{\Omega}=&\, \frac{1}{2}\sum_{\delta_n =\pm}\sum_{\delta_m = \pm}\bigg[\left(\OpCoeff{\delta_n}{0}{\NSeq}{\sigma}\OpCoeff{0}{0} {\LSeq{\mu}}{\MSeq{\mu}}\ket{\LSeq{\mu}, \MSeq{\mu}}-\delta_n \delta_m \, \OpCoeff{\delta_n}{\delta_m}{\NSeq}{\sigma}\OpCoeff{0}{\delta_m}{\LSeq{\mu}}{\MSeq{\mu}}\ket{\LSeq{\mu}, \MSeq{\mu}+\delta_m}\right)\otimes\ket{\NSeq+\delta_n}\bigg]\nonumber
\\
& \otimes \ket{\Omega \notin \{n, \mu\}},
\end{align}

\begin{align}
\LSeqOp_{\mu} \cdot \crossvL{2}{2}=&\,\frac{i}{2}\sum_{\delta_n = \pm}\sum_{\delta_m = \pm}\bigg[\delta_n \OpCoeff{0}{\delta_m}{\LSeq{\mu}}{\MSeq{\mu}}\left(\delta_m \delta_n\,\, \OpCoeff{\delta_n}{0}{\NSeq}{\sigma+\delta_m}\OpCoeff{0}{\delta_m}{\NSeq}{\sigma}+ \OpCoeff{\delta_n}{\delta_m}{\NSeq}{\sigma}\OpCoeff{0}{0}{\NSeq}{\sigma}\right)\ket{\LSeq{\mu}, \MSeq{\mu}+\delta_m}\nonumber 
\\
&- \delta_n \, \, \OpCoeff{0}{0}{\LSeq{\mu}}{\MSeq{\mu}}\OpCoeff{\delta_n}{-\delta_m}{\NSeq}{\sigma+\delta_m}\OpCoeff{0}{\delta_m}{\NSeq}{\sigma}\ket{\LSeq{\mu}, \MSeq{\mu}}\bigg]\otimes \ket{n+\delta_n}\otimes \ket{\Omega \notin \{n, \mu\}}.
\end{align}

\item This operator changes $\LSeq{\mu}$ and can also change $\MSeq{\mu}$, which results in a change in $\sigma$
\begin{align}
\nSeqBody{\mu}\cdot \LSeqOp_{2}\ket{\Omega} =&\, -\frac{1}{2}\sum_{\delta_\ell = \pm}\sum_{\delta_m = \pm}\Big(\OpCoeff{0}{0}{n}{\sigma}\OpCoeff{\delta_\ell}{0}{\LSeq{\mu}}{\MSeq{\mu}}\ket{\LSeq{\mu}+\delta_\ell, \MSeq{\mu}} \nonumber
\\
&-\delta_\ell\delta_m \,\OpCoeff{0}{\delta_m}{n}{\sigma}\OpCoeff{\delta_\ell}{\delta_m}{\LSeq{\mu}}{\MSeq{\mu}}\ket{\LSeq{\mu}+\delta_\ell, \MSeq{\mu}+\delta_m}\Big) \otimes \ket{\Omega \notin \{\mu \}}
\end{align}

\item This operator can only change $\MSeq{\mu}$ and therefore $\sigma$
\begin{align}
\LSeqOp_\mu \cdot \LSeqOp_2 &=-\frac{1}{2}\sum_{\delta_m}\bigg[\OpCoeff{0}{0}{\NSeq}{\sigma}\OpCoeff{0}{0}{\LSeq{\mu}}{\MSeq{\mu}}\ket{\LSeq{\mu}, \MSeq{\mu}}+\OpCoeff{0}{\delta_m}{\NSeq}{\sigma}\OpCoeff{0}{\delta_m}{\LSeq{\mu}}{\MSeq{\mu}}\ket{\LSeq{\mu}, \MSeq{\mu}+\delta_m}\bigg] \otimes \ket{\Omega \notin \{\mu \}}
\label{eq:L3dotL2}
\end{align}
\end{itemize}

\item Operators that appear in $\boldsymbol{\CE}_{1 \zeta_1}\cdot \boldsymbol{\CE}_{\mu \zeta_\mu}$ 

Many of the operators in this operator will be very similar to those appearing in $\boldsymbol{\CE}_{2 \zeta_2}\cdot \boldsymbol{\CE}_{\mu \zeta_\mu}$.

\begin{itemize}
\item These operators can only change $\LSeq{\mu}$
\begin{align}
\nSeqBody{1}\cdot \nSeqBody{\mu}\ket{\Omega} &= \sum_{\delta_\ell = \pm}\OpCoeff{\delta_\ell}{0}{\LSeq{\mu}}{\MSeq{\mu}}\ket{\LSeq{\mu}+\delta_\ell, \MSeq{\mu}}\otimes \ket{\Omega \notin \{ \mu \}}
\\[2ex]
\nSeqBody{1}\cdot \crossvL{\mu}{\mu}\ket{\Omega} &= -\frac{i}{2} \sum_{\delta_\ell=\pm}\sum_{\delta_m = \pm} \delta_\ell \,\, \OpCoeff{\delta_\ell}{-\delta_m}{\LSeq{\mu}}{\MSeq{\mu}+\delta_m}\OpCoeff{0}{\delta_m}{\LSeq{\mu}}{\MSeq{\mu}}\ket{\LSeq{\mu}+\delta_\ell, \MSeq{\mu}} \otimes \ket{\Omega \notin \{ \mu \}}
\label{Eq.N1dN3}
\end{align}

\item This operator leaves the state unchanged
\begin{align}
\nSeqBody{1}\cdot \LSeqOp_\mu \ket{\Omega} &=\OpCoeff{0}{0}{\LSeq{\mu}}{\MSeq{\mu}}\ket{\Omega}
\end{align}

\item These operators change $\LSeq{\mu}$. They can also change $\MSeq{\mu}$ and $\NTot$, either in a manner that preserves $\sigma$ (which involves changing $\MSeq{\nu}$ or $\NTot$ in the opposite manner) or in a manner that changes $\sigma$:
\begin{align}
\nSeqBody{\mu}\cdot \crossvL{1}{1} \ket{\Omega} =&\, -\frac{i}{2} \sum_{\delta_\ell = \pm} \delta_\ell \,\sum_{\delta_m = \pm}\bigg[\,  \OpCoeff{\delta_\ell}{-\delta_m}{\LSeq{\mu}}{\MSeq{\mu}+\delta_m}\OpCoeff{0}{\delta_m}{\LSeq{\mu}}{\MSeq{\mu}}\ket{\LSeq{\mu}+\delta_\ell, \MSeq{\mu}}\otimes \ket{\Omega \notin \{\mu\}} \nonumber
\\
&+ \OpCoeff{\delta_\ell}{\delta_m}{\LSeq{\mu}}{\MSeq{\mu}}\ket{\LSeq{\mu}+\delta_m, \MSeq{\mu}+\delta_m}\otimes \bigg\{\OpCoeff{0}{-\delta_m}{\LTot}{\NTot}\ket{\LTot, \MTot, \NTot-\delta_m}
\otimes \ket{\Omega \notin \{L, \mu\}} \nonumber 
\\
&+\sum_{\nu \neq \mu}\OpCoeff{0}{-\delta_m}{\LSeq{\nu}}{\MSeq{\nu}}\ket{\LSeq{\nu}, \MSeq{\nu}-\delta_m}
\otimes \ket{\Omega \notin \{\mu, \nu\}} - \OpCoeff{0}{\delta_m}{\NSeq}{\sigma}  \ket{\Omega \notin \{\mu\}} \bigg\}\bigg]
\\[2ex]
\nSeqBody{\mu}\cdot \LSeqOp_{1} \ket{\Omega} =&\, \frac{1}{2} \sum_{\delta_\ell = \pm} \,\sum_{\delta_m = \pm} \bigg[\,  \OpCoeff{\delta_\ell}{0}{\LSeq{\mu}}{\MSeq{\mu}}\OpCoeff{0}{0}{\LSeq{\mu}}{\MSeq{\mu}}\ket{\LSeq{\mu}+\delta_\ell, \MSeq{\mu}}\otimes \ket{\Omega \notin \{ \mu \}} \nonumber 
\\
&+ \delta_\ell \delta_m\,\,\OpCoeff{\delta_\ell}{\delta_m}{\LSeq{\mu}}{\MSeq{\mu}}\ket{\LSeq{\mu}+\delta_m, \MSeq{\mu}+\delta_m}\otimes \bigg\{\OpCoeff{0}{-\delta_m}{\LTot}{\NTot}\ket{\LTot, \MTot, \NTot-\delta_m}
\otimes \ket{\Omega \notin \{ L, \mu \}} \nonumber
\\
&+\sum_{\nu \neq \mu}\OpCoeff{0}{-\delta_m}{\LSeq{\nu}}{\MSeq{\nu}}\ket{\LSeq{\nu}, \MSeq{\nu}-\delta_m}
\otimes \ket{\Omega \notin \{ \mu,\nu \}} - \OpCoeff{0}{\delta_m}{\NSeq}{\sigma}  \ket{\Omega \notin \{ \mu \}} \bigg\} \bigg]
\\[2ex]
\crossvL{\mu}{\mu}\cdot \crossvL{1}{1}=&\, -\frac{1}{2} \sum_{\delta_\ell = \pm} \,\sum_{\delta_m = \pm}\bigg[\,  \left(\cas{\LSeq{\mu}}\OpCoeff{\delta_\ell}{0}{\LSeq{\mu}}{\MSeq{\mu}}-\delta_\ell, \OpCoeff{\delta_\ell}{\delta_m}{\LSeq{\mu}}{\MSeq{\mu}-\delta_m}\OpCoeff{0}{-\delta_m}{\LSeq{\mu}}{\MSeq{\mu}}\right) \ket{\LSeq{\mu}+\delta_\ell, \MSeq{\mu}}\otimes \ket{\Omega \notin \{ \mu \}} \nonumber 
\\
&+\left(\OpCoeff{\delta_\ell}{0}{\LSeq{\mu}}{\MSeq{\mu}+\delta_m}\OpCoeff{0}{\delta_m}{\LSeq{\mu}}{\MSeq{\mu}}+ \delta_\ell \delta_m\,\,\OpCoeff{\delta_\ell}{\delta_m}{\LSeq{\mu}}{\MSeq{\mu}}\OpCoeff{0}{0}{\LSeq{\mu}}{\MSeq{\mu}}\right)\ket{\LSeq{\mu}+\delta_m, \MSeq{\mu}+\delta_m} \nonumber
\\
&\otimes \bigg\{\OpCoeff{0}{-\delta_m}{\LTot}{\NTot}\ket{\LTot, \MTot, \NTot-\delta_m}
\otimes \ket{\Omega \notin \{ \LTot, \mu \}} \nonumber
\\
&+\sum_{\nu \neq \mu}\OpCoeff{0}{-\delta_m}{\LSeq{\nu}}{\MSeq{\nu}}\ket{\LSeq{\nu}, \MSeq{\nu}-\delta_m}
\otimes \ket{\Omega \notin \{ \mu,\nu \}} - \OpCoeff{0}{\delta_m}{\NSeq}{\sigma}  \ket{\Omega \notin \{ \mu \}} \bigg\}\bigg]
\end{align}
Notice that, for all three operators appearing above, all the terms except the last one preserve $\sigma$. 
Additionally, these operators cannot change $\NSeq$, as $\nSeqBody{2}$ does not appear.
\item This action of the operator $ \crossvL{\mu}{\mu}\cdot\LSeqOp_1$ can be written as
\begin{align}
\crossvL{\mu}{\mu}&\cdot\LSeqOp_1\ket{\Omega}= 
\crossvL{\mu}{\mu}\cdot\LpTotOp\ket{\Omega}-\underbrace{\crossvL{\mu}{\mu}\cdot\LSeqOp_2\ket{\Omega}}_{\text{Eq.~}\ref{eq:Cross33L2}}-\sum_{\nu \neq \mu}\underbrace{\crossvL{\mu}{\mu}\cdot\LSeqOp_\nu\ket{\Omega}}_{\text{Eq.~}\ref{eq:v3dotv4crossL4}}
\end{align}
where we have used $\nSeqBody{\mu}\cdot\crossvL{\mu}{\mu}=0$, and that $\crossvL{\mu}{\mu}\cdot\LSeqOp_\nu = \LSeqOp_\nu \cdot \crossvL{\mu}{\mu}$; the last piece to evaluate is
\begin{align}
\crossvL{\mu}{\mu}\cdot\LpTotOp\ket{\Omega} =&\, \frac{i}{2}\sum_{\delta_\ell = \pm}\sum_{\delta_m = \pm}\bigg[\delta_\ell\,\, \OpCoeff{0}{0}{\LTot}{\NTot}\OpCoeff{\delta_\ell}{-\delta_m}{\LSeq{\mu}}{\MSeq{\mu}+\delta_m}\OpCoeff{0}{\delta_m}{\LSeq{\mu}}{\MSeq{\mu}}\ket{\LSeq{\mu}+\delta_\ell, \MSeq{\mu}}\ket{\LTot, \MTot, \NTot}\nonumber
\\
&-\OpCoeff{0}{-\delta_m}{\LTot}{\NTot}\left(\delta_m \, \OpCoeff{\delta_\ell}{0}{\LSeq{\mu}}{\MSeq{\mu}+\delta_m}\OpCoeff{0}{\delta_m}{\LSeq{\mu}}{\MSeq{\mu}}+\delta_\ell \, \OpCoeff{\delta_\ell}{\delta_m}{\LSeq{\mu}}{\MSeq{\mu}}\OpCoeff{0}{0}{\LSeq{\mu}}{\MSeq{\mu}}\right) \nonumber
\\
&\times \ket{\LSeq{\mu}+\delta_\ell, \MSeq{\mu}+\delta_m}\ket{\LTot, \MTot, \NTot-\delta_m}\bigg]\otimes \ket{\Omega \notin \{L, \mu\}} 
\end{align}

\item These operators only change $\MSeq{\mu}$, in either a $\sigma$-preserving or $\sigma$-violating way. Note that in order to preserve $\sigma$, either $\MSeq{\nu}$ or $\NTot$ has to also be changed.
\begin{align}
\LSeqOp_\mu \cdot \crossvL{1}{1} \ket{\Omega}=&\, i \OpCoeff{0}{0}{\LSeq{\mu}}{\MSeq{\mu}}\ket{\Omega} +\frac{i}{2} \sum_{\delta_m=\pm} \delta_m \OpCoeff{0}{\delta_m}{\LSeq{\mu}}{\MSeq{\mu}}\ket{\LSeq{\mu}, \MSeq{\mu}+\delta_m} \nonumber 
\\
&\otimes \bigg\{ \OpCoeff{0}{-\delta_m}{\LTot}{\NTot}\ket{\LTot, \MTot, \NTot-\delta_m} \ket{\Omega \notin \{L, \mu\}} \nonumber
\\
&+\sum_{\nu\neq\mu}\OpCoeff{0}{-\delta_m}{\LSeq{\nu}}{\MSeq{\nu}}\ket{\LSeq{\nu},\MSeq{\nu}-\delta_m} \ket{\Omega \notin \{\mu, \nu\}} - \OpCoeff{0}{\delta_m}{\NSeq}{\sigma} \ket{\Omega \notin \{\mu\}} \bigg\}
\\[2ex]
\LSeqOp_{\mu}\cdot \LSeqOp_1 \ket{\Omega}=&\, \LSeqOp_{\mu}\cdot \LpTotOp \ket{\Omega}-\cas{\LSeq{\mu}}\ket{\Omega}-\underbrace{\LSeqOp_{\mu}\cdot \LSeqOp_2}_{\text{Eq.~}\ref{eq:L3dotL2}}-\sum_{\nu \neq \mu}\underbrace{\LSeqOp_{\mu}\cdot \LSeqOp_{\nu}}_{\text{Eq.~}\ref{eq:L3dotL4}}
\end{align}
where
\begin{align}
\LSeqOp_{\mu}\cdot \LpTotOp \ket{\Omega} =&\, -\bigg[\OpCoeff{0}{0}{\LSeq{\mu}}{\MSeq{\mu}}\OpCoeff{0}{0}{\LTot}{\NTot}\ket{\Omega} \nonumber
\\
&+\frac{1}{2} \sum_{\delta m = \pm} \OpCoeff{0}{\delta_m}{\LSeq{\mu}}{\MSeq{\mu}}\OpCoeff{0}{-\delta_m}{\LTot}{\NTot}\ket{\LSeq{\mu}, \MSeq{\mu}+\delta_m}\ket{\LTot, \MTot, \NTot-\delta_m}\ket{\Omega \notin \{L, \mu \}} \bigg] 
\end{align}
\end{itemize}

\item Operators that appear in $\boldsymbol{\CE}_{1 \zeta_1}\cdot \boldsymbol{\CE}_{2 \zeta_2}$
\begin{itemize}
\item These operators only change $\NSeq$
\begin{align}
\nSeqBody{1}\cdot \nSeqBody{2}\ket{\Omega}&=\sum_{\delta_n = \pm}\OpCoeff{\delta_n}{0}{\NSeq}{\sigma}\ket{\NSeq+\delta_n}\ket{\Omega \notin \{n\}}
\\
\nSeqBody{1}\cdot \crossvL{2}{2}&=-\frac{i}{2}\left[\sum_{\delta_n = \pm}\sum_{\delta_m=\pm}\delta_n \, \, \OpCoeff{\delta_n}{\delta_m}{\NSeq}{\sigma-\delta_m}\OpCoeff{0}{-\delta_m}{\NSeq}{\sigma}\ket{\NSeq+\delta_n}\right]\ket{\Omega \notin \{n\}}
\label{Eq.N1dN2}
\end{align}

\item These operators change $\NSeq$ and can also change $\sigma$, by changing either $\MSeq{\mu}$ or $\NTot$:
\begin{align}
\nSeqBody{2}\cdot\crossvL{1}{1}\ket{\Omega} =&\, \frac{i}{2}\sum_{\delta_n =\pm} \delta_n\, \ket{\NSeq+\delta_n} \otimes \sum_{\delta_m = \pm} \bigg[ \OpCoeff{\delta_n}{\delta_m}{\NSeq}{\sigma}\bigg\{\OpCoeff{0}{\delta_m}{\LTot}{\NTot}\ket{\LTot, \MTot, \NTot+\delta_m} \ket{\Omega \notin \{n, L\}} \nonumber 
\\
&+ \sum_\mu \OpCoeff{0}{\delta_m}{\LSeq{\mu}}{\MSeq{\mu}}\ket{\LSeq{\mu},\MSeq{\mu}+\delta_m} \ket{\Omega \notin \{n, \mu\}}  \bigg\} - \OpCoeff{\delta_n}{\delta_m}{\NSeq}{\sigma-\delta_m} \OpCoeff{0}{-\delta_m}{\NSeq}{\sigma} \ket{\Omega \notin \{n\}} \bigg]
\\[2ex]
\nSeqBody{2}\times \LSeqOp_{1}\ket{\Omega} =&\, \frac{1}{2}\sum_{\delta_n =\pm}\delta_n \ket{\NSeq+\delta_n}\sum_{\delta_m = \pm} \bigg[\delta_m \,\,\OpCoeff{\delta_n}{\delta_m}{\NSeq}{\sigma}\bigg\{\OpCoeff{0}{\delta_m}{\LTot}{\NTot}\ket{\LTot, \MTot, \NTot+\delta_m} \ket{\Omega \notin \{n, L\}} \nonumber 
\\
&+ \sum_\mu \OpCoeff{0}{\delta_m}{\LSeq{\mu}}{\MSeq{\mu}}\ket{\LSeq{\mu},\MSeq{\mu}+\delta_m}\ket{\Omega \notin \{n, \mu \}} \bigg\} - \OpCoeff{\delta_n}{0}{\NSeq}{\sigma} \OpCoeff{0}{0}{\NSeq}{\sigma}\ket{\Omega \notin \{ n\}} \bigg]
\end{align}

\item This operator does not change any quantum number
\begin{align}
\nSeqBody{1}\cdot \LSeqOp_{2}\ket{\Omega} = - \OpCoeff{0}{0}{n}{\sigma}\ket{\Omega}
\end{align}

\item This operator changes $\NSeq$ and can also change $\sigma$
\begin{align}
\hspace{-0.4in} \bigg[\crossvL{2}{2}\cdot \crossvL{1}{1}+ i \csc^2\thetaDef\, \nSeqBody{2}\cdot\crossvL{1}{1}\bigg]\ket{\Omega} =&\, \frac{1}{2}\sum_{\delta_n = \pm} \ket{\NSeq+\delta_n} 
\\
&\otimes \sum_{\delta_m = \pm}\bigg[\left(\OpCoeff{\delta_n}{0}{\NSeq}{\sigma+\delta_m}\OpCoeff{0}{\delta_m}{n}{\sigma}+\delta_n \delta_m\,\, \OpCoeff{\delta_n}{\delta_m}{\NSeq}{\sigma}\OpCoeff{0}{0}{n}{\sigma}\right) \nonumber
\\
&\times \bigg\{ \OpCoeff{0}{\delta_m}{\LTot}{\NTot}\ket{\LTot, \MTot, \NTot+\delta_m} \ket{\Omega \notin \{n, L\}} \nonumber
\\
&+\sum_{\mu}\OpCoeff{0}{\delta_m}{\LSeq{\mu}}{\MSeq{\mu}}\ket{\LSeq{\mu}, \MSeq{\mu}+\delta_m}\ket{\Omega \notin \{n, \mu\}} \bigg\} \nonumber
\\
&+\left(\delta_n \OpCoeff{\delta_n}{\delta_m}{\NSeq}{\sigma - \delta m}\OpCoeff{0}{-\delta_m}{\NSeq}{\sigma}-\cas{\NSeq}\OpCoeff{\delta_n}{0}{\NSeq}{\sigma}\right) \ket{\Omega \notin \{n\}} \bigg] \nonumber
\end{align}

\item Similar to the previous operator, this operator changes $\NSeq$ and can also change $\sigma$
\begin{align}
\bigg[\crossvL{2}{2}\cdot \LSeqOp_{1}+ i \csc^2\thetaDef\, \nSeqBody{2}\cdot \LSeqOp_1\bigg]\ket{\Omega} =&\, -\frac{i}{2}\sum_{\delta_n = \pm}\ket{\NSeq+\delta_n}\sum_{\delta_m = \pm}\bigg[\left(\delta_m\,\,\OpCoeff{\delta_n}{0}{\NSeq}{\sigma+\delta_m}\OpCoeff{0}{\delta_m}{n}{\sigma} + \delta_n \,\, \OpCoeff{\delta_n}{\delta_m}{\NSeq}{\sigma}\OpCoeff{0}{0}{n}{\sigma}\right) \nonumber
\\
&\times \bigg\{\OpCoeff{0}{\delta_m}{\LTot}{\NTot}\ket{\LTot, \MTot, \NTot+\delta_m}\ket{\Omega \notin \{n, L\}} \nonumber
\\
&+\sum_{\mu}\OpCoeff{0}{\delta_m}{\LSeq{\mu}}{\MSeq{\mu}}\ket{\LSeq{\mu}, \MSeq{\mu}+\delta_m}\ket{\Omega \notin \{n,\mu\}}\bigg\} \nonumber 
\\
&-\left(\delta_n \,\,\OpCoeff{0}{0}{n}{\sigma}\OpCoeff{\delta_n}{\delta_m}{\NSeq}{\sigma - m}\OpCoeff{0}{-\delta_m}{\NSeq}{\sigma}\right)\ket{\Omega \notin \{n\}}\bigg]
\end{align}
where the triple $\CC$ product arises from the relation that
\begin{align}
\left(-\LpTotOpComp{z}+ \sum_\mu \LSeqOpComp{z}\right)\ket{\Omega} &= \sigma \ket{\Omega}=\OpCoeff{0}{0}{\NSeq}{\sigma}\ket{\Omega}.
\end{align}

\item These operators can only change projected quantum numbers, in either a $\sigma$-preserving or $\sigma$-violating way
\begin{align}
\bigg[\LSeqOp_{2}\cdot \crossvL{1}{1}- i \cot\thetaDef\csc\thetaDef\, \nSeqBody{2}\cdot \LSeqOp_1\bigg]\ket{\Omega}
=&\,\frac{i}{2}\sum_{\delta_m = \pm}\bigg[\delta_m\,\,\OpCoeff{0}{\delta_m}{\NSeq}{\sigma} \bigg\{\OpCoeff{0}{\delta_m}{\LTot}{\NTot}\ket{\LTot, \MTot, \NTot+\delta_m}\ket{\Omega \notin \{L\}} \nonumber
\\
&+\sum_{\mu}\OpCoeff{0}{\delta_m}{\LSeq{\mu}}{\MSeq{\mu}}\ket{\LSeq, \MSeq{\mu}+\delta_m}\ket{\Omega \notin \{\mu\}}\bigg\} -\OpCoeff{0}{0}{n}{\sigma}\ket{\Omega}\bigg]
\\[2ex]
\bigg[\LSeqOp_{2}\cdot \LSeqOp_{1}+ i \cot\thetaDef\csc\thetaDef\, \nSeqBody{2}\cdot \crossvL{1}{1}\bigg]\ket{\Omega} =&\, \frac{1}{2}\sum_{\delta_m = \pm}\bigg[\OpCoeff{0}{\delta_m}{\NSeq}{\sigma} \bigg\{\OpCoeff{0}{\delta_m}{\LTot}{\NTot}\ket{\LTot, \MTot, \NTot+\delta_m}\ket{\Omega \notin \{L\}}\nonumber 
\\
&+\sum_{\mu}\OpCoeff{0}{\delta_m}{\LSeq{\mu}}{\MSeq{\mu}}\ket{\LSeq{\mu}, \MSeq{\mu}+\delta_m}\ket{\Omega \notin \{\mu\}}\bigg\} \nonumber
\\
&+\left(\left(\OpCoeff{0}{0}{n}{\sigma}\right)^2-\cas{\NSeq}\right)\ket{\Omega}\bigg]
\end{align}

\end{itemize}
\end{itemize}

These are all the operators that appear in all possible electric bilinears. The quantum numbers that can be changed by each bilinear, or even each term, can be intuited by the occurrence of $\nSeqBody{\kappa}$ and $\LSeqOp_\kappa$.

\subsection{Operators appearing in Magnetic Loops}
The operators appearing in the magnetic loop variables must also be translated from the magnetic basis into the mixed-basis. We follow the same procedure as above, and show the state that results from acting the operators on the basis state
\begin{align}
\ket{\Omega} &=\left(\ket{\NSeq}\otimes\left(\prod_{\nu=3}^{\NL} \ket{\LSeq{\nu}, \MSeq{\nu}}\right)\otimes\ket{ \LTot, \MTot, \NTot}\right).
\end{align}
The only operators that appear in magnetic loop variables that can change the mixed-basis quantum numbers are dot products and triple scalar products of the directions $\mathbf{n}_\kappa$.
The results for all possible expressions are given below:
\begin{itemize}
\item Dot Products of Unit Vectors
\begin{align}
\nSeqBody{\kappa}\cdot \nSeqBody{\kappa'}\ket{\Omega}:
\begin{cases}
\text{Eq.~}\eqref{Eq.N1dN2} \qquad &\kappa = 1, \kappa' = 2 \\
\text{Eq.~}\eqref{Eq.N1dN3} \qquad &\kappa = 1, \kappa' = \mu\\
\text{Eq.~}\eqref{Eq.N2dN3} \qquad &\kappa = 2, \kappa' = \mu\\
\text{Eq.~}\eqref{Eq.N3dN4} \qquad &\kappa = \mu, \kappa' = \nu
\end{cases}
\end{align}

\item Triple Products of Unit Vectors
\begin{align}
\nSeqBody{1}\cdot \left(\nSeqBody{2} \times \nSeqBody{\mu}\right)\ket{\Omega} =&\, -\frac{i}{2} \sum_{\substack{\delta_n = \pm\\\delta_\ell = \pm}}\sum_{\delta_m = \pm} \delta_m \delta_n \delta_\ell \, \, \OpCoeff{\delta_n}{\delta_m}{\NSeq}{\sigma}\OpCoeff{\delta_\ell}{\delta_m}{\LSeq{\mu}}{\MSeq{\mu}}\ket{\NSeq+\delta_n}\ket{\LSeq{\mu}+\delta_\ell, \MSeq{\mu}+\delta_m} \otimes \ket{\Omega \notin \{n, \mu\}} 
\end{align}

\begin{align}
\nSeqBody{1}\cdot \left(\nSeqBody{\mu} \times \nSeqBody{\nu}\right)\ket{\Omega} =&\, -\frac{i}{2}\sum_{\substack{\delta_{\LSeq{\mu}}=\pm\\\delta_{\LSeq{\nu}}=\pm}}\sum_{\delta_m = \pm} \delta_m \delta_{\LSeq{\mu}}\delta_{\LSeq{\nu}}\,\,\OpCoeff{\delta_\LSeq{\mu}}{\delta_m}{\LSeq{\mu}}{\MSeq{\mu}}\OpCoeff{\delta_\LSeq{\nu}}{-\delta_m}{\LSeq{\nu}}{\MSeq{\nu}} \nonumber
\\
&\times \ket{\LSeq{\mu}+\delta_\LSeq{\mu}, \MSeq{\mu}+\delta_m}\ket{\LSeq{\nu}+\delta_\LSeq{\nu}, \MSeq{\nu}-\delta_m}  \ket{\Omega \notin \{\mu, \nu\}}
\end{align}

\begin{align}
\hspace{-0.3in} \nSeqBody{2}\cdot \left(\nSeqBody{\mu}\times \nSeqBody{\nu}\right)\ket{\Omega}=&\, \cos \thetaDef\, \nSeqBody{1}\cdot \left(\nSeqBody{\mu}\times \nSeqBody{\nu}\right)-\cos \thetaSeq{\mu}\, \nSeqBody{1}\cdot \left(\nSeqBody{2}\times \nSeqBody{\nu}\right)+\cos \thetaSeq{\nu}\, \nSeqBody{1}\cdot \left(\nSeqBody{2}\times \nSeqBody{\mu}\right) \nonumber 
\\
=&\, -\frac{i}{2}\sum_{\substack{\delta_n = \pm\\\delta_\LSeq{\mu}=\pm\\\delta_\LSeq{\nu}=\pm}}\sum_{\delta_m=\pm} \delta_m\,\, \bigg\{\delta_\LSeq{\mu}\delta_\LSeq{\nu}\OpCoeff{\delta_\LSeq{\mu}}{\delta_m}{\LSeq{\mu}}{\MSeq{\mu}}\OpCoeff{\delta_\LSeq{\nu}}{-\delta_m}{\LSeq{\nu}}{\MSeq{\nu}}\OpCoeff{\delta_n}{0}{n}{\sigma}\ket{\NSeq+\delta_n}\ket{\LSeq{\mu}+\delta_\LSeq{\mu}, \MSeq{\mu}+\delta_m}\ket{\LSeq{\nu}+\delta_\LSeq{\nu}, \MSeq{\nu}-\delta_m} \nonumber 
\\
&+\delta_n\,\,\OpCoeff{\delta_n}{\delta_m}{\NSeq}{\sigma}\ket{\NSeq+\delta_n}\bigg[\delta_\LSeq{\mu}\OpCoeff{\delta_\LSeq{\mu}}{\delta_m}{\LSeq{\mu}}{\MSeq{\mu}}\OpCoeff{\delta_\LSeq{\nu}}{0}{\LSeq{\nu}}{\MSeq{\nu}}\ket{\LSeq{\mu}+\delta_\LSeq{\mu}, \MSeq{\mu}+\delta_m}\ket{\LSeq{\nu}+\delta_\LSeq{\nu}, \MSeq{\nu}}\nonumber 
\\
&-\delta_\LSeq{\nu}\OpCoeff{\delta_\LSeq{\nu}}{\delta_m}{\LSeq{\nu}}{\MSeq{\nu}}\OpCoeff{\delta_\LSeq{\mu}}{0}{\LSeq{\mu}}{\MSeq{\mu}}\ket{\LSeq{\nu}+\delta_\LSeq{\nu}, \MSeq{\nu}+\delta_m}\ket{\LSeq{\mu}+\delta_\LSeq{\mu}, \MSeq{\mu}}\bigg]\bigg\} \otimes \ket{\Omega \notin \{n, \mu, \nu\}}
\end{align}
For the last triple product, it will be easiest to first break it apart. In particular,
\begin{align}
\nSeqBody{\mu}&\cdot \left(\nSeqBody{\nu}\times \nSeqBody{\lambda}\right)=\cos \thetaSeq{\mu}\, \nSeqBody{1}\cdot \left(\nSeqBody{\nu}\times \nSeqBody{\lambda}\right)+\cos \thetaSeq{\nu}\, \nSeqBody{1}\cdot \left(\nSeqBody{\lambda}\times \nSeqBody{\mu}\right)+\cos \thetaSeq{\lambda}\, \nSeqBody{1}\cdot \left(\nSeqBody{\mu}\times \nSeqBody{\nu}\right)
\end{align}
where
\begin{align}
\cos \thetaSeq{\lambda}\, \nSeqBody{1}\cdot \left(\nSeqBody{\mu}\times \nSeqBody{\nu}\right)\ket{\LSeq{\mu}, \MSeq{\mu},\LSeq{\nu}, \MSeq{\nu},\LSeq{\lambda}, \MSeq{\lambda}}=&\,-\frac{i}{2}\sum_{\substack{\delta_\LSeq{\mu}=\pm\\\delta_\LSeq{\nu}=\pm\\\delta_\LSeq{\lambda}=\pm}}\sum_{\delta_m=\pm} \delta_m \delta_\LSeq{\mu}\delta_\LSeq{\nu}\bigg\{ \OpCoeff{\delta_\LSeq{\mu}}{\delta_m}{\LSeq{\mu}}{\MSeq{\mu}}\OpCoeff{\delta_\LSeq{\nu}}{-\delta_m}{\LSeq{\nu}}{\MSeq{\nu}}\OpCoeff{\delta_\LSeq{\lambda}}{0}{\LSeq{\lambda}}{\MSeq{\lambda}} \nonumber
\\
&\times \ket{\LSeq{\mu}+\delta_\LSeq{\mu}, \MSeq{\mu}+\delta_m,\LSeq{\nu}+\delta_\LSeq{\nu}, \MSeq{\nu}-\delta_m,\LSeq{\lambda}+\delta_\LSeq{\lambda}, \MSeq{\lambda}}\bigg\}
\end{align}
\end{itemize}
As expected, none of these operators can affect the total angular momentum (or total color charge) quantum numbers.
\end{document}